\documentclass{elsart}
\usepackage{amsfonts}\usepackage{amsbsy}
\usepackage{mathrsfs}
\usepackage{graphicx}
\def\cal{\mathcal}

\makeatletter
\def\ps@copyright{\let\@mkboth\@gobbletwo
  \def\@oddhead{}%
  \let\@evenhead\@oddhead
  \let\@evenfoot\@oddfoot
}

\def\lsim{\stackrel{\scriptstyle <}{\phantom{}_{\sim}}}
\def\gsim{\stackrel{\scriptstyle >}{\phantom{}_{\sim}}}
\def\be{\begin{eqnarray}}
\def\ee{\end{eqnarray}}
\def\om{\omega}
\def\re{{\rm Re}}
\def\im{{\rm Im}}
\usepackage{graphicx}
\usepackage{colordvi,feynmp}
\makeatletter
\newdimen\@bls
\@bls=\baselineskip
\newbox\slashbox \setbox\slashbox=\hbox{$/$}
\newbox\Slashbox \setbox\Slashbox=\hbox{\large$/$}
\def\pFMslash#1{\setbox\@tempboxa=\hbox{$#1$}
  \@tempdima=0.5\wd\slashbox \advance\@tempdima 0.5\wd\@tempboxa
  \copy\slashbox \kern-\@tempdima \box\@tempboxa}
\def\pFMSlash#1{\setbox\@tempboxa=\hbox{$#1$}
  \@tempdima=0.5\wd\Slashbox \advance\@tempdima 0.5\wd\@tempboxa
  \copy\Slashbox \kern-\@tempdima \box\@tempboxa}
\def\FMslash{\protect\pFMslash}
  
\makeatother
\begin{document}
\begin{fmffile}{dblur}
\begin{frontmatter}


\title{Hadron Liquid with a Small Baryon Chemical Potential
at Finite Temperature}

\author{D. N.~Voskresensky$^{1,2}$}
\maketitle

\noindent                        
$^{1}${\it\small Gesellschaft f\"ur Schwerionenforschung mbH, Planckstr. 1,
64291 Darmstadt, Germany}
\\
$^{2}${\it\small Moscow Institute for Physics and Engineering, 
Kashirskoe sh. 31, Moscow 115409, Russia}

\begin{abstract}
We discuss general properties of
a system of  
heavy fermions (including antiparticles) interacting with  rather
light  bosons.  First, we consider one diagram of $\Phi$.
The fermion chemical potential
is assumed to be small, $\mu_f \lsim T$. Already for the low temperature,  
$T\ll \mbox{min} (T_{\rm bl.f}, m_{b})$,
the fermion mass shell proves to be partially blurred
due to multiple fermion rescatterings on virtual bosons,
$m_{b}$ is the boson mass, $T_{\rm bl.f}$ $(\ll m_f)$ 
is the typical temperature corresponding to a
complete blurring of
the gap between fermion-antifermion continua,  $m_f$ is the
fermion mass.
As the result,
the ratio of the number
of fermion-antifermion pairs to 
the number provided by the ordinary Boltzmann distribution becomes 
larger than unity
($R_N >1$). 
For  
$T\gsim m_{b}^* (T)$ (hot hadron liquid, blurred boson continuum),
$m_{b}^* (T)$ is the effective boson mass,   the abundance of all particles
dramatically increases.
Bosons behave
as quasi-static impurities, on which heavy fermions undergo multiple
rescatterings. 
The soft thermal loop approximation
solves the problem.
The effective fermion mass $m_f^* (T)$
decreases with the temperature increase. 
 For $T\gsim T_{\rm bl.f}$ 
fermions are essentially relativistic particles.
Due to the
interaction of the
boson with
fermion-antifermion pairs, $m_{b}^* (T)$ decreases
leading to the possibility of the
``hot Bose condensation'' for $T>T_{\rm cb}$. The phase transition might be of the
second order or of 
the first order depending on the species under consideration.
We study  in detail properties of the system of  spin $1/2$
heavy fermions interacting
with  substantially lighter scalar neutral bosons (e.g.,
$N\sigma $ system). 
Correlation effects of 
higher order diagrams 
of $\Phi$ 
are evaluated resulting in a suppression
of vertices for $T\gsim m_b^* (T)$.
The abundance of high-lying baryon resonances 
proves to be  of the same 
order, as the nucleon-antinucleon abundance, or might be even  
higher for some species. Further 
we discuss the system of heavy
fermions  interacting with more light vector bosons
(e.g., $N\omega$ and $N\rho$)
and then, with pseudo-scalar bosons (e.g., $N\pi$).
For the fermion -- vector boson system 
correlation effects are incorporated by keeping the Ward identity. 
In case of the 
fermion -- pseudo-scalar boson
system correlation effects are rather small.
Finally, we allow for all interactions. We estimate $R_N \sim 1.5$ for $T\sim 
m_{\pi}/2$;
$T_{\rm bl.f}$ proves to be near $T_{\rm cb}$; both values are in the vicinity of
the pion mass $m_{\pi}$.
\end{abstract}

\end{frontmatter}

PACS number(s): 25.65.+f, 25.75.Nq, 25.75.-q, 25.70.Lm

keywords: hadron matter, medium effects, temperature, heavy ion collisions

\vspace*{5mm}

\section{~Introduction} 
In heavy ion collisions
at ultra-high 
energies a huge
amount of secondary particles is produced, cf.  \cite{EXP}.
At 
$\sqrt{s}\simeq 130$~GeV at RHIC the ratio of produced negative
pions to protons is
 $\pi^- /p \simeq 9.5$, the ratio of negative kaons to negative
pions is $K^- /\pi^- \simeq 0.15$,
the ratio of antiprotons to protons is $\bar{p}/p \simeq 0.71$, and
at $\sqrt{s}\simeq 200$~GeV at RHIC $\pi^- /p \simeq 9.3$, 
$K^- /\pi^- \simeq 0.16$, $\bar{p}/p \simeq 0.74$. The set of
available data is summarized in  tables in \cite{RL}.
 
Heavy ion collisions  at RHIC create a non-equilibrium
fireball consisting of strongly interacting secondary particles. 
Due to multiple collisions of
particles the system rapidly
reaches a quasi-equilibrium. Then, the
fireball is characterized by a temperature $T(t, \vec{r})$ in the center of
mass frame and by a velocity. 
It is commonly believed that at an initial stage a produced
fireball is constructed of quarks and gluons. Then during the expansion of the
fireball 
there 
occurs a process of hadronization. 
The hadronized fireball contains
mainly pions and also heavier particles like kaons, $\sigma$
($f_0 (600~\mbox{MeV})$), $\rho (769~\mbox{MeV})$, 
$\omega (782~\mbox{MeV})$ and other mesons, and nucleons, $\Lambda (1116~
\mbox{MeV})$,
$\Sigma (1192~\mbox{MeV})$ hyperons, $\Delta (1232~\mbox{MeV})$ 
isobars and heavier baryon
and antibaryon resonances in a smaller amount. At the so called
freeze-out stage
resonances decouple, their momentum distributions freeze and with these 
distributions
one observes particles at infinity.
Statistical equilibrium ideal resonance
gas model, cf. \cite{BR} and refs therein, 
and somewhat different models, e.g. \cite{RL}, 
that introduces effective
non-equilibrium
occupancy parameters, being applied to the fireball break up stage,  allow
to fit well observed particle ratios.  
Typical values of the non-strange
baryon chemical potential used to fit mentioned RHIC data vary 
in the range  $\mu_{\rm bar} \simeq 20\div 40$~MeV, 
and the break up temperature varies in the interval 
$T\simeq 140\div 170$~MeV. At LHC 
($\sqrt{s}\simeq 5500$~GeV) one expects a
tiny baryon chemical potential ($\sim 1$~MeV), cf. \cite{BR}. 
This system can already
be called  {\em the hot
vacuum}.

Since  
secondary particle production rates are very large inside the fireball, 
one may expect an importance of
effects, which are beyond the scope of the 
usually used  simple approximation of elastic collisions of free
particles and even  of a more involved
quasiparticle approximation. 
To describe a system
of strongly coupled resonances one needs to develop
an approach including both
particle widths
and the dispersion, taking into account particle feed-back effects. Although 
the kinetic
description of resonances within the self-consistent so called 
$\Phi$-derivable scheme has been constructed (cf. refs \cite{IKV}
and refs therein) it looks very complex. Therefore, in order to understand most
important signatures of processes one needs further simplifications.

The problem  of the behavior of the heated pion-nucleon vacuum (zero total
baryon charge)
was risen by Dyugaev in
ref. \cite{D93}. 
A very intuitive consideration was
sketched in  an analogy with the description of the electron-phonon
interaction in doped 
semiconductors. Even at
zero temperature the tail 
of the electron wave function  penetrates deeply into the band gap
due to multiple electron-phonon collisions \cite{E}. 
Dyugaev conjectured  that in the nuclear problem 
nucleons may play the same role as electrons, and
pions, as phonons.
Within the given analogy,
in order to construct a qualitative picture of the phenomenon,
ref. \cite{D93} considered 
nucleons and antinucleons, as
non-relativistic particles interacting with pions by
the non-relativistic pion-nucleon coupling,
$f_{\pi N}\bar{\psi}\vec{\sigma}
\vec{q}\cdot \vec{\tau}\vec{\pi}\psi$. 
Ref.  \cite{D93} also conjectured 
existence of the ending temperature for the hadron world,  
above which  the system can't be
anymore in the hadron state due to anomalous production of
fermion-antifermion pairs (cf. Hagedorn picture, its recent application 
to RHIC energies see in 
\cite{BB03}).
This critical temperature  ($T_{\rm dec}$) 
was estimated to be   $\lsim 250$~MeV.

In this paper we 
focus on the description of the heated quasi-equilibrium hadron
liquid having a small or even zero  total baryon charge (the hadron
vacuum).
More precisely we will exploit that the net 
baryon density is much less than the 
density of produced baryons
(and antibaryons). This allows us to neglect particle-hole
effects compared to particle-antiparticle effects, that essentially simplifies
the consideration. The limit is safely fulfilled for $\mu_f \lsim T$,
where $\mu_f$ is the fermion chemical potential.
We develop a general relativistic approach and match it to the
non-relativistic one. We use the term ``hadron liquid'' rather than ``hadron
gas'' to stress crucial role of strong coupling. Besides we will see that
the pion-nucleon-antinucleon attraction is essentially enhanced in the p-wave.  
Thereby a short range correlation 
naturally comes into play.

To better understand the physics of the phenomenon 
we first consider properties of an idealized hadron
liquid
consisting of strongly interacting fermions of one kind and bosons of one kind.
It is assumed that fermions are essentially heavier particles than bosons.
We argue for the following qualitative picture.

There are several temperature regimes.
The regime $T\ll {\mbox{min}}\{m_b^2 /m_f , T_{\rm bl.f}\}$ corresponds to a {\em slightly
heated 
hadron liquid} (or the hadron vacuum, if the fermion chemical potential
$\mu_f =0$). 
$T_{\rm bl.f}$ is the temperature, at which 
the gap between fermion-antifermion continua becomes completely blurred. 
Typically $T_{\rm bl.f}(g)$
is of the order of $\sim m_{\pi}$ for relevant values of the fermion-boson 
coupling constant $g$, 
$m_{\pi}=140$~MeV is the pion mass.
Bosons are almost free particles in this temperature regime. 
Fermion distributions begin to deviate
from Boltzmann distributions due to
multiple collisions of each fermion on 
bosons. This deviation increases with the temperature
increase. 

If there exists a
temperature interval  
$m_{b}^{2}/m_f \lsim T\ll \mbox{min} \{T_{\rm bl.f}, m_{b}\}$ 
({\em a warm hadron liquid, partially blurred fermion continuum}),  
then the fermion mass shell is  already partially blurred
due to multiple rescatterings of the fermion on bosons. 
The quasiparticle approximation for fermions fails, if the fermion-boson 
coupling constant $g$ is rather large (e.g., $g\sim 10$ 
for $\sigma$, $\om$ and $\rho$ meson -- nucleon ($N$) interaction).
As the result, 
fermion distributions become  essentially enhanced
compared to the ordinary Boltzmann distribution. 
For realistic hadron parameters
the regime of {\em{a warm hadron liquid}}
can be realized 
only for pions, not for $\sigma$, $\om$ and $\rho$ due to their large masses. 
However the enhancement
is not too strong for pions, since $g\sim 1$ (rather than 
$\gg 1$) in the latter case.

With further 
increase of the
temperature, hadron effective masses substantially decrease. 
For  $m_f \gg T\gsim m_b^{*} (T)$ ({\em hot hadron liquid, blurred
boson continuum}), where
$m_b^{*}(T, g)$ is the effective boson mass depending on $g$, 
heavy  rather rapid fermions abundantly
produce effectively
less massive and slower virtual (off shell) bosons (boson tadpole diagram)
and undergo multiple rescatterings on them, as on
quasi-static impurities. 
Due to the width effect (from multiple quasielastic rescatterings)
the fermion  propagator completely
looses the former quasiparticle pole shape it had in a dilute medium.
For $m_f \gg T\gsim T_{\rm bl.f}$ 
fermions become essentially relativistic particles.
The hot hadron liquid
comes to the regime of {\em the blurred fermion continuum.}
The fermion sub-system represents then {\em a rather dense packing of
fermion-antifermion pairs.} 
Since also $T>m_b^* (T)$, this  state is the state 
of {\em the blurred hadron continuum} 
(blurred
continua for both boson and fermion sub-systems).
The fermion-antifermion density,
$\rho_{f,\bar{f}}$ grows exponentially with the temperature
in a wide temperature interval. Bosons rescatter on fermion-antifermion pairs
(fermion-antifermion loop diagram) and 
due to that decrease their effective masses. 
At a temperature $T>T_{\rm cb}$, 
the effective  scalar boson mass may vanish
and a  {\em hot Bose condensation}
(HBC) may set in by the second order phase transition. We call it HBC, since 
the condensate appears for the temperature
larger than a critical temperature. 
For vector and pseudo-scalar bosons (if the latter interact with fermions
via pseudo-vector
coupling) the HBC may arise by the first order phase transition
at finite value of  $m^*_b (T_{\rm cb})$. Moreover, for scalar and vector bosons
HBC occurs in the s-wave state, whereas for pseudo-scalar bosons with
the pseudo-vector
coupling to fermions the HBC may arise in the p-wave state.
For realistic values of hadron parameters, the problem of the
determination of 
$T_{\rm bl.f}$, $T_{\rm cb}$ and $m^*_b (T)$ is the coupled-channel problem.
As the result of its solution,
$T_{\rm cb}$ proves to be close to $T_{\rm bl.f}$.
In spite of large values
of bare masses of $\sigma$, $\om$, $\rho$ mesons and nucleons,
numerically both values
$T_{\rm bl.f}$ and $T_{\rm cb}$ prove to be in the vicinity of 
the pion mass $m_{\pi}$.
At such a temperature the resulting density of fermion-antifermion pairs is 
estimated as $\lsim \rho_0$, where $\rho_0 \simeq 0.5m_{\pi}^3$ 
is the nuclear saturation density, $\hbar =c=1$. 

With subsequent increase of the temperature, for $T>T_{\rm bl.f}$
the baryon-antibaryon density continues to increase, then it
reaches a maximum (the value $\rho^{max}$ 
may several times
exceed $\rho_0$) 
and then may even begin to drop down. 

The strange particle production, as well as the production of other baryon
resonances,  are significantly enhanced
with the increase of the temperature. It is known that at high
energies of 
heavy ion collisions the experimentally observed
kaon to pion ratio
becomes energy independent. It is usually associated
with the quark deconfinement, cf. \cite{Gad,BR}.
However such a behavior can be also
naturally explained
within the pure hadron picture. 
In this sense one may speak about {\em{ a
quark-hadron duality}}: 
observable can be explained in terms of both the quark-gluon
degrees of freedom, cf. \cite{AF}, and only hadron degrees of freedom
\cite{R02}.
Taking into account multiple scattering effects,
the number of hadron degrees of freedom is significantly enhanced 
simulating the
same effects, as from deconfined quarks.

In reality hadron and quark-gluon degrees of freedom may
interact. Incorporating  
the quark structure of hadrons and the possibility of
hadron states in the quark matter,
one may expect
strong thermal fluctuation effects of the
quark-gluon origin in the hadron phase, and strong fluctuation effects 
of the hadron origin in the quark-qluon state. 
\footnote{This might be in an analogy
to that was recently found for the
phase transition to the di-quark condensate state \cite{V03}. The fluctuation 
region might be very broad there.} Thereby, instead of a sharp first or second 
order hadron-quark phase transition
one may expect the existence of a broad region of 
{\em{a hadron-quark continuity}},
cf. arguments for the crossover 
from lattice simulations \cite{Fodor}.
Thus,
more likely, at such conditions
the system state represents {\em a strongly correlated boiled
hadron-quark-gluon
porridge (HQGP)} rather than the pure hadron or pure quark-gluon state. 
The pure quark-gluon phase probably occurs at a higher temperature,
$T\gsim (1.5\div 2) m_{\pi}$, and a pure hadron phase, at 
$T\lsim (0.7 \div 1) m_{\pi}$. Below, discussing a high temperature regime  
we artificially disregard the quark-gluon effects
postponing their study to the future work. 

In the low
temperature regime, $T\ll \mbox{min}(m_b , T_{\rm bl.f})$,
fermion energies of our interest are
$p_0-m_f \simeq -m_b -O (m_f T/m_b ,\sqrt{m_f T})$, virtual 
boson energies are $q_0 \simeq  
m_b +O (m_f T/m_b ,\sqrt{m_f T})$, and fermion and
virtual boson momenta are of the same order
$|\vec{p}|\sim |\vec{q}|\sim \sqrt{2m_f T}$.
In the high temperature regime, $T\gsim m_b^{*}(T)$, 
the quantity $J$, which we
further call {\em the intensity of the multiple scattering}, 
is much larger than the 
temperature squared,
if the coupling
constant $g$ is rather high (e.g., $g \simeq 10$ for the
$N\sigma$ interaction). 
Then, typical departures of fermion energies
from the mass shell are substantially higher than those for bosons. 
Typical fermion momenta increase with the temperature
from non-relativistic values ($|\vec{p}|\sim \sqrt{2Tm_f }$) to relativistic
ones ($|\vec{p}|\sim m_f$). These values
are significantly higher 
than typical 
boson momenta $|\vec{q}|\sim \{\sqrt{2Tm_b^{*}(T)}, T\}$. 
It allows one to consider fermions, as hard particles, and bosons, as soft
ones, that greatly simplifies the analysis and, actually,  solves the problem. 
We call such an approximation {\em{the soft 
thermal loop}} (STL) approximation.

We argue for a huge stopping power 
in the course of  highly energetic heavy ion collisions
due to mentioned above multiple collisions. After  a short 
non-equilibrium stage the system continues to
live rather long at a quasi-equilibrium in the center of mass frame
undergoing a slow expansion into vacuum. Due to a large hadron density
and a softening of the vector meson spectrum the dilepton production is
expected to be enhanced, that can be measured.
Distributions of particles radiating at the break up stage of the fireball
are also enhanced compared to Boltzmann distributions at given
temperature,
since at least a part of a large number of virtual particle degrees of freedom
concentrated in the fireball before its break up can be transformed into 
distributions of particles measured at infinity (however the 
value of the enhancement depends
on the dynamical mechanism of the break up, cf. \cite{VS89,MSTV}). 

The paper is organized as follows.
In sect.  \ref{Approach}
we introduce a  $\Phi$ approach of Baym \cite{Baym}
and its application
to the description of the system of
coupled fermions and bosons. Technical details of the formalism: relations 
between Green
functions and self-energies and the scheme of the particle-antiparticle
separation
are deferred to the Appendix
\ref{Formalism}.
Sects \ref{Interaction}
and \ref{neutral bosons} are key sections of our study.
In
sect. \ref{Interaction} we construct the description of the 
coupled fermion -- boson system
within the simplest diagram of $\Phi$-functional. The main approximation we
use is that fermions are assumed to be effectively
significantly heavier than relevant 
bosons. We first focus our discussion on 
the low temperature limit $T\ll (m_{b}, T_{\rm bl.f})$ 
and then on a high temperature limit
$m_f\gg T\gsim m_{b}^{*}(T)$.
In sect. \ref{neutral bosons} 
we study an example of a 
fermion -- boson system
coupled by the Yukawa
(spin $\frac{1}{2}$-fermion -- scalar neutral boson -- fermion) interaction.
We include
higher
baryon resonances into consideration and evaluate correlation
effects. 
The scheme is then applied to other interactions,
between fermions and vector bosons (sect. \ref{vector}), and fermions and
pseudo-scalar  bosons coupled by the  pseudo-vector 
coupling (sect. \ref{pseudo-vector}). 
In sect. \ref{Porridge} we schematically discuss the behavior of the state of
{\em the hadron porridge,}
when all interactions between different particle species are included.
In Appendix 
\ref{non-rel}
we present relevant non-relativistic limit  formulas 
for heavy fermions.

Throughout the paper we use units $\hbar =c=1$ and the temperature is measured
in energetic units.

\section{$\Phi$-derivable Approach}\label{Approach}
\subsection{Dyson equation}
Let us consider a system of  interacting fermions and bosons. 
It is described by a coupled channel system of Dyson equations
for the fermion ($f$) and boson ($b$) Green functions
\be\label{Dyson}
\widehat{G}_i =\widehat{G}_i^0 +\widehat{G}_i^0 \widehat{\Sigma}_i
\widehat{G}_i, \,\,\, i=\{f,b\}. 
\ee
Here $\widehat{G}_i^0$ are free fermion and boson Green functions,
$\widehat{G}_i$ are full  Green functions,
and $\widehat{\Sigma}_i$ are fermion and
boson self-energies. All quantities are operators in the spin space. 
For
the description of an arbitrary non-equilibrium  system all the values in 
(\ref{Dyson})
are expressed in terms of the non-equilibrium diagram technique.
For that aim one may use contour, or matrix $\{-,+\}$ notations, cf. 
\cite{LP81,IKV}. For the further convenience we prefer the latter. 
For the sake of brevity we will often suppress
tensor indices and sometimes sign $\{-,+\}$ indices
using symbolic hat-operator notation.

\subsection{$\Phi$ functional and self-energies}

As it is known, perturbation theory  fails to describe collective
phenomena. On the other hand,  coupled Dyson equations (\ref{Dyson}) 
can't be solved exactly
and approximation schemes are required. Among different approximation 
approaches
the self-consistent $\Phi$-derivable method seems to be 
promising. It
keeps exact conservation laws 
and exact sum rules. For the quark-gluon plasma it provides the quantitative
prediction of thermodynamic characteristics, 
which match lattice results down to
3 times deconfinement critical temperature,
$T\sim 3T_{\rm dec}$, cf. \cite{BJ}.

Assume that fermions are coupled to bosons with the help of a
two-fermion -- one-boson interaction.
Then,  the $\Phi$-functional is 
given by the series
of diagrams
\be\label{phi}
\includegraphics[width=7cm,clip=true]{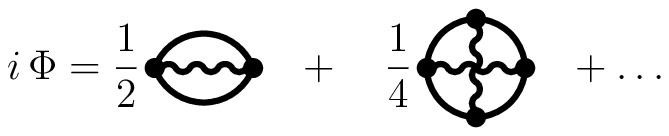}
\ee
where the bold solid line corresponds to the fermion/antifermion full Green
function and the bold wavy line, to the boson/antiboson full Green
function, small dots denote free vertices. 

Fermion and boson self-energies are obtained by the variation
of the functional $\Phi$ over the corresponding Green function
(the cut of the line
in (\ref{phi}))
\be\label{sigm-f}
-i\widehat{\Sigma}_{f,b} =\mp \frac{\delta i\Phi}{\delta (i\widehat{G}_{f,b})}\times
\left\{
\begin{array}{ll}
2\quad&\mbox{for neutral bosons,}\\[1.5mm]
1\quad&\mbox{for fermions and charged bosons.}
\end{array}\right.
\ee
The fermion self-energy is determined by the diagram
\be\label{selfzfex}
\includegraphics[width=2cm,clip=true]{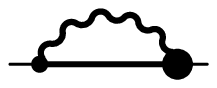}
\ee
and the boson  self-energy is given by the diagram
\be\label{selfzbex}
\includegraphics[width=2cm,clip=true]{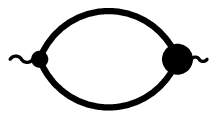}
\ee
The fat dot in (\ref{selfzfex}), (\ref{selfzbex}) symbolizes the full vertex.
In case of the  $\Phi$-derivable approximation, that deals with  finite number of
diagrams in (\ref{phi}), the fat vertex symbolizes the full vertex related to
the given $\Phi$.

\subsection{The simplest $\Phi$-diagram}

Let us restrict ourselves by the consideration of the simplest $\Phi$ 
(the first
diagram (\ref{phi})). 
Then the fermion self-energy takes the form
\be\label{selfzf}
\includegraphics[width=2cm,clip=true]{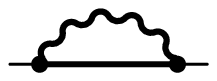}
\ee
and the boson self-energy reads 
\be\label{selfzb}
\includegraphics[width=2cm,clip=true]{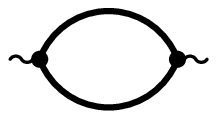}
\ee
All the multi-particle rescattering processes
\be\label{ladpr}
\includegraphics[width=7cm,clip=true]{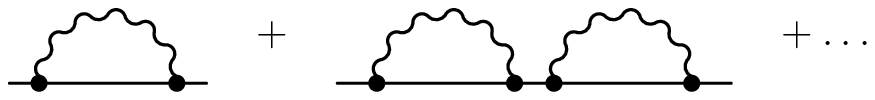}
\ee
are then included, whereas processes with the crossing of boson lines
(correlation effects) like
\be\label{cr}
\includegraphics[width=3cm,clip=true]{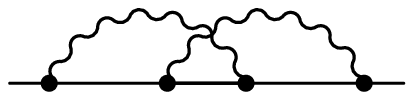}
\ee
are not incorporated. 

Using diagrammatic rules and relations of Appendix \ref{Formalism}
between ``$\{-,+\}$'' and retarded (``$R$'') and advanced (``$A$'') Green
functions (\ref{rel}), for the self-energy 
we find 
\be\label{sigm-R} 
-i\widehat{\Sigma}_f^{R} (p)&=&\int \widehat{V}_0 (q)
\widehat{G}_f^{-+} (p+q)
\widehat{V}_0 (-q)\widehat{G}_b^{R}(q)\frac{d^4 q}{(2\pi)^4}\nonumber \\
&+&\int \widehat{V}_0 (q)\widehat{G}_f^{R} (p+q)
\widehat{V}_0 (-q)\widehat{G}_b^{-+}(q)\frac{d^4 q}{(2\pi)^4}\nonumber \\
&-&\int \widehat{V}_0 (q)i\widehat{G}_f^{-+} (p+q)
\widehat{V}_0 (-q) 2\im 
\widehat{G}_b^{R}(q)\frac{d^4 q}{(2\pi)^4}\,,
\ee
where we took into account that $\int \widehat{V}_0 (q)\widehat{G}_f^{R} (p+q)
\widehat{V}_0 (-q) 
\widehat{G}_b^{R}(q)\frac{d^4 q}{(2\pi)^4}=0$, $\widehat{V}_0$
is the bare vertex. E.g.,  for the coupling of the $s_f =1/2$
fermion with the scalar ($s_b =0$) neutral boson the interaction
term in the Lagrangian density is $L_{int}=-
g_s \bar{\Psi}\phi\Psi$ and $\widehat{V}_0 =g_s$.

Then, separating 
boson particle and antiparticle
contributions (using $(+)$ and $(-)$ sub-scripts 
in notation of Appendix \ref{Formalism})
we obtain
\be\label{sigm-R0} 
-i\widehat{\Sigma}_f^{R}(p) &=&\int \Theta (q_0)
\widehat{V}_0 (q)\widehat{G}_f^{-+} (p+q)
\widehat{V}_0 (-q)\widehat{G}_{b,(+)}^{R}(q)\frac{d^4 q}{(2\pi)^4}\nonumber \\
&+&\int \Theta (q_0)\widehat{V}_0 (-q)\widehat{G}_f^{-+} (p-q)
\widehat{V}_0 (q)\widehat{G}_{b,(-)}^{A}(q)\frac{d^4 q}{(2\pi)^4}\nonumber \\
&+&\int \Theta (q_0)\widehat{V}_0 (q)\widehat{G}_f^{R} (p+q)
\widehat{V}_0 (-q)\widehat{G}_{b,(+)}^{-+}(q)\frac{d^4 q}{(2\pi)^4}\nonumber \\
&+&\int \Theta (q_0)\widehat{V}_0 (-q)\widehat{G}_f^{R} (p-q)
\widehat{V}_0 (q)\widehat{G}_{b,(-)}^{+-}(q)\frac{d^4 q}{(2\pi)^4}\nonumber \\
&+&\int \Theta (q_0)\widehat{V}_0 (q)i\widehat{G}_f^{-+} (p+q)
\widehat{V}_0 (-q) 
\widehat{A}_{b,(+)}(q)\frac{d^4 q}{(2\pi)^4}\,,\nonumber \\
&-&\int \Theta (q_0)\widehat{V}_0 (-q)i\widehat{G}_f^{-+} (p-q)
\widehat{V}_0  (q)
\widehat{A}_{b,(-)}(q)\frac{d^4 q}{(2\pi)^4}\,, 
\ee
$\Theta (x)$ is the step function, $\widehat{A}_{i,(+)}$ and 
$\widehat{A}_{i,(-)}$ are spectral functions for $i=b$ (or $i=f$)
particle and
antiparticle respectively, see Appendix \ref{Formalism}.

\section{Heavy fermions and less massive bosons
within simplest $\Phi$}\label{Interaction}

\subsection{Fermion and boson self-energies}
We will continue to deal with the simplest $\Phi$ 
(the first
diagram (\ref{phi})). 
Assume that {\em{boson occupations are
essentially higher than fermion ones}}. This condition is obviously
fulfilled
for low temperatures $T\ll m_b$,
since in our case
$m_f > m_b$. Thus the ratio of fermion to boson occupancies is $r\propto 
\mbox{exp}[-(m_f -m_b )/T]\ll 1$. For a higher
temperature
$T\gsim m_b^{*}(T)$, for
$m_f^{*}(T)\gg
m_b^{*}(T)$, one estimates $r\propto 
\mbox{exp}[-(m_f^* -m_b^* )/T]\ll 1$. For a still higher temperature,
$m_f\gg T> m_f^{*}(T)$,
there remains a power law suppression, $r\propto m_b^{*}(T)/T <1$.
Thereby, in the whole temperature interval of our interest we may retain in 
(\ref{sigm-R0}) only terms proportional to boson occupations. Then
eq. (\ref{sigm-R0})
yields
\be\label{sigm-R0-sim} 
\widehat{\Sigma}_f^{R}(p) &\simeq&\int \frac{d^3 q}{(2\pi)^3}
\int_{0}^{\infty} \frac{d q_0}{2\pi}
\widehat{V}_0 (q)\widehat{G}_f^{R} (p+q)
\widehat{V}_0 (-q)\widehat{A}_{b,(+)}(q)n_{b,(+)}(q_0 )\nonumber \\
&+& \int \frac{d^3 q}{(2\pi)^3}
\int_{0}^{\infty} \frac{d q_0}{2\pi}\widehat{V}_0 (-q)\widehat{G}_f^{R} (p-q)
\widehat{V}_0 (q)\widehat{A}_{b,(-)}(q)n_{b,(-)}(q_0 ).
\ee
From this expression we recover the fermion width (see eq. (\ref{A-G})):
\be\label{sigm-R0-simwid} 
\widehat{\Gamma}_f^{R}(p) &\simeq&\int \frac{d^3 q}{(2\pi)^3}
\int_{0}^{\infty} \frac{d q_0}{2\pi}
\widehat{V}_0 (q)\widehat{A}_f^{R} (p+q)
\widehat{V}_0 (-q)\widehat{A}_{b,(+)}(q)n_{b,(+)}(q_0 )\nonumber \\
&+& \int \frac{d^3 q}{(2\pi)^3}
\int_{0}^{\infty} \frac{d q_0}{2\pi}\widehat{V}_0 (-q)\widehat{A}_f^{R} (p-q)
\widehat{V}_0 (q)\widehat{A}_{b,(-)}(q)n_{b,(-)}(q_0 ).
\ee
In eqs (\ref{sigm-R0-sim}), (\ref{sigm-R0-simwid}) in order not to complicate
the consideration we omitted
contributions of quantum fluctuations, which do not depend on 
boson occupations. 
These pieces need a proper renormalization. Within the $\Phi$ derivable
approach the renormalization procedure was derived in \cite{KH}.
In principle, mentioned terms are responsible for important effects.
E.g., widths of the $\Delta$ isobar and the $\rho$ meson in vacuum
are due to such effects.
Here we are interested in temperature effects. Therefore we for
the sake of simplicity
consider
only thermal contributions assuming that the latter being  dominating terms.

The boson self-energy related to the first diagram of $\Phi$
(\ref{phi}) is as follows:
\be\label{sigm-R0-first} 
\re \widehat{\Sigma}_b^{R}(q) &\simeq&-2 \mbox{Tr}
\int \frac{d^4 p}{(2\pi)^4}
\widehat{V}_0 (q)\left[ \re \widehat{G}_f^{R} (p+q)+\re \widehat{G}_f^{R} (p-q)
\right]\nonumber \\
&\times&
\widehat{V}_0 (-q) \im \widehat{G}_{f}^R (p)n_{f}(p_0 ),
\ee
\be\label{sigm-R0-first1}
\widehat{\Gamma}_b^R (q) &\simeq& 4 \mbox{Tr}
\int \frac{d^4 p}{(2\pi)^4}
\widehat{V}_0 (q)\im \widehat{G}_f^{R} (p+q)
\widehat{V}_0 (-q)\im \widehat{G}_{f}^R (p)\nonumber \\
&\times&
[n_{f}(p_0 )-n_{f}(p_0 +q_0)]  .
\ee
We used eqs (\ref{mpeqS})
and that $n_{f}(p_0 +q_0)(1-n_{f}(p_0 ))=
n_{b}(q_0 )[n_{f}(p_0 )-n_{f}(p_0 +q_0)]$. 

We stress that {\em{fermion and boson Green functions}} entering above
self-energies {\em{are full Green functions,
although calculated within only one diagram of
$\Phi$.}} 

We will further study
two temperature regimes
of low and high temperature.
Below we demonstrate that for $m_f \gg T\gsim m_b^* (T)$   {\em{a high
temperature regime}} is realized. The boson continuum is then blurred.
The intensity of multiple scattering is not exponentially small anymore. 
For $T\gsim T_{\rm bl.f}$ 
the fermion effective mass essentially decreases with the increase of the
temperature and
the density of fermion-antifermion pairs is enhanced accordingly. 
Thus the fermion continuum becomes 
blurred.
As the reaction on the increase of the number of fermion-antifermion 
pairs for $T\gsim T_{\rm bl.f}$,  the boson effective mass being  significantly
reduced compared to the bare boson mass,
if initially were $m_b \gg T_{\rm bl.f}$. As a rough estimate we obtain 
$m_b^* (T_{\rm bl.f})\sim 
T_{\rm bl.f}$ for $T\sim T_{\rm bl.f}$.

For
$T\ll T_{\rm bl.f}$, 
the fermion-antifermion density is exponentially 
small and, thus,
$m_b^* (T) \simeq m_b$. Since $m_b^* (T_{\rm bl.f})\sim 
T_{\rm bl.f}$, if $m_b > T_{\rm bl.f}$, then
we also have $T\ll m_b$.
If $m_b < T_{\rm bl.f}$,
the fermion-antifermion density is exponentially 
suppressed for $T\ll m_b$.
We call the temperature regime $T\ll \mbox{min}(m_b ,
T_{\rm bl.f})$  
{\em{ the low temperature regime}}. 

\subsection{Low temperature limit (slightly heated 
and then, warm hadron liquid). 
Urbach law vs. 
Boltzmann law}

For temperatures $T\ll \mbox{min}(m_b ,
T_{\rm bl.f})$, for 
$m_b$, being essentially smaller than $m_f$, 
fermion occupations ($\propto e^{-m_f /T}$) are much less than
boson ones  ($\propto e^{-m_b /T}$), and we may use simplified eqs
(\ref{sigm-R0-sim}), (\ref{sigm-R0-simwid}). 

In the quasiparticle approximation
the fermion spectral function (\ref{A-G}) reads
\be\label{Af-qp}
\widehat{A}_{f}^{q.p} (p)
\simeq \widehat{\Lambda}_f^0 (p)\cdot
2\pi \delta \left( p_0^2 -\epsilon_p^2 -\frac{1}{4}\mbox{Tr}[
\widehat{\Lambda}_f^0 (p)
\re \widehat{\Sigma}_f^{R} (p)]\right),
\ee
$\epsilon_p =+\sqrt{m_f^2 +\vec{p}^{\,2}}$. In the very same approximation the boson spectral function
becomes
\be\label{Askb-qp}
\widehat{A}_{b,(\pm)}^{j; \,q.p} \simeq \widehat{\Lambda}_{b} (q)\cdot
2\pi \delta \left(q_0^2 -\om_q^2 -\frac{1}{N_b^{j}}\mbox{Tr}[
\widehat{\Lambda}_b^{j} (q)
\re \widehat{\Sigma}_b^{j;R} (q)]\right),
\ee
$   q_0 >0 $, $\om_q =\sqrt{m_b^2
  +\vec{q}^{\,\,2}}$,
$\widehat{\Lambda}^0_f$, $\widehat{\Lambda}_{b}^{j}$ are spin operators.
 Index $j$ counts scalar bosons ($j=s$)
and vector transversal ($j=t$) and longitudinal ($j=l$)
bosons. 
$\widehat{\Lambda}_{b}^{s}=1$ for scalar and
pseudo-scalar bosons, whereas for vector bosons one deals separately with
the transversal spectral function ($\widehat{\Lambda}_{b}^t =T^{\mu\nu}$) and  
the longitudinal spectral function ($\widehat{\Lambda}_{b}^l =L^{\mu\nu}$),
see eqs (\ref{frvec-m}), 
(\ref{tr-l}) below. $\widehat{\Lambda}^0_f$ is
introduced in (\ref{zer-f}).

The quasiparticle approximation is valid, if the particle width is much smaller
than all other relevant quantities in the given energy momentum region.
In the  quasiparticle term (\ref{Af-qp})
we may neglect the contribution of the $\re \widehat{\Sigma}_f
\propto e^{-m_b /T}$. In (\ref{Askb-qp})
we may omit the fermion particle-antiparticle
loop term $\re \widehat{\Sigma}_b\propto e^{-2m_f /T}$
(since $T\ll m_b <m_f$). After that these
spectral functions are reduced to free ones.

Outside the validity of the quasiparticle approximation,  
in case of still
rather small fermion width, from (\ref{A-G}) one obtains a regular
contribution 
\be\label{Af}
\widehat{A}_{f}^{\bf reg} (p) \simeq
\widehat{\Gamma}_{f}(p)\left(\re(\widehat{G}_f^{R}(p))^{-1}\right)^{-2} 
+
O[\widehat{\Gamma}_{f}^2 ]. 
\ee
$\widehat{\Gamma}_{f}$ is the fermion width operator introduced
by eq. (\ref{A-G}).
If integrated, the first term should be understood in the sense of the
principal value.
The contribution $\propto O[\widehat{\Gamma}_{f}^2 ]$ 
gets an additional exponentially small factor and can be dropped. 
Also $\re\widehat{G}_f^{R}(p)\simeq \re\widehat{G}_f^{0,R}(p)$
in the approximation we use.
The full
fermion spectral function is the sum of the quasiparticle and regular terms
related to different energy-momentum regions, cf. \cite{SRS}. 
The boson spectral function still can be
considered within the quasiparticle approximation, since 
$|\re \Sigma_b|\ll |\re \Sigma_f|$ for energies and
momenta relevant for the low temperature case, that we now discuss.
Also the consideration is essentially simplified, if one describes fermions 
within
the non-relativistic approximation, see Appendix \ref{non-rel}. 
This approximation is, indeed, fulfilled
in the low temperature limit, since particle energies are near the
mass-shell and 
typical thermal momenta are small.

First approximation to find the fermion width is to use
quasiparticle spectral functions (\ref{Af-qp}),
(\ref{Askb-qp}) suppressing there small self-energy dependent terms, i.e.
reducing these spectral functions to 
 spectral functions of free particles.
Then, replacing these spectral functions into (\ref{sigm-R0-simwid}) we obtain
\be\label{gamf-gen}
&&\widehat{\Gamma}_{f}^{q.p} \simeq
\int \Theta (q_0) \delta (q_0 -\om_q )\frac{d q_0}{(2\pi )^2}
\frac{d^3 q}{2\om_q}
\nonumber\\
&&\times 
\left[ n_{b,(+)} (q_0 )\frac{\delta (p_0 +q_0 -\epsilon_{\vec{p}+\vec{q}})}
{2\epsilon_{\vec{p}+\vec{q}}}
\widehat{V}_0 (q)\widehat{\Lambda}_f^0 (p +q)
\widehat{V}_0 (-q)\right. \nonumber\\
&&\left. +n_{b,(-)} (q_0 )
\frac{\delta (p_0 -q_0 -\epsilon_{\vec{p}-\vec{q}})}
{2\epsilon_{\vec{p}-\vec{q}}}\widehat{V}_0 (-q)
\widehat{\Lambda}_f^0 (p -q )
\widehat{V}_0 (-q)\right]\widehat{\Lambda}_b^0 (q).
\ee
Integrating (\ref{gamf-gen}) in $q_0$ we find
\be\label{gamf1-gen}
&&\widehat{\Gamma}_{f}^{q.p} (p)\simeq
\frac{1}{4}\int \frac{d^3 q}{(2\pi)^2}\frac{\mbox{exp}
[-(\om_q +\mu_b) /T]}{\om_q
  \epsilon_{\vec{p}+\vec{q}}} 
\delta (p_0 +\om_q -\epsilon_{\vec{p}+\vec{q}})\widehat{S}^0 (q) \nonumber\\
&&+\frac{1}{4}\int \frac{d^3 q}{(2\pi)^2}\frac{\mbox{exp}
[-(\om_q -\mu_b )/T]}{\om_q
  \epsilon_{\vec{p}-\vec{q}}} 
\delta (p_0 -\om_q -\epsilon_{\vec{p}-\vec{q}})\widehat{S}^0 (-q) ,
\ee
where
$\widehat{S}^0 (q)=\widehat{V}_0 (q)\widehat{\Lambda}_f^0 
(p+q)
\widehat{V}_0 (-q)\widehat{\Lambda}_b^0 (q)$ and we used that 
$\widehat{\Lambda}_b^0 (q) =\widehat{\Lambda}_b^0 (-q)$ and 
that 
$T\ll m_b$. 
The contribution
of the second term in (\ref{gamf1-gen}) to the fermion occupations is
$e^{-2m_b /T}$ times smaller than that of the first term for typical fermion
energies and momenta. Thus the second term can be omitted. 
The first term corresponds to the energy
$p_0 
\simeq m_f -m_b <m_f$, for $T\rightarrow 0$.

We may present the fermion 3-momentum 
distribution (\ref{3oc}) as the sum of two contributions
\be\label{foclowsum}
\widehat{n}_{f,(\pm)} (\vec p )=\widehat{n}_{f,(\pm)}^0 (\vec p )+
\delta \widehat{n}_{f,(\pm)} (\vec p ).
\ee
First term is obtained, if one substitutes (\ref{Af-qp}) into (\ref{3oc}):
\be\label{foclow-0}
\widehat{n}_{f,(\pm)}^0 (\vec p )\simeq
\frac{\gamma_0 \widehat{\Lambda}_f^0 (p)}
{2\epsilon_{p}}n_{{\rm Bol},(\pm)}(\vec p ),
\ee
where $n_{{\rm Bol},(\pm)}(\vec p )= \mbox{exp}[-(\epsilon_{p} \pm\mu_f ) /T]$
are Boltzmann occupations of particles and antiparticles.
We dropped exponentially suppressed contribution of
$\re \Sigma_f \propto e^{-m_b /T}$.
Replacing the quasiparticle
width term (\ref{gamf1-gen}) into eq. (\ref{Af}) we evaluate the regular
contribution to the spectral function. With the help of eq. (\ref{3oc}) we find the
term $\delta \widehat{n}_{f,(\pm)} (\vec p )$.
Since $|\vec{p}|, |\vec{q}|\ll m_f$,
we may put $\epsilon_{\vec{p}+\vec{q}}\simeq m_f$,
if the latter term does not enter the exponent. In the exponent we use
the expansion 
$\epsilon_{\vec{p}+\vec{q}}\simeq \epsilon_{p}+\vec{p}\vec{q}/\epsilon_{p}
+\vec{q}^{\,2} /(2\epsilon_{p} )$.
Moreover, 
we will 
exploit that $p_0^2 -\epsilon_p^2  \simeq  2 \epsilon_p (p_0 -\epsilon_p )$.
Taking off the integral in $|\vec{q}|=\sqrt{\omega_q^2 -m_b^2}$, using the
$\delta$ function 
we arrive at the expression
\be\label{foclow1}
&&\delta \widehat{n}_{f,(\pm)} (\vec p )\simeq 
\frac{1}{64\pi^2 \epsilon_p^3}\int_{\epsilon_p -m_b}^{\epsilon_p} 
\gamma_0\widehat{S}^0 [\widehat{\Lambda}_f^0
  (\widehat{p})]^2 |\vec{q}_{*}|dp_0 \int_{-1}^{1}dx\nonumber \\
&&\times\mbox{exp}\left[ -\left(
    \frac{\epsilon_p}{T}+\frac{|\vec{p}||\vec{q}_{*}|x}{\epsilon_p
      T}+\frac{\vec{q}_{*}^{\,\,2}}{2\epsilon_p T}\right)\right] .
\ee
Here the boson momentum $|\vec{q}_{*}|\simeq \sqrt{(\epsilon_p -p_0 )^2 -m_b^2}$.
We see that typical momenta of our interest are $|\vec{p}|\sim |\vec{q}|\sim \sqrt{m_f
T}\ll m_f$.
Then 
integrating  the rest over the angle 
we obtain
\be\label{foclow}
\delta \widehat{n}_{f,(\pm)} (\vec p )\simeq 
\frac{m_f^3}
{8\pi^2\epsilon_{p}^3}\widehat{I}_0 (\frac{\epsilon_{p}}{m_b})
 n_{{\rm Bol},(\pm)}(\vec p ),
\ee
with
\be\label{I0} 
\widehat{I}_0 (x)&=& \frac{1}{4m_f^3}
\int_{1}^{x} \gamma_0\widehat{S}^0 [\widehat{\Lambda}_f^0
  (\widehat{p})]^2 \frac{\mbox{sinh}(y\eta)}{y\eta} e^{-y^2 /2}
\sqrt{z^2 -1}\frac{d z}{z^2},  \\
y &=&m_b \sqrt{z^2 -1}/\sqrt{\epsilon_p T} ,
\quad \eta =|\vec{p}|/\sqrt{\epsilon_p T},\quad z=(\epsilon_p -p_0 )/m_b .
\nonumber
\ee
One can  see that the integral
is cut off at $z^2 \sim 1+m_f T/m_b^2$ (corresponding to $y\sim 1$). 

{\em{ Eq. (\ref{foclow}) is the key expression of this sub-section.}}
Supposing for a rough estimate
that $\widehat{S}^0$ does not depend on
$q$  we obtain an estimation
\be
\delta n_{f,(\pm)} (\vec p )\sim 
(g^2 /(24\pi^2) ) (m_f T /m_b^2 )^{3/2}n_{\rm Bol,(\pm)}
 \ee
for   $T\ll
m_b^2 /m_f $, where
$g$ is the typical value of the coupling constant. Notice that we continue to
consider the case $T\ll \mbox{min}(m_b , T_{\rm bl.f})$.
We see that  the correction to the ordinary Boltzmann distribution is small,
$\delta n_{f,(\pm)} (\vec p )\ll n_{\rm Bol,(\pm)}$ for
$T\ll
m_b^2 (24\pi^2 )^{2/3}/[g^{4/3} m_f ]$. We may
call such a temperature regime
{\em ``a slightly heated hadron liquid''}.

For $\mbox{min}(m_b , T_{\rm bl.f})\gg T\gsim
m_b^2 /m_f$  (we may call this regime {\em 
``a warm hadron liquid''}) we get
\be
\delta n_{f,(\pm)} (\vec p )\sim (g^2/(16\pi^2) ) n_{\rm Bol,(\pm)}
\mbox{ln}\frac{m_f
  T}{m_b^2},
 \ee 
i.e., for $g\sim 10$ of our interest the fermion distribution 
could be  up to
several times (depending on $g$) enhanced compared to the ordinary Boltzmann 
distribution. 

To find the regular contribution  to the fermion width 
one replaces (\ref{Af}) into 
(\ref{sigm-R0-simwid}). Then
we obtain an integral equation
\be\label{deltagam}
&&\widehat{\Gamma}_{f}^{\bf reg}\simeq 
\int \Theta (q_0)n_b (q_0 ) \delta (q_0 -\om_q )\frac{d q_0}{(2\pi )^3} 
\frac{d^3 q}{2\om_q}\nonumber\\
&&\times \widehat{V}_0 (q)\widehat{\Gamma}_{f}^{\bf reg}
(p+q)[\re\widehat{G}_f^{0,R}(p+q)]^2 \widehat{V}_0 (-q)
\widehat{\Lambda}_b^0 (q)\nonumber\\
&&+
\int \Theta (q_0)n_b (q_0 ) \delta (q_0 -\om_q )\frac{d q_0}{(2\pi )^3} 
\frac{d^3 q}{2\om_q}\nonumber\\
&&\times \widehat{V}_0 (-q)\widehat{\Gamma}_{f}^{\bf reg}
(p-q)[\re\widehat{G}_f^{0,R}(p-q)]^2 \widehat{V}_0 (q)
\widehat{\Lambda}_b^0 (-q).
\ee
Eq. (\ref{deltagam}) is much more involved than the quasiparticle term, 
eq. (\ref{gamf-gen}). 
In the perturbative
regime ($\delta n_{f,(\pm)} (\vec p )\ll n_{\rm Bol}$)
eq. (\ref{deltagam})
can be solved iteratively yielding small corrections to the
quasiparticle estimate. 
However in the limit $T\gsim
m_b^2 /m_f $ the
perturbative consideration is valid only, if the 
expansion parameter, 
being $\sim g^2 /(4\nu \pi^2 )$,
 is much smaller than unity. The value $\nu$ depends 
on the choice of the fermion-boson
coupling. Typically, one has $\nu \sim 1$. 
For realistic values of the meson-nucleon coupling constant
(e.g., $g\sim 10$ for $\sigma$, $\om$, $\rho$) 
{\em multiple  rescatterings of the heavy fermion
on light bosons should be taken into account in all orders (for
$T\gsim
m_b^2 /m_f$).}
In the physics of solids a similar
 effect (Urbah law) is well known for the case of the electron-phonon interaction
in semiconductors. A long tail of the
electron wave function arises
inside the band gap of the semiconductor \cite{E}.
For massless phonons the effect is stronger
than for massive bosons (if couplings in both cases are of the
same order of magnitude). 
The limiting case  $T\gg m_b^2 /m_f$ estimation
({\em 
a warm hadron liquid}), if done with an appropriate
vertex $\widehat{V}_0$, is relevant for
phonons. The description of {\em a slightly heated hadron
liquid} is different from that for massless phonons.

Concluding this sub-section,  we have shown that
{\em even at low temperatures (for ``a warm hadron liquid'')
fermion particle-antiparticle densities might be essentially
enhanced 
compared to quasiparticle ones (Boltzmann law) due to
multiple rescatterings of fermions in the thermal bath of bosons.} 

To calculate particle distributions explicitly we need to know  
the explicit form of the spin structure operator
$\gamma_0 \widehat{S}^0 (q)
[\widehat{\Lambda}_f^0 (p)]^2$, that
depends on the choice of the coupling between particle species. 
Since the  fermion energy departs only little from the mass shell, 
$|\delta p_0 |\ll p_0$, and the fermion momentum is also small $|\vec{p}|\sim 
\sqrt{2m_f T}\ll m_f$, from the very beginning one could use
the non-relativistic approximation for heavy fermions. We did not do it
in order not to spoil our general relativistic approach,
which we further use to describe the high temperature regime.

\subsection{
 High temperature limit (``hot hadron liquid''
). Multiple rescatterings}
Now we will consider a high temperature regime, $T\gsim
m_b^{*}(T)$.  In this case the boson continuum is blurred.
The temperature exceeds the effective 
gap between particle-antiparticle continua
and the corresponding antiparticle density is rather high thereby.
As we argue below, in a wide temperature range
the departure of the fermion energy from the mass
shell $\delta p_0 \sim \sqrt{J}$ (see eq. (\ref{typf}) 
below) is much larger than that for bosons, 
$\delta q_0\sim \mbox{max}\{ m_b -m_b^* (T), T\}$, and typical fermion momenta
$|\vec{p}|\sim \sqrt{2 m_f (T)T}$ are much higher than typical boson
momenta $|\vec{q}|\sim \mbox{max}\{\sqrt{2 m_b^* (T)T},\,T\}$.
Therefore we are able to drop a $q$-dependence 
of fermion Green functions in 
(\ref{sigm-R0-sim}). Then eq. (\ref{sigm-R0-sim}) is simplified as
\be\label{sigmJ}
&&\widehat{\Sigma}_f^{R}(p)\simeq \widehat{J}\cdot 
\widehat{G}_f^{R}(p)\\
&&\equiv
\int \frac{d^3 q}{(2\pi)^3}\int_{0}^{\infty} \frac{d q_0}{2\pi}
\left[\widehat{V}_0 (q)\widehat{G}_f^{R}(p)\widehat{V}_0 (-q) 
\widehat{A}_{b,(+)}(q) n_{b,(+)}(q_0 )\right.\nonumber\\
&&\left. +
\widehat{V}_0 (-q)\widehat{G}_f^{R}(p)\widehat{V}_0 (q)
\widehat{A}_{b,(-)}(q) n_{b,(-)}(q_0 )\right],\nonumber
\ee
where
\be\label{Jhat}
\widehat{J}&=&\int \frac{d^3 q}{(2\pi)^3}\int_{0}^{\infty} \frac{d q_0}{2\pi}
\left[\widehat{V}_0 (q)\widehat{\Lambda}_f (p)\widehat{V}_0 (-q)
\widehat{A}_{b,(+)}(q) n_{b,(+)}(q_0 )\right.\nonumber \\
&&\left. +\widehat{V}_0 (-q)\widehat{\Lambda}_f (p)\widehat{V}_0 (q)
\widehat{A}_{b,(-)}(q) n_{b,(-)}(q_0 )\right]\widehat{\Lambda}_f^{-1} (p).
\ee
If we for a moment ignored a complicated spin-structure, we could associate
the quantity $\widehat{J}$ with a tadpole diagram
\be\label{ladprtad}
\includegraphics[width=2cm,clip=true]{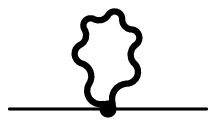}
\ee
This diagram describes fluctuations of virtual (off-mass shell) bosons.
For bosons, which number is not
conserved, we have $\mu_b =0$ and $n_{b,(+)}(q_0 )=n_{b,(-)}(q_0 )$. 
We formally
presented
$\widehat{G}_f^{R}(p)=\widehat{\Lambda}_f (p)G_f^{R}(p)$, where the
spin structure term $\widehat{\Lambda}_f$ is separated
from the term $G_f^{R}(p)$
related to dynamical
degrees of freedom, cf. eqs (\ref{spinstrTG}), (\ref{spinstr})  below.
Then the dynamical part of the
fermion Green function $G_f^{R}(p)$ 
decouples from the integral. For the case of a
hard external fermion having a large 3-momentum
compared with the typical momentum transfer in the loop we may use
{\em the soft thermal loop} (STL)
approximation. The latter is opposite to the hard thermal loop approximation
of the soft external particle with a small momentum
compared with the typical momentum transfer in
the loop. The hard thermal loop approximation is  widely
used in the description of the quark-gluon plasma, cf. \cite{BJ}
and refs therein. 
In the STL approximation we drop the dependence of the internal fermion Green
function in
the loop on the internal momentum transfer.
Please notice that only in the high temperature limit
$T\gsim
m_b^{*}(T)$ it is possible to drop the $q$-dependence
of the fermion Green function. 
Considering the low temperature limit we were forced to retain the
$q$-dependence of the fermion Green function in the calculation of the fermion
self-energy, since there $|\vec{q}|\sim |\vec{p}|$ 
for typical values of momenta, as it followed from 
eqs (\ref{gamf1-gen}), (\ref{I0}).
Besides, in the low temperature limit in case of {\em a slightly heated hadron
liquid}
we used
an expansion of the full fermion Green function near its non-perturbed value
$\widehat{G}^{0,R}_f$. In the high temperature limit the full
fermion Green function is obtained straight
from the Dyson equation (\ref{Dyson}). The latter equation is
greatly simplified  in the STL approximation and reads
\be\label{Dscat}
[\widehat{G}^{0,R}_f(p)]^{-1}\widehat{G}_f^{R}(p)=1+
\widehat{J}\cdot 
\widehat{G}_f^{R}(p)\cdot\widehat{G}_f^{R}(p).
\ee
{\em{This is the key equation of this sub-section.}}
A perturbative analysis of eq. (\ref{Dscat}) is possible only for $J \ll T^2$
(see the corresponding estimate after eq. (\ref{tblrel}) below).
However the latter limit is not realized within the high temperature regime
for the case of a strong coupling ($g\sim 10$), 
see eq. (\ref{Jexp-lim2})
below.

The operator $\widehat{\Lambda}_f$ and the quantity 
$\widehat{G}^R$ are complicated functions of
invariants. To determine them we
present the fermion
self-energy in the most general form as
\be\label{spinstrT}
\widehat{\Sigma}_f =\Sigma_1 \FMslash{p}+\Sigma_2 m_f +\Sigma_3 \FMslash{u},
\ee
$u^{\mu}$ is the 4-velocity of the frame. The Green function takes the form
\be\label{spinstrTG}
\widehat{G}_f =G_{11} \,\FMslash{p}+G_{12} m_f +G_{13} \,\FMslash{u}\equiv 
\widehat{\Lambda}_f G_f .
\ee
In the rest frame $u^{\mu}=
(1,\vec{0})$ and  (\ref{spinstrTG}) is simplified as
\be\label{spinstr}
\widehat{G}_f (p)=G_1 (p_0 ,\vec{p} )\FMslash{p}+G_2 (p_0 ,\vec{p})m_f 
+ G_3 (p_0 ,\vec{p})\vec{p}\vec{\gamma} .
\ee
This equation shows that in general
in the rest frame the Green function depends separately 
on $p_0$ and $\vec{p}$.

The term $G_3$ 
yields only small corrections to thermodynamical quantities
in a wide temperature region.
Moreover, the fermion-antifermion density, which we are interested in below,
does not depend at all
on the term $\propto
\vec{p}\vec{\gamma}$
in the fermion Green function. Thus  $G_3$ can be omitted.
Also in many cases one may put
$G_1 \simeq G_2 $ with appropriate accuracy,
that 
further simplifies the consideration, 
since then $\widehat{G}_f (p)\simeq G_1 
\widehat{\Lambda}_f^0 (p)$ is determined by only one quantity ($G_1$). 

Let us consider spin-zero bosons, $\widehat{V}_0 =g_s$.
Then 
the Dyson equation (\ref{Dscat}) has clear diagrammatic interpretation.
It can be presented as follows
\be\label{scat}
\includegraphics[width=7cm,clip=true]{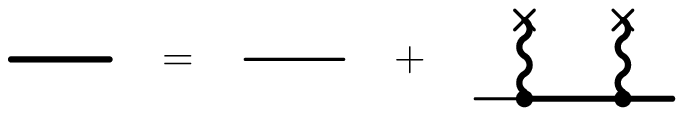}
\ee
describing the fermion propagation in an external field $\phi_c^{\rm ext}$,
$\phi_c^{\rm ext} \cdot (\phi_c^{\rm ext})^* = J/g^2_s$, 
\be\label{Jnothat}
J=g^2_s \int \frac{d^3 q}{(2\pi)^3}\int_{0}^{\infty} \frac{d q_0}{2\pi}
\left[ A_{b,(+)}(q) n_{b,(+)}(q_0 )+
A_{b,(-)}(q) n_{b,(-)}(q_0 )\right] ,
\ee
compare with (\ref{Jhat}). 
As follows from eq. (\ref{sigmJ}),
in difference with the standard Dyson equation in the external field the
r.h.s. of eq. (\ref{scat}) contains two full fermion Green functions and
two lines of the external field.
Eq. (\ref{scat}) shows that the
propagating heavy fermion undergoes multiple quasi-elastic rescatterings on 
pairs  of quasi-static boson
impurities. Impurities are quasi-static in the sense of the above used STL
approximation.
In the given approximation the quantity $J$ does not depend on the external
frequency and 3-momentum.
The value $J$ is proportional
to the density of impurities. Thus, 
it demonstrates
{\em the intensity of the multiple elastic scattering}. 
To better understand this,  one may compare  the 
non-relativistic spin-averaged  limit expression
\be\label{nonr}
{\overline{[\Sigma_f^{R}(p)]}}^{\,\rm n.rel}=\frac{1}{4}\mbox{Tr}\left[ 
\widehat{\Sigma}_f^{R}(p)\frac{(\FMslash{p}+m_f )}{2m_f}\right]=
\overline{J}^{\rm n.rel}\cdot 
[G_f^{R}(p)]^{\rm n.rel},
\ee
cf. the first
line of eq. (\ref{sigmJ}), 
with the quasiclassical non-relativistic
equation \cite{MSTV,IKHV00},
\be\label{gas}
\overline{\Sigma_f}^{\,\rm n.rel} \simeq 4\pi 
\rho_b (2m_b^* )^{-1} \overline{F_f}^{\rm n.rel} (0).
\ee
Here 
$\Sigma_f^{\rm n.rel}$ is the particle 
(heavy fermion in our case) non-relativistic self-energy,
\be
\overline{J}^{\,\rm n.rel}=\frac{1}{4}
\mbox{Tr}\frac{\widehat{V}_s  (\FMslash{p}+m_f )\widehat{V}_s
(\FMslash{p}+m_f )}{4m_f^2}J, 
\ee
$F_f^{\,\rm n.rel} (0)$ is the non-relativistic fermion forward
scattering amplitude 
in the medium of independent static scattering centers, 
\be
4\pi 
\overline{F}_f^{\,\rm n.rel} (0)=g^2\frac{1}{4}
\mbox{Tr}\frac{\widehat{V}_s  (\FMslash{p}+m_f )\widehat{V}_s
(\FMslash{p}+m_f )}{4m_f^2} [G_f^{R}(p)]^{\rm n.rel},
\ee
$\rho_b$
is the density of centers
(in our case the density of quasi-static boson impurities,
which we introduce as $\rho_b 
\equiv 2m_b^* |\phi_c^{\rm ext} |^2$ ). 
In more detail different non-relativistic limit expressions
for fermions are discussed in
Appendix \ref{non-rel}.

In general case, e.g. for vector bosons,
eq. (\ref{scat}) has only a symbolic meaning. It is 
the operator equation for several values  of {\em{intensities of multiple 
quasi-elastic scattering}}
and for coupled functions $G_1^R$, $G_2^R$ and $G_3^R$,
which determine the fermion Green function. 

The STL approximation may allow to develop a simplified kinetic 
description of the non-equilibrium system with the help of 
the 3-momentum fermion distribution
function. Such a kinetic scheme could be then spread out to describe coherent
di-lepton radiation processes in an analogy to the kinetic description of the
Landau-Pomeranchuk-Migdal effect, cf. \cite{KV96}.
However these problems are beyond the scope of this paper.

\section{System of heavy spin $1/2$
fermions and less massive scalar neutral
bosons. }\label{neutral bosons}

To avoid complications with spin-isospin degrees of
freedom, as the simplest example,  
we will consider a system of
spin $1/2$ ($s_f =1/2$) fermions and  spin zero
($s_b=0$) neutral bosons coupled by the Yukawa
interaction, 
\be\label{Lints}
L_{int} =-g_s \bar{\psi}\phi\psi.
\ee
In this case $\widehat{\Lambda}_f^0 =(\FMslash{p}+m_f )$,
$\widehat{\Lambda}_b^0 =1$,
$\widehat{V}_0 =g_s$,
$(\widehat{G}_f^{0,R})^{-1}=\FMslash{p}-m_f +i0$, 
and we  assume, as before, $m_f > m_b $.

Results of this section
can be applied for the description of the 
$N\sigma$ sub-system. 
In subsequent sections we
consider $N\omega$,  $N\rho$ and $N\pi$ systems
and summarize results.

\subsection{Low temperature limit}

From the first term of eq.
(\ref{gamf1-gen}) we find 
\be\label{gamf1}
&&\widehat{\Gamma}_{f}^{q.p} (p_0 -\epsilon_{p})\simeq
\frac{g^2_s (\gamma_0 +1 )}{4\pi } 
\mbox{exp}\left[-(\epsilon_{p}-p_0 ) /T\right]
\frac{\mbox{sinh }(y\eta)}{y\eta} e^{-y^2 /2}\\
&&\times
\sqrt{(p_0 -\epsilon_{p})^2 -m_b^{\,2} }, \quad y=
\frac{\sqrt{(p_0 -\epsilon_{p})^2 -m_b^{\,2}}}{\sqrt{\epsilon_{p}T}}, 
\quad \eta =\frac{|\vec{p}|}{\sqrt{\epsilon_{p}T}},\quad p_0 <\epsilon_{p}.
\nonumber
\ee
Using that typically both $y\eta \sim 1$ and $y\sim1$, we estimate
typical fermion and boson momenta $|\vec{p}|\sim |\vec{q}|\sim \sqrt{m_f
T}\ll m_f$. 
We used  that  $|\vec{p}|\ll m_f$, $|\vec{q}|\ll m_f$ and
we have put $\epsilon_{\vec{p}+\vec{q}}\simeq m_f$,
everywhere except  the exponent. In the exponent we used
the expansion 
$\epsilon_{\vec{p}+\vec{q}}\simeq \epsilon_{p}+\vec{p}\,\vec{q}/ \epsilon_{p}
+\vec{q}^{\,\,2} /(2\epsilon_{p} )$. 
Also we dropped the  term $\propto \vec{\gamma}$ 
entering the
$\widehat{S}^0$ operator,
since it does not
contribute to the fermion density and to other relevant quantities, 
as it is seen after the
corresponding angular integration.

With the help of 
eqs (\ref{Af-qp}) and (\ref{3oc}),
and also (\ref{Af}), (\ref{gamf1}),  we obtain two contributions to
the 3-momentum fermion distribution:
\be\label{3ocU}
\widehat{n}_{f,(\pm)} (\vec p )\simeq
\frac{(\epsilon_p +m_f\gamma_0 )}{2\epsilon_{p}}n_{{\rm Bol},(\pm)}(\vec p )
+
\frac{1}
{8\pi^2}\widehat{I}_{0s} (\frac{\epsilon_{p}}{m_b}
  ) n_{{\rm Bol},(\pm)}(\vec p ),
\ee 
\be\label{i0s}
\widehat{I}_{0s} (x) =g_s^2
(\gamma_0 +1 )
\int_{1}^{x} \frac{\mbox{sinh }(y\eta)}{y\eta} e^{-y^2 /2}
\sqrt{z^2 -1}\frac{d z}{z^2}\equiv g_s^2 (\gamma_0 +1 )I_{0s} (x),
\ee
cf. eqs (\ref{foclow-0}), (\ref{foclow}), (\ref{I0}).
Typical values of $z=(\epsilon_p -p_0
)/m_b$ in  $I_{0s}$ are determined by an estimate
$y^2 /2 -y\eta \sim 1$.  The characteristic
averaged value of $|\vec{p}|$ is $\sim \sqrt{2m_f T}$.
Thereby, $\eta \sim 1$ and thus $y\sim 1$.
For $|\vec{p}|\gg \sqrt{2m_f T}$ typically $y\sim \eta \gg 1$. In both cases  
$|\vec{p}|\sim |\vec{q}|\simeq \sqrt{(p_0 -\epsilon_p )^2 -m_b^2}$.
Cutting off the integral at given value of $z$ we evaluate
\be\label{sap}
I_{0s} (x)\sim -\frac{\sqrt{\bar{x}^2 -1}}{\bar{x}}+
\mbox{ln} (\sqrt{\bar{x}^2 -1}+\bar{x}). 
\ee
Here $\bar{x}\simeq 
(1+m_f T /m_b^{\,2} )^{1/2}$ for $|\vec{p}|\lsim \sqrt{2m_f T}$
of our interest,
$I_{0s} (x)\simeq \frac{2^{3/2}}{3}(\bar{x}-1)^{3/2}\rightarrow 0$ for 
$\bar{x}\rightarrow 1$ and 
$I_{0s} (x)\simeq  \mbox{ln} (\bar{x})$ for $\bar{x}\gg 1$.

The ratio of the fermion/antifermion density to the corresponding
density calculated 
with the Boltzmann distribution, 
$n_{{\rm Bol},(\pm)}(\vec p )= \mbox{exp}[-(\epsilon_{p} \pm\mu_f ) /T]$, is
as follows
\be\label{r-rat}
\frac{\rho_{f,(+)}}{\rho_{{\rm Bol},(+)}}=\frac{\rho_{f,(-)}}
{\rho_{{\rm Bol},(-)}}=
1+\frac{ g_s^2}{4\pi^2}I_{0s} \left(\frac{m_f }{m_b}\right), 
\ee
where we introduced the density of the ideal relativistic non-degenerate 
(Boltzmann) gas
\be\label{r-Bol}
\rho_{{\rm Bol},(\pm)}&=&N_f \int \frac{d^3 p}{(2\pi)^3}\mbox{exp}\left( -(
\epsilon_p \mp \mu_f )/T\right)\nonumber \\
&=&N_f \left(\frac{m_f T}{2\pi}\right)^{3/2}
\mbox{exp}\left( -(m_f \mp \mu_f )/T\right)\nonumber \\ 
&\equiv&\rho_{{\rm Bol}}^{sym}\mbox{exp}\left( \pm \mu_f /T\right) ,
\ee
$N_f =2s_f +1 $ is the degeneracy factor, $N_f =2$ for $s_f =1/2$ fermions. 

Now we may make an attempt to solve eq. (\ref{deltagam}) in general case.
We will use that $p_0$ is
near $m_f$, and $q_0$, $|\vec{q}|$, $|\vec{p}|\ll m_f$. 
Introducing convenient variables we present 
\be\label{gamf1phi}
\widehat{\Gamma}_{f}^{\rm reg} (\widetilde{\omega})\simeq
\frac{g^2_s (\gamma_0 + 1)}{4\pi } 
\mbox{exp}\left(-|\widetilde{\omega}| /T\right)\phi (\widetilde{\omega} ),
\quad \widetilde{\omega} =-p_0 +\epsilon_p ,
\ee
where  $\phi (\widetilde{\omega} )$
satisfies the integral equation:
\be\label{phiint}
&&\phi (\widetilde{\omega} )=\sqrt{\widetilde{\omega}^2 -m_b^2}+
\frac{g_s^2}{4\pi^2 }I_{1s}, \\
&&I_{1s}(\widetilde{\omega})=\Theta (\widetilde{\omega}-2m_b)
\int^{\mbox{min}[\widetilde{\omega}-m_b ,m_b\sqrt{1+m_f T/m_b^2 }]}_{
m_b}
d\widetilde{\epsilon}\,\frac{\phi (\widetilde{\epsilon} ) 
\sqrt{(\widetilde{\omega}-\widetilde{\epsilon})^2 -m_b^2 } }
{\widetilde{\epsilon}^2}, 
\nonumber
\ee
with $\widetilde{\epsilon}=\widetilde{\omega}-\omega_q $.
In (\ref{phiint}) we separated the term leading to eq. (\ref{gamf1})
and the residual term.
We used that $e^{-y^2 /2}(y\eta)^{-1}\mbox{sinh }(y\eta)\simeq 1$ for typical
energies and momenta of our interest (related to $y(\widetilde{\omega} )
\lsim 1$) and we also
cut off the integration in $\widetilde{\epsilon}$
using that $y (\widetilde{\epsilon})\lsim 1$ for typical 
$|\vec{p}|\lsim \sqrt{2m_f T}$. For $|\vec{p}|\gg \sqrt{2m_f T}$ one should
replace $\sqrt{1+m_f T/m_b^2}$ in expression for $I_{1s}$
by $\sqrt{1+\vec{p}^{\,2} /m_b^2}$. 

We may try to solve eq. (\ref{phiint}) iteratively.
First term in  (\ref{phiint}) 
yields eq. (\ref{gamf1}). Replacing this term 
into $I_{1s}$ we obtain next term of $\phi$, etc.

For $y(\widetilde{\omega} )
\lsim 1$
we find $\widetilde{\omega}- m_b \ll m_b$ and
\be 
\phi \simeq \sqrt{\widetilde{\omega}^2 -m_b^2}
,\quad for \quad T\ll m_b^2 /m_f ,
\ee
and  for $\widetilde{\omega}\gg m_b$, 
\be
\phi \simeq 
\widetilde{\omega} \left[ 1+\frac{g_s^2}{4\pi^2} \mbox{ln}\left(\mbox{min}
\left\{ 
\frac{\widetilde{\omega}}{m_b}, \sqrt{\frac{Tm_f }{m_b^2}} \right\}\right)\right], 
\quad \mbox{for} \quad
T\gg m_b^2 /m_f . 
\ee
Provided $g_s^2 /(4\pi^2)\gsim 1$,  {\em corrections due to multiparticle
rescatterings of the fermion are substantial already
for sufficiently low temperature 
$\mbox{min}( m_b ,T_{\rm bl.f}) \gg T\gsim m_b^2 /m_f$}. Then,
one should go beyond the iterative procedure in order to get 
an appropriate quantitative result. {\em{Thus quasiparticle approximation may fail
already for rather small temperatures, if the case of a warm hadron
liquid is realized.}}

For the $N\sigma$ interaction
we estimate $m_N \simeq g_\sigma f_{\pi}$, $f_{\pi}\simeq 93$~MeV,
$g_\sigma \simeq 10$, $m_{\sigma}\simeq (500\div 600)$~MeV. 
Therefore the limiting case
$T\ll m_b^2 /m_f$ is always
realized for relevant low temperatures $T\ll T_{\rm bl.f}\sim m_{\pi}\ll 
m_{\sigma}$. (We will further argue that the value $T_{\rm bl.f}$
is close to the value of the pion mass $m_{\pi}\simeq 140$~MeV.) 
Then we may use  the quasiparticle estimation of the nucleon width. 
For the $N\bar{N}\sigma$ system for zero total baryon number (``sym''), 
for typical thermal momenta and at $T\sim m_{\pi}/2$, we estimate
\be\label{ratio-distBold}
n_p^{sym} (\vec{p})=n_n^{sym} (\vec{p})=n_{\bar{p}}^{sym} (\vec{p})
=n_{\bar{n}}^{sym} (\vec{p})
\simeq (1.07\div 1.1) n_{\rm Bol}^{sym},\nonumber 
\ee
$n_p^{sym} (\vec{p})$, $n_n^{sym}(\vec{p})$, $n_{\bar{p}}^{sym} (\vec{p})$
and $n_{\bar{n}}^{sym} (\vec{p})$ are proton, neutron, antiproton and
antineutron 3-momentum distributions. The same estimate is
valid also for particle densities: 
\be\label{ratio-distBol}
\rho_p^{sym} &=&\rho_n^{sym} =\rho_{\bar{p}}^{sym} =\rho_{\bar{n}}^{sym} 
\simeq (1.07\div 1.1) \rho_{{\rm Bol}}^{sym}. 
\ee
We see that particle  3-momentum distributions and 
densities
are enhanced up to 
$\simeq 1.1$ times for $T\sim m_{\pi}/2$
compared to the standard Boltzmann particle distribution and the density. 

Concluding, in this sub-section we have demonstrated that {\em{already at low
temperatures
the heavy fermion
3-momentum distribution
is enhanced compared to the ordinary Boltzmann distribution.}}
This is the consequence of rescatterings 
of the fermion on virtual bosons (cf. diagram (\ref{selfzf}) and its reduction to (\ref{ladprtad})). It results
in {\em{a partial blurring of the gap between
fermion-antifermion continua, if the case of ``a warm hadron liquid''
is realized.}}

\subsection{High temperature limit}

\subsubsection{Analytic solution for fermion Green functions in STL
  approximation }

Assuming that typical values of $|\vec{p}|\sim \sqrt{m_f T}$ are rather small
($\ll m_f$) and to avoid more cumbersome expressions  we further drop the
term $G_3 \vec{p}\,\vec{\gamma}$ 
in eq. (\ref{spinstr}).
Using (\ref{spinstr}) and (\ref{zer-f})
we may as follows
rewrite the Dyson equation (\ref{Dscat}) for the fermion sub-system
derived in the STL approximation:
\be\label{Dys-coup1}
G_1^R =G_f^{0,R}+J_s G_f^{0,R}\left[(G_1^R)^2 p^2 +(G_2^R)^2 m_f^2 +
2 G_1^R G_2^R m_f^2\right],
\ee
\be\label{Dys-coup2}
G_2^R =G_f^{0,R}+J_s G_f^{0,R}\left[(G_1^R)^2 p^2 +(G_2^R)^2 m_f^2 +
2 G_1^R G_2^R p^2\right].
\ee
We introduced the quantity $J_s$ related to the operator $\widehat{J}_s$
from eq. (\ref{sigmJ}) as $\widehat{J}_s \cdot 
\widehat{G}_f^{R}(p)= J_s \widehat{G}_f^{R}(p)$, that yields
\be\label{sigmJ-s}
&&J_s =g_s^2
\int \frac{d^3 q}{(2\pi)^3}\int_{0}^{\infty} \frac{d q_0}{2\pi}
\left[ A_{b,(+)}(q) n_{b,(+)}(q_0 )+
A_{b,(-)}(q) n_{b,(-)}(q_0 )\right],
\ee
in complete analogy with  eq. (\ref{Jnothat}).
As it follows from (\ref{A-G}), the scalar boson spectral function is 
\be\label{Asb}
A_{b,(\pm )}=\frac{\Gamma_{b,(\pm )}}{M^2_{b,(\pm )}+\Gamma_{b,(\pm )}^2 /4},
\ee
$\widehat{\Gamma}_{b,(\pm)}=\Gamma_{b,(\pm)}$,
$\re\widehat{\Sigma}_{b,(\pm)}=\re\Sigma_{b,(\pm)}$.
For scalar neutral bosons $A_{s ,(+)}(q)=A_{s ,(-)}(q)$
and $n_{s ,(+)}(q_0 )=n_{s ,(-)}(q_0 )$.

Resolving (\ref{Dys-coup1}), (\ref{Dys-coup2}) we find 
\be\label{G1s}
G_1 =G_2 (1 +2G_2 J_s )^{-1},
\ee
\be\label{G2sol}
&&4J_s^2 G_2^4 +8J_s G_2^3 +\left(5-\frac{p^2}{m_f^2}+\frac{4J_s}
{m_f^2}\right)G_2^2\nonumber \\
&&- \left(\frac{p^2 -m_f^2}{J_s m_f^2}-\frac{4}{m_f^2}\right) G_2 
+\frac{1}{J_s m_f^2}=0.
\ee
Analytic solution of the fourth power eq. (\ref{G2sol}) for $G_2$ looks 
cumbersome.
To simplify the consideration we  drop $G_2^4$ and $G_2^3$ terms in 
(\ref{G2sol}) (accuracy of this approximation is discussed below)
and find the corresponding solution:
\be\label{tblrel}
G_2^R \simeq 
\frac{p^2 -m_f^2 -4J_s \pm \sqrt{\left(p^2 -m_f^2 -4J_s \right)^2 -4m_f^2 J_s 
\left(5-\frac{p^2}{m_f^2}+\frac{4J_s}
{m_f^2}\right)}}
{2m_f^2 J_s \left(5-\frac{p^2}{m_f^2}+\frac{4J_s}{m_f^2}\right)} .
\ee
This equation has the pole-like solution only for $(p^2 -m_f^2 )^2\gg 16 m_f^2
J_s$.
It is obtained by 
taking the corresponding  branch (taking
negative sign in front of the square root in
(\ref{tblrel})) and expanding the  square root term in the parameter
\be\label{ze}
\zeta =16m_f^2 J_s /(p^2 -m_f^2 )^2 .
\ee
In the framework of the quasiparticle approximation typical energies are  $p_0
-m_f \sim T$ and $\zeta =4J_s /T^2$. Thus, the quasiparticle approximation
works only for $J_s \ll T^2$. In the leading order in $\zeta$ we obtain
$G_2^R \simeq (p^2 -m_f^2 )^{-1}=G_{0}^R$. In next order eq. (\ref{tblrel})
yields
\be\label{Gap}
G_1 \simeq G_2 \simeq G_0 +G_0 4m_f^2 J_s G_0^2
\ee
in complete agreement with eq. (\ref{Dys-coup1}). The proper value of the $\im
G^R$ is recovered with the help of the standard replacement $m^2_f \rightarrow 
m^2_f -i0$.
In the limit case $\zeta \gsim 1$, $G_2^R$ is already completely regular function.

From (\ref{tblrel}) we obtain that
for $12m_f^2 J_s +4J_s p^2 -(p^2 -m_f^2 )^2 >0$:
\be\label{img2}
\im G_2^R =-\frac{\sqrt{12m_f^2 J_s +4J_s p^2 -(p^2 -m_f^2 )^2}}{2m_f^2 J_s 
(5-\frac{p^2}{m_f^2}+\frac{4J_s}
{m_f^2})}.
\ee 
For
$12m_f^2 J_s +4J_s p^2 -(p^2 -m_f^2 )^2 \leq 0$ one gets $\im G_2^R
\rightarrow 0$.
In the energy-momentum region, where  the pole solution
is absent, 
the condition $\im G_2^R \neq 0$ determines those
energies and momenta, which
contribute to different fermion characteristics, e.g.,
to the fermion 3-momentum 
distribution. 

\subsubsection{Intensity of multiple scattering}
Now we may evaluate the intensity of multiple scattering
$J_s$.
Supposing that 
scalar bosons are good quasiparticles in the relevant energy-momentum region,
from (\ref{sigmJ-s}) 
we find 
\be\label{Jexp}
&&J_s =2g_s^2 \int \frac{d^3 q}{(2\pi)^3}\int_{0}^{\infty}
\frac{d q_0 \delta \left(q_0^2  -\vec{q}^{\,\,2}-m_b^2 -\re \Sigma_b^{R,(2)}
(q_0 , \vec{q})\right)}{e^{q_0 /T}-1}\\
&&=\frac{g_s^2}{2\pi^2}\int_{0}^{\infty}\frac{ \vec{q}^{\,\,2} d |\vec{q}|} 
{ \left[
m_b^{*2}(T)+ \beta_s\vec{q}^{\,\,2}
\right]^{1/2}}\frac{1}{\mbox{exp}\left[\left(
m_b^{*2}(T)+ \beta_s\vec{q}^{\,\,2}\right)^{1/2}
/T\right]- 1} ,\nonumber
\ee
where in the second line we adopted the simple form of the boson spectrum
\be\label{branch}
\om^2_s (\vec{q},T) \simeq m_b^{*2}(T)
+\beta_s (T)\vec{q}^{\,\,2} +O(\vec{q}^{\,\,4})
\ee
which
we reproduce below, see eq. (\ref{branch1}).
In a wide region of temperatures of our interest $\beta_s$
proves to be rather close to unity.

In the limiting case of a high temperature typical values of momenta are
$|\vec{q}|\sim T 
\gg m_b^{*}(T)$ and we obtain
\be\label{Jexp-lim2}
J_s \simeq \frac{g_s^2  T^2}{12\beta_s^{3/2}}
\,,\quad \mbox{for} \quad T\gg m_b^{*}(T) .
\ee

Although the STL approximation is valid only in the high temperature limit,
let us present also the estimate of $J_s$ in the
limit 
$T\ll m_b^{*}(T)$
in order to
show how much this quantity is then suppressed,
\be\label{Jexp-lim1}
J_s \simeq \frac{g_s^2 T^{3/2}\sqrt{m_b^{*}} }
{2^{3/2} \pi^{3/2}\beta_s^{3/2}} 
\mbox{exp}\left( -m_b^{*}/T\right) \,,\quad \mbox{for} \quad T\ll m_b^{*}(T).
\ee
\begin{figure*}
\includegraphics[clip=true,width=8cm]{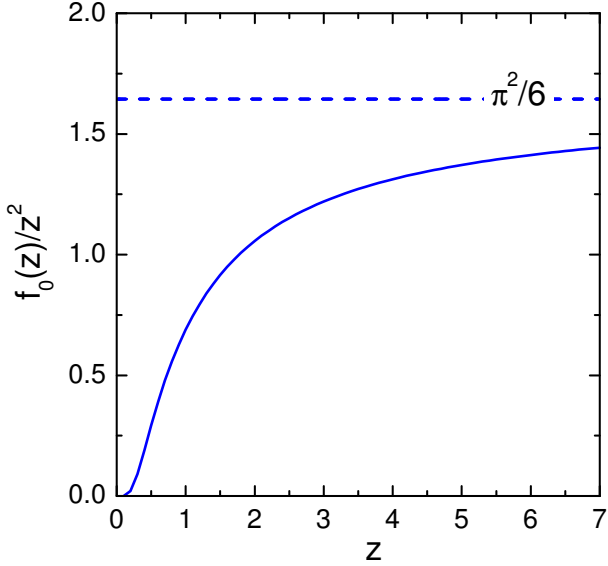}
\caption{$f_0 (z)/z^2$, cf. eq. (\ref{f0}). Dash line demonstrates asymptotic behavior for $z\gg 1$.}
\label{fig:f0.eps}
\end{figure*}
Numerical evaluation of the integral (\ref{Jexp}) is demonstrated in Fig. \ref{fig:f0.eps},
\be\label{f0}
J_s =\frac{g_s^2 (m_b^*)^2 z^2}{12\beta_s^{3/2}}r_s (z), 
\quad r_s (z)=\frac{6f_0 (z)}{\pi^2 z^2}, 
\quad z=\frac{T}{m_b^*}.
\ee
The horizontal line shows the asymptotic behavior (\ref{Jexp-lim2}), $r_s
(z)\rightarrow 1$ for $z\rightarrow \infty$.
Below we will use a simplified eq. (\ref{Jexp-lim2}) to do analytic estimates for $T\gg m_b^{*}(T)$.
In this case the quasiparticle approximation (valid for $\zeta \ll 1$) would work only for
$g_s^2 \ll 3\beta_s^{3/2}$. The latter inequality is not fulfilled for the
$\sigma$ meson ($g_s \sim 10$).

Following (\ref{scat}), (\ref{f0}) 
we may evaluate the density of virtual (off-shell) bosons
in the system, cf. eq. (\ref{ladprtad}), 
\be
\rho_b =2m_b^* (T)(\phi_c^{\rm ext})^2 =m_b^* (T)T^2 r_s /
(6\beta_s^{3/2} ). 
\ee
For $T\simeq m_b^* (T)\simeq m_{\pi}$, $\beta_s \simeq 1$,
using numerical value of $J_s$ shown in
Fig.  \ref{fig:f0.eps} we obtain $\rho_b\simeq 0.1\rho_0 $. 
It is worthwhile to notice that virtual bosons contribute to
thermodynamic quantities, like phonons and diffusion modes do in the condensed matter physics.
The calculation of their contribution to the energy, pressure, entropy, etc is 
however rather non-trivial task, see corresponding expressions for
thermodynamic quantities in 
Appendix \ref{non-rel}.

\subsubsection{Non-relativistic fermion 
distributions and  density of fermion-antifermion pairs.}

Assume $g_s \gsim 1$, $T\gsim m_b^* (T)$
and $4J_s \ll
m_f^2$. Then  typical 4-momenta of fermions are
\be\label{typf}
p^2 \simeq m_f^2 \pm O\left( \sqrt{J_s}m_f\right),
\ee
as we will show it below, see after (\ref{hdist}),
(\ref{ratd1}). With the condition (\ref{typf}) fulfilled, 
we have $|p_0 -m_f |\ll m_f$ and $|\vec{p}|\sim \sqrt{2m_f T}\ll m_f$
for typical energies and momenta. This
means
that conditions for  the applicability
of the non-relativistic approximation for fermions are satisfied.

If the condition (\ref{typf}) is fulfilled, we have $2J_s G_2 \ll 1$. 
Then from (\ref{G1s}) it follows that
$G_1 \simeq G_2$. 
Eqs (\ref{Dys-coup1}) and
(\ref{Dys-coup2}) coincide, if one 
replaces $p^2 \simeq m_f^2$ in squared brackets in (\ref{Dys-coup1}),
(\ref{Dys-coup2}) (but not in the pole term $G_f^{0,R}$).
The same equation also 
follows from (\ref{G2sol}), if one neglects there a small $
4J_s^2 G_2^4 +8J_s G_2^3$ term, according to eq. (\ref{typf}). 

In the non-relativistic limit for fermions eq. (\ref{tblrel}) is simplified as
\be\label{mainsol}
&&G_1^R \simeq G_2^R \simeq \frac{p^2 -m_f^2
\pm \sqrt{(p^2 -m_f^2 +i0)^2 -2\gamma_s^2}}{\gamma_s^2},\\
&&\gamma_s^2 =8J_s m_f^2 .\nonumber
\ee
We suppressed terms of the order of $O( J_s /m_f^2 )$.

We find from (\ref{mainsol}) that the condition $4\sqrt{J_s} m_f \ll
|p^2 -m_f^2 |$ that would permit us to use the quasiparticle approximation is not  satisfied 
for typical values of fermion energies and momenta given by (\ref{typf}). Therefore,
{\em{in the energy-momentum range of our interest
we definitely deal with off-mass shell fermions described by regular
Green functions.}}

The accuracy of the approximation $G_1 \simeq G_2$ made to get (\ref{mainsol}),
cf. eq. (\ref{Jcr}) below, is rather appropriate for temperatures $T\lsim T_{\rm bl.f}$.
The parameter of
the expansion is $2J_s G_2^R (T\leq T_{\rm bl.f})\leq \sqrt{J_s}/m_f
 \leq \frac{1}{2\sqrt{3}}$, see eq. (\ref{Jbl}) below. 

The solution of eq. (\ref{mainsol}) that yields
$\im G_f^R >0$
should be
omitted  as unphysical one. Thus,
for 
the energy-momentum region of our interest
$|p^2 -m_f^2|\lsim \sqrt{2\gamma_s^2}$
there remains only the lower sign  ($-$) solution:
\be\label{mainsol-lov}
&&\re G_1^R \simeq \re G_2^R \simeq \frac{p^2 -m_f^2}{\gamma_s^2},\nonumber\\
&&\im G_1^R \simeq \im G_2^R\simeq -
\frac{\sqrt{2\gamma_s^2 -(p^2 -m_f^2 )^2}}{\gamma_s^2}\Theta \left(
2\gamma_s^2 -(p^2 -m_f^2 )^2\right).
\ee
As follows from 
(\ref{mainsol-lov}), the fermion spectral function
satisfies the full sum rule (\ref{fsum-r}), if both regions $p^2 >m_f^2$
and $p^2 <m_f^2$ are taken into account. Thus, {\em{although we did
approximations, their consistency is preserved.}}
The Green function (\ref{mainsol-lov})
is  the regular function, opposite to the pole solution $G_f^{0,R}$.

Replacing (\ref{mainsol-lov}) into (\ref{3oc}) we find the 
3-momentum fermion distribution
\be\label{hdist}
\widehat{n}_{f,(\pm)} (\vec p )=\int^{2m_f \sqrt{J_s}/\epsilon_p}_
{-2m_f \sqrt{J_s}/\epsilon_p}\frac{d\xi}{\pi}
\frac{\sqrt{2\gamma_s^2 -4\epsilon_p^2 \xi^2}}{\gamma_s^2}
\frac{(\xi +\epsilon_p )  +m_f\gamma_0}{\mbox{exp}[
(\xi +\epsilon_p \mp\mu_f )/T]+ 1},
\ee
where we introduced the variable $\xi =p_0 -\epsilon_p$ and used that $|\xi| 
<2m_f \sqrt{J_s}/\epsilon_p \ll 
\epsilon_p \sim m_f$. We also have dropped the term 
$\propto \vec{p}\,\vec{\gamma}$,
since it does not contribute to the particle density $\rho_{f,(\pm)}$ 
due to the
angular integration and since $\mbox{Tr}\gamma_0 \vec{\gamma}=0$. 
Doing further the replacement
$\xi =-2m_f \sqrt{J_s}/\epsilon_p +Ty$ and using that $\epsilon_p >m_f \gg T$,
we obtain
\be\label{hdist1}
\widehat{n}_{f,(\pm)} (\vec p )&\simeq&
\frac{(\epsilon_p
m_f^{-1}+\gamma_0 )\sqrt{\epsilon_p}\,T^{3/2}}{2\pi m_f^{1/2}J_s^{3/4}}
I_2 \left(
\frac{4m_f \sqrt{J_s}}{\epsilon_p T}\right)\nonumber \\
&\times&\mbox{exp}\left[-\frac{\epsilon_p
- 2m_f \sqrt{J_s}/\epsilon_p \mp \mu_f}{T}\right],
\ee
\be\label{I1}
I_2 (x)=\int^{x}_{0} dy e^{-y}\sqrt{y-y^2 x^{-1}},\quad 
x=4m_f\sqrt{J_s} / (\epsilon_p T),
\ee
\be\label{I1lim}
I_2 (x)&\simeq& \frac{\pi }{8}x^{3/2}(1-\frac{x}{2}) 
\,,\quad \mbox{for} \quad x\ll 1,\\
I_2 (x)&\simeq& \frac{\sqrt{\pi}}{2} \,,\quad \mbox{fo}r \quad x\gg 1. \nonumber
\ee
As we have mentioned, the condition
$x\simeq 4\sqrt{J_s}/ T \ll 1$ 
is not fulfilled within the high temperature limit, which we are interested
here, for $g_s
\gg 1$, 
cf. estimate of $J_s$, eq. (\ref{Jexp-lim2}). Thereby, we show
the result in this limit
only for the completeness of the consideration.

Replacing (\ref{hdist1}) into (\ref{den}) we find
the fermion-antifermion density for one fermion
species:
\be\label{rho-to} 
\rho_{f,\bar{f}}&\equiv&\rho_{f,(+)}+\rho_{f,(-)}
\simeq 
\left[
\mbox{exp}(\mu_f /T)+\mbox{exp}(-\mu_f /T)\right]\nonumber \\
&\times&
\frac{2T^{3/2}}{\pi J_s^{3/4}
m_f^{1/2}} 
\int \frac{d^3 p}{(2\pi)^3}\mbox{exp}\left(\frac{x}{2}-
\frac{\epsilon_p }{T}\right)\epsilon_p^{1/2} I_2 (x) .
\ee
Subsequent integration yields
\be\label{ratd}
\frac{\rho_{f,(\pm)}}{\rho_{{\rm Bol},(\pm)}}\simeq \frac{T^{3/2}}{\pi J_s^{3/4}}
I_2 \left(
\frac{4 \sqrt{J_s}}{T}\right)\mbox{exp}\left[ \frac{2\sqrt{J_s}}{T}\right] ,
\ee
\be\label{ratd1}
\rho_{f,(\pm)}\simeq \frac{m_f^{3/2}T^{3}}{\sqrt{2}\pi^{5/2} J_s^{3/4}}
I_2 \left(
\frac{4 \sqrt{J_s}}{T}\right)\mbox{exp}\left[ 
\frac{-m_f \pm \mu_f +2\sqrt{J_s}}{T}\right] .
\ee
Integrating 
we have set in
$\epsilon_p \simeq m_f +\vec{p}^{\,\,2} /(2m_f)$
in the exponential factor and $\epsilon_p \simeq m_f$ in other values. 
At this instant we are able to support our 
above used estimate of typical fermion momenta given by (\ref{typf}).

In the limit $\sqrt{J_s}/T \rightarrow 0$ (i.e. for $g_s\rightarrow 0$, cf.
eq. (\ref{Jexp-lim2}))
the ratio (\ref{ratd})
of the particle/antiparticle density to the density
of the Boltzmann gas (given by eq. (\ref{r-Bol}))
tends to the unity. 
With the growth of the parameter $\sqrt{J_s}/T$ the ratio (\ref{ratd})
monotonously increases. Thus the density of fermion-antifermion pairs is
exponentially increased compared to the standard Boltzmann value.

The result (\ref{ratd1}) can be interpreted with the help  of two new
relevant quantities
\be\label{effermm}
m_{f, (+)}^* (T) =m_f \left(1-\frac{\mu_f}{m_f} -\frac{2\sqrt{J_s}}{m_f}\right),
\quad 2\sqrt{J_s}\ll m_f ,
\ee
and
\be\label{a-effermm}
m_{f, (-)}^* (T) =m_f \left(1+\frac{\mu_f}{m_f} -
\frac{2\sqrt{J_s}}{m_f}\right),\quad 2\sqrt{J_s}\ll m_f .
\ee
These quantities have the meaning of {\em{effective fermion and antifermion
masses.}} However, contrary to the
usually introduced effective masses, quantities (\ref{effermm}),
(\ref{a-effermm}) enter only the exponent in the expression (\ref{rho-to}).
We see that 
$m_{f, (+)}^* (T)$ and $m_{f, (-)}^* (T)$
decrease with increase of the intensity of the multiple scattering
$J_s$. The latter value   
rises with the temperature, cf. eq. (\ref{Jexp-lim2}). 
In case of the non-zero baryon
density
the antifermion
effective mass proves to be 
slightly higher than the  fermion effective mass. {\em{Thus, for $\mu_f \neq 0$
the fermion mass-shell is blurred a bit more intensively
than 
the antifermion mass-shell.}}

Were
eq. (\ref{ratd}) correct also
for sufficiently large values of $J_s$, 
we could estimate a value 
\be\label{Jcr}
J_s =J_{s, \rm n.rel}^{({\rm bl.f})} (T_{\rm bl.f}^{s, \rm n.rel}) 
\simeq (m_f -\mu_f )^2 /4 ,  
\ee
at which the effective fermion mass (\ref{effermm}) would vanish.
$T_{\rm bl.f}^{s, \rm n.rel}$ would be then
the typical temperature demonstrating a complete
blurring of the gap between
fermion and antifermion continua. Within the non-relativistic approach 
from
(\ref{Jexp-lim2}) and (\ref{Jcr}) 
we would get
\be\label{Tcssn}
T_{\rm bl.f}^{s,\rm n.rel} \simeq
\sqrt{3}\beta_s^{3/4} g_s^{-1}(m_f -\mu_f)r_s^{-1/2} (T_{\rm bl.f}^{s,\rm n.rel}).
\ee
Here the correction factor $r_s$ takes into account deviation of the
numerical value of the integral (\ref{Jexp}) from its asymptotic value (\ref{Jexp-lim2}).

However for
$T>T_{\rm bl.f}^{s,\rm n.rel}$ 
the non-relativistic approximation for fermions is definitely
incorrect, since the exponential factor in (\ref{ratd1}) arose from fermion
occupations, which in any case should be less than unity,  whereas for
$T=T_{\rm bl.f}^{s,\rm n.rel}$ this factor has already reached the unity.
In 
reality, as we show below, the non-relativistic approximation
fails at still smaller temperatures. Thereby,  we supplied here
corresponding 
artificial values by an additional index 
``${\rm n.rel}$''. 

The absolute maximum value of the density, which could be achieved
in the region of the validity of the
non-relativistic approximation,
can be estimated with the help of the replacement of the exponential
factor in (\ref{ratd1}) by unity. Then 
we evaluate 
\be\label{ratd1-max}
\rho_{f,(\pm)}^{\rm max}
\simeq \frac{m_f^{3/2}T^{3}}{2^{3/2}\pi^{2} J_s^{3/4}}
\rightarrow \frac{3^{3/4}\beta_s^{9/8} m_f^{3/2}T^{3/2}}{\pi^2 g_s^{3/2}} ,
\ee
where we have used eq. (\ref{Jexp-lim2}). For $T\simeq m_{\pi}$, $g_s \simeq
10$, $\beta_s \simeq 1$, $m_f \simeq 6.7 m_{\pi}$ we estimate
$2\rho_{f,\bar{f}}^{\rm max}\simeq \rho_0$.

\subsubsection{
Relativistic fermion distributions and  density of 
fermion-antifermion pairs. Blurred hadron continuum }

An exponential smallness of fermion 3-momentum
distributions disappears, if typical
energies satisfy the condition $p_0 \pm \mu_f \lsim T\ll m_f$. Thereby,
and since we consider $|\mu_f| \ll T$, to
keep the $\mu_f$ dependence in equations
becomes even less important in this energy regime.
Thus we further consider the case of the hadron vacuum.
$\im G_2 $ is non-zero at $p_0$ near $0$ only for $J_s >m_f^2 /12$,
as it follows from eq. (\ref{img2}). This means
that such small energies are present with a high probability only for 
$J_s \gsim m_f^2 /12$.
For $p_0\ll m_f$ the 
non-relativistic approximation, which we used above dealing 
with the fermion energies $p_0$ near
the mass shell,
becomes invalid.  Thus, {\em for 
$J_s \gsim m_f^2 /12$ we should adopt the fully
relativistic approach} in order to incorporate the region
of small fermion energies.
If $p_0 , q_0 \ll \sqrt{J_s}$ 
 ($q_0$ is the boson energy variable in the diagrams (\ref{selfzf}), (\ref{ladprtad})),
one may drop the energy
dependence of the fermion Green function at all in the calculation
of the fermion 3-momentum distribution, where typical energies $p_0$ are $\sim T$. 
The typical fermion 3-momentum
is there $|\vec{p}|\sim m_N \gg |\vec{q}|\sim (T, \sqrt{2m_b^* T})$, 
see eq. (\ref{rhobl}) below. 
Thereby the STL approximation continues to hold in the given regime.

The condition of appearance of a non-trivial contribution to the imaginary
part of $G_2$ for small values $p_0 \lsim T$,
\be\label{Jbl}
J_s^{\rm bl.f} (T_{\rm bl.f}^s)\simeq
m_f^2 /12 
\ee 
determines the characteristic temperature of the complete
blurring of the gap
between fermion-antifermion continua. Within the fully relativistic approach
for fermions,
from (\ref{Jexp-lim2}), (\ref{f0}), (\ref{Jbl}) we evaluate the typical temperature
of the blurring of the fermion vacuum,
\be\label{Tcss}
T_{\rm bl.f}^s \simeq \beta_s^{3/4} g_s^{-1}m_f r_s^{-1/2} (T_{\rm bl.f}^s ),\quad \mbox{for} \quad 
T_{\rm bl.f}^s \gsim m_b^* (T_{\rm bl.f}^s),
\ee
for the heavy
fermion - scalar boson system under consideration.
If one assumes $r_s (T_{\rm bl.f}^s )\simeq r_s (T_{\rm bl.f}^{s,\rm n.rel}
)$, 
the quantity (\ref{Tcss}) is $1/\sqrt{3}$ times 
smaller than the artificial value (\ref{Tcssn}) estimated above beyond the
region of the validity of the non-relativistic approximation for fermions.
We would like to draw attention to the fact that 
the limit  $T_{\rm bl.f}^s \ll m_b^* (T_{\rm bl.f}^s)$ is never realized,
since $J_s$ is exponentially suppressed in this case, cf. (\ref{Jexp-lim1}).
The  opposite limit $T_{\rm bl.f}^s \gg m_b^* (T_{\rm bl.f}^s)$
can be realized, only if the bare boson mass is rather small, namely 
$T_{\rm bl.f}^s \gsim
m_b$, cf. next two subsections. Otherwise we have 
\be\label{tblsigma}
T_{\rm bl.f}^s \sim m_b^* (T_{\rm bl.f}^s),
\ee
and the problem of the determination of $T_{\rm bl.f}^s$ is then a 
coupled-channel
problem of a simultaneous evaluation of  quantities $T_{\rm bl.f}^s$ and  
$m_b^* (T_{\rm bl.f}^s)$. 

Now we may find the fermion 3-momentum
distribution for $J_s >J_s^{\rm bl.f}\simeq 
m_f^2 /12$ 
(i.e., for $T> T_{\rm bl.f}^s$). 
Let us first consider 
only the contribution of the energy region $p_0\ll \sqrt{J_s}$. 
For typical values
$p_0 \sim T$, for the case $T\ll \sqrt{J_s}$,
we may put $p_0 =0$
in the expression for $\im G_2 $.
Then we obtain the additional contribution of this energy region to the
3-momentum fermion distribution
\be\label{nrelzero}
&&\delta\widehat{n}_{f, (\pm)}(\vec{p}) \simeq - \int_{0}^{\sqrt{J_s}}
\frac{dp_0 \im G_2^R (p_0 =0)}{\pi}
\frac{( p_0 +m_f\gamma_0 )}{e^{p_0 /T}+1} \nonumber\\
&&\simeq -\frac{\im G_2^R (p_0 =0)}{\pi}
\left(T^2 \frac{\pi^2}{12} +
\gamma_0 T m_f \mbox{ln}2 \right),
\ee
and to the fermion-antifermion density (one species
of fermion)
\be\label{rhobl}
&&\delta\rho_{f,(\pm)}(p_0 \lsim T)
=\frac{T^2}{6\pi}\int_0^{|\vec{p}_{\rm max}|}\vec{p}^{\,2} 
d|\vec{p}|
\frac{\sqrt{12m_f^2 J_s -4J_s \vec{p}^{\,2} -(\vec{p}^{\,2} 
+m_f^2 )^2}}
{2m_f^2 J_s 
(5+\frac{\vec{p}^{\,2}}{m_f^2}+\frac{4J_s}{m_f^2})},\\
&&\vec{p}_{\rm max}^{\,2} =-m_f^2 -2J_s
+\sqrt{4J_s^2 +16J_s m_f^2},\nonumber
\ee
where we used eq. (\ref{img2}).
We see that for 
$J_s
> J_s^{\rm bl.f}$ 
{\em the fermion sub-system represents a rather dense packing of
fermion-antifermion pairs}
($J_s \geq J_s^{\rm bl.f}$ corresponds to the 
filling of a Fermi sea,
$\vec{p}_{\rm max}=0$ for $J_s = J_s^{\rm bl.f}$). This is rather
similar to the standard Fermi distribution
at zero temperature but in our case fermion width effect is significant and
the Fermi momentum $|\vec{p}_{\rm max}|$ has a different value.
Moreover, effective fermion and antifermion Fermi seas exist simultaneously
in our case.

In order to come to the quadratic equation from the fourth order one (see
eq. (\ref{G2sol})) we assumed
that $2J|G_2 | \ll 1$. For $p_0 \simeq 0$ and, e.g., for $J_s \simeq
2J_{s}^{({\rm bl.f})} $ 
we have $|\vec{p}_{\rm max}|\simeq m_f /\sqrt{3}$
and, as it follows from eq. (\ref{tblrel}),
$2J|\re G_2 | \simeq 1/3 $. For $J_s \gg m_f^2$, $|\re G_2^R |\sim 1/
(2J_s )$
and $2J|\re G_2 | \simeq 1$. Thus, 
we may continue to 
use eqs (\ref{tblrel}), (\ref{img2}) also in relativistic energy region
(even for $J_s \gg m_f^2$ for qualitative estimates).

For $T$ in the vicinity of  $T^s_{\rm bl.f}$ ($J_s -J_s^{\rm bl.f}\ll  
J_s^{\rm bl.f}$)
using that $|\vec{p}_{\rm max}|=0$ for 
$J_s =J_s^{\rm bl.f}$ and expanding   in (\ref{rhobl}) all quantities in
small difference $J_s -J_s^{\rm bl.f}$
we obtain $\vec{p}_{\rm max}^{\,2}=36(J_s -J_s^{\rm bl.f})/7$ and
\be\label{rhobl-1}
\delta \rho_{f,(\pm)}(p_0 \lsim T)
=\frac{81\sqrt{3} T^2 (J_s -J_s^{\rm bl.f})^2}
{112\sqrt{7}m_f^3}\Theta (J_s -J_s^{\rm bl.f}).
\ee
Notice that the total value 
$\rho_{f,(\pm)}$ does not tend to zero for 
$J_s =J_s^{\rm bl.f}\simeq m_f^2 /12$. Indeed, one should still add to (\ref{rhobl-1})
the contribution of the energy region near the mass-shell, which we have
estimated above within the non-relativistic approximation, see  (\ref{ratd1}).

Suppose $g_s \simeq 10$, as for the $\sigma NN$ interaction, and
$m_f =m_N \simeq g_s f_\pi$. 
Assuming $m_b^* (T)$ being significantly less than  $T$ and using (\ref{Jexp-lim2}) thereby,
we would get  $T_{\rm bl.f}^{s,0}\simeq 93$~MeV for $\beta_s \simeq 1$, 
see (\ref{Tcss}). 
From (\ref{ratd1})
for two fermion (nucleon) species and for
$T=T_{\rm bl.f}^{s,0}\simeq 93$~MeV, 
we then estimate $\rho_{N,\bar{N}}=
2\rho_{f,\bar{f}}(p_0 \sim m_f)\simeq 4\rho_{f,(+)}(p_0 \sim m_f)\simeq
0.01\rho_0$. 
This is a tiny quantity.
It means that in reality we still have 
$m^*_b (T)\simeq m_b $
at such a temperature. 
Thus the estimate 
$T_{\rm bl.f}^{s,0}\simeq 93$~MeV is not relevant, e.g., for the
$N\sigma$ system, since the
bare mass of $\sigma$
is $(500\div 600)$~MeV, i.e. much higher than $93$~MeV. We introduced a
superscript $''0''$ to indicate this artificial feature.

The $\sigma$ meson has  a large bare mass. 
Therefore we actually deal here with a coupled-channel 
problem, see eq. (\ref{tblsigma}). 
In presence of nucleon-antinucleon pairs the effective mass of the
$\sigma$ meson decreases that permits an extra
production of pairs. 
The value 
of the nucleon pair density $\rho_{N,\bar{N}}$, at which the nucleon continuum
is blurred, proves to be much
smaller than the density that is necessary to reach the deconfinement
transition at such a low temperature.  
Thus we should incorporate the factor $r_s <1$, cf.  eq. (\ref{f0}). 
We will correct above estimate after
evaluation of the value $m_{\sigma}^* (T)$, see below.
Also, as we shall see below, with inclusion of
correlation effects the effective coupling constant $g^*_s$ becomes smaller than $g_s$
that results in an increase of the value $T_{\rm
  bl.f}^{\sigma }$.

For artificially large values $J_s (T)\gg m_f^2$ using (\ref{rhobl}),
(\ref{Jexp-lim2})
we estimate
$\vec{p}_{\rm max}^{\,\,2}\simeq 3m_f^2$ and
\be\label{rhoblart}
\rho_{f,(\pm)}\simeq \frac{3 T^2 m_f^4 }
{128 J_s^{3/2}}\rightarrow \frac{3^{5/2}\beta_s^{9/4}m_f^4}{16 g_s^3 T}.
\ee
The contribution of the energy region near the mass shell
evaluated  above within the non-relativistic approximation should be omitted
in this limit. Thereby we replaced $\delta \rho_{f,(\pm)}$ to $\rho_{f,(\pm)}$.
Also one should bear in mind that in derivation (\ref{rhobl})
we have put $p_0 =0$
in the estimate of $\im G$,
although for $J_s (T)\gg m_f^2$ the whole energy region is populated. Moreover, we
suppressed $G_3$ term 
in (\ref{spinstr}) that might be 
incorrect for $\vec{p}^{\,\,2}\gsim m_f^2$.
Thus (\ref{rhoblart}) can be considered only as a very rough estimate.

Even at such high temperatures (for $J_s (T) \gg m_f^2$)
we find no end point for
the hadron world conjectured
in \cite{D93} 
(the value $d\rho_{f,(\pm)} /dT$ 
has no singularity at finite $T$ in our case). 
Furthermore, we see that
$\rho_{f,(\pm)}$ may  even decrease with the temperature
increase in this energy  region. 
Maximum available density of fermions can be very roughly
estimated equating
(\ref{rhobl-1}) plus (\ref{ratd1}) (in the latter equation
we take into account that the exponential factor should not exceed unity, see
eq. (\ref{ratd1-max})), 
and on the other hand (\ref{rhoblart}), 
from where we find $J_s (\rho_f^{\rm max}) 
\simeq 0.4 m_f^2 $, $T\simeq 204~$MeV, 
and $2\rho_{f,\bar{f}}^{\rm max}
\simeq
10\rho_0$ for $g_s \simeq 10$. 
Although estimates of many works show that 
the  quark 
deconfinement transition may occur at much  smaller temperature for given high
density,
all of them are done within simplified assumptions. E.g., one often
compares pressures
of the quark-gluon and hadron gases to conclude about the possibility of
the deconfinement transition.
We found that {\em{the hadron phase represents in reality
a strongly correlated state,
where the number of effective 
hadron degrees of freedom is dramatically increased with the temperature}}
following the increase of $J_s (T)$.
Thereby,  one may expect a smoothening of the transition. More likely, 
in this case
{\em{ the system
up to rather high temperatures may
represent a strongly correlated hadron-quark-gluon state
rather than the pure quark-gluon or the pure hadron state.}}

Note that calculating $\widehat{\Sigma}_f$, cf. eq. (\ref{sigm-R0}), 
we omitted terms proportional to fermion occupations.
These terms are as small as the ratio of the contribution of quantum
fluctuations to thermal fluctuations:
\be
\frac{\int A_b n_f (p_0 +q_0 )\frac{d^4 q}{(2\pi )^4}}{
\int A_b n_b (q_0)\frac{d^4 q}{(2\pi )^4}}
<
\frac{\int A_b \frac{d^4 q}{(2\pi )^4}}{
\int A_b n_b (q_0)\frac{d^4 q}{(2\pi )^4}}.
\ee
We also notice
that the value $J_{s}^{({\rm bl.f})}$ evaluated within the relativistic
approach 
is 
smaller than the quantity $J_{s, \rm n.rel}^{({\rm bl.f})}$ 
that would follow from  the non-relativistic estimation
outside the  region of its validity.

\subsubsection{Boson spectrum in the regime of non-relativistic fermions. 
Possibility of hot Bose condensation}\label{Bose}
We discussed the behavior of fermion Green functions and 
self-energies.
Now let us evaluate the boson self-energy and find the boson spectrum.  
We will continue to exploit the high temperature limit ($T\gsim m^*_b (T)$)
using fermion Green functions obtained in the STL approximation.
Let us also assume, as before,  that  $G_3 =0$,
$G_1^R \simeq G_2^R$, and that boson 4-momenta
$q_0 ,|\vec{q} |$ are rather small ($q_0 ,|\vec{q} |\ll m_f $, 
see estimation below).  
Then, dropping  in 
eq. (\ref{sigm-R0-first})
a small
term, which does not depend on thermal fermion occupations, 
and expanding  (\ref{sigm-R0-first}) in 
$q_0$, $|\vec{q}|$ we find
\be\label{sigm-R0-fap} 
&&\re \Sigma_b^{R} (q) \simeq -16g^2_s 
\int \frac{d^3 p}{(2\pi)^3}\int_{0}^{\infty}\frac{dp_0}{(2\pi)}
 (p^2 +m_f^2 )
\im G_2^{R} (p)\nonumber\\
&&\times [\re G_2^{R} (p)n_{f,(+)}(p_0)+\re G_2^{R} (-p)
n_{f,(-)}(p_0)] \nonumber\\
&&-8g^2_s 
\int \frac{d^3 p}{(2\pi)^3}\int_{0}^{\infty}\frac{dp_0}{(2\pi)}
\left[ 
\frac{d^2\re G_2^{R} (p)}{dp^{\mu}dp^{\nu}}n_{f,(+)}(p_0)
+\frac{d^2\re G_2^{R} (-p)}{dp^{\mu}dp^{\nu}}n_{f,(-)}(p_0)
\right]
\nonumber\\
&&\times q^{\mu}q^{\nu}
(p^2 +m_f^2 )\im G_2^{R} (p)
\nonumber\\
&&-16g^2_s 
\int \frac{d^3 p}{(2\pi)^3}\int_{0}^{\infty}\frac{dp_0}{(2\pi)}
\left[ 
\frac{d\re G_2^{R} (p)}{dp^{\mu}} n_{f,(+)}(p_0)
+\frac{d\re G_2^{R} (-p)}{dp^{\mu}} n_{f,(-)}(p_0)
\right]\nonumber\\
&&\times q^{\mu}\cdot qp
\im G_2^{R} (p)
.
\ee
In our case  according to (\ref{tblrel}) $\re G_2^{R} (p)= 
\re G_2^{R} (-p)$ and eq. (\ref{sigm-R0-fap}) is still simplified.

Let us first follow the approximation 
of non-relativistic fermions, $2\sqrt{J_s}\ll m_f$.
With the help of eq. (\ref{mainsol-lov}), 
doing the replacement $\xi =p_0 -\epsilon_p$, for $\xi
\ll \epsilon_p$, and then introducing
the variable $\xi = -2m_f \sqrt{J_s}/\epsilon_p +Ty$ we get
\be\label{sigm-R0-fin} 
&&\re \Sigma_b^{R}\simeq 
-\frac{2g^2_s \rho_{f,\bar{f}}}{\sqrt{J_s}} 
\left[ 1-\frac{T}{2\sqrt{J_s}}
I_3 \left(\frac{4\sqrt{J_s}}{T}\right)I_2^{-1} 
\left(\frac{4\sqrt{J_s}}{T}\right)\right] \nonumber \\ 
&&+\alpha_s (q_0^2 -\frac{1}{2}\vec{q}^{\,2}),
\ee
where 
we also used that $|\vec{p}|\sim \sqrt{2m_f T}\ll m_f$
and we have 
put $\epsilon_p \simeq m_f$ everywhere except the exponential factor 
$\mbox{exp}(-\epsilon_p /T)$.
The quantity $\rho_{f,\bar{f}}$ is the fermion-antifermion  density
for one fermion species, given by eq. (\ref{ratd}), the parameter
\be\label{alpha}
\alpha_s =\frac{2g^2_s \rho_{f ,\bar{f}}}
{J_s m_f}
\ee
is associated with the renormalization of the boson quasiparticle wave
function, cf. eq. (\ref{branch}),
\be\label{I3}
I_3 (x)=\int^{x}_{0}dy e^{-y} y\sqrt{y-y^2 /x} .
\ee
In the limit cases we get
\be\label{I3ap}
&&I_3 (x)\simeq \frac{\pi}{16}x^{5/2}(1-\frac{5x}{8})
\,,\quad \mbox{for} \quad x\ll 1,\nonumber \\
&&I_3 (x)\simeq \frac{3\sqrt{\pi}}{4}\,,\quad \mbox{for} \quad x\gg 1. 
\ee
Thus
\be\label{sigm-R0-finlim1} 
\re \Sigma_b^{R}(q)&\simeq&-\frac{g_s^2 \rho_{f,\bar{f}} }{
T} +\alpha_s (q_0^2 -\frac{1}{2}\vec{q}^{\,2}) ,
 \quad \mbox{for} \quad 
x\simeq \frac{4\sqrt{J_s}}{T}\ll 1, 
\ee
whereas in the high temperature limit under consideration
\be\label{sigm-R0-finlim2} 
\re \Sigma_b^{R}(q)&\simeq& 
-\frac{2 g_s^2 \rho_{f,\bar{f}} }{
\sqrt{J_s}}+\alpha_s (q_0^2 -\frac{1}{2}\vec{q}^{\,2}), \quad \mbox{for} \quad x
\simeq \frac{4\sqrt{J_s}}{T}\gg 1. 
\ee
As follows from eq. (\ref{Jexp-lim2}), the limit $x\gg 1$ is indeed fulfilled, if coupling is
strong
($g_s \gg 1$).
In the approximation (\ref{oc-sim}), $\mu_f /T \ll 1$, the real part of the
boson self-energy does not depend on $\mu_f$ up to $O(\mu_f^2 /T^2)$ terms.
Comparing $q$-independent and $q$-dependent terms in
eqs (\ref{sigm-R0-finlim1}) and (\ref{sigm-R0-finlim2}) we see that
expansions hold up to rather large values of $q_0^2$ and $\vec{q}^{\,\,2}$:
$q_0^2 , \vec{q}^{\,\,2} \ll \sqrt{J_s}m_f\sim m_f^2 /\sqrt{12}$ 
for $T\sim T_{\rm bl.f}^s$.

Assuming the
validity of the quasiparticle approximation for bosons
($\im \Sigma_b^{R,(0)}\rightarrow
0$), 
we find the spectrum of boson excitations
that takes the form (\ref{branch}) with 
\be\label{branch1}
\beta_s =(1-\frac{1}{2}
\alpha_s )/(1-\alpha_s ).
\ee
The wave function renormalization parameter $\alpha_s$
yields corrections to the $\om^2$ 
and $\vec{q}^{\,2}$ terms. 

The value
\be\label{mbr}
m_b^{*2}(T)\simeq (1-\alpha_s)^{-1}
\left[ m_b^2 -
\frac{2g_s^2 \rho_{f,\bar{f}}}{\sqrt{J_s}} \left( 1-\frac{TI_3 (x)}{ 
2\sqrt{J_s}I_2 (x)
}\right)\right]
\ee
has the meaning of the squared 
effective boson mass. For 
the most interesting case, $x\gg 1$, we obtain
\be\label{mbr-l}
m_b^{*2}(T)\simeq 
m_b^2 (1-\alpha_s)^{-1}\left(1-
\frac{2g_s^2 \rho_{f,\bar{f}}}{m_b^2\sqrt{J_s}}\right), 
\quad 4\sqrt{J_s}\gg T .
\ee
Thus the 
effective boson mass achieves zero at  some critical temperature $T_{\rm cb}$,
being determined by the condition
\be\label{Jcb}
\sqrt{J_{cs}^{(b)} (T_{\rm cb}^s )} =\frac{2g_s^2 \rho_{f,\bar{f}} (T_{\rm cb}^s 
)}{m_b^2}.
\ee
Assuming that the non-relativistic approximation for fermions is 
fulfilled up to $T\sim T_{\rm cb}^s$ we  estimate
\be\label{Jcb2}
\rho_{f,\bar{f}} (T_{\rm cb}^s )=
\frac{m_b^2 T_{\rm cb}^s}{4\sqrt{3}\beta_s^{3/4} g_s}.
\ee
For $T_{\rm cb}^s \gsim T^s_{\rm bl.f}$
we have $\alpha_s\propto m_b^2 /m_f^2 \ll 1$ and $\beta_s \simeq
1+\frac{1}{2}
\alpha_s $.
To get expression (\ref{Jcb2})
 we used 
eq. (\ref{Jexp-lim2}). Note that at $T=T_{\rm cb}^s$ the correction factor
$r_s =1$.
For $m_{\sigma}\simeq 600$~MeV, $T^s_{\rm cb}\simeq
m_{\pi}$, $g_s \simeq 10$, $\alpha_s =0$ we estimate $\rho_{N,\bar{N}}\simeq 0.7\rho_0$.

For $T>T_{\rm cb}^s$ the value $m_b^{*2}(T)$ becomes negative
leading to the instability of the spectrum. 
The
stability is recovered due to the appearance of 
the $s$-wave Bose condensation of the classical
scalar field. Such a condensation can be called 
{\em a hot Bose condensation} (HBC), since it arises for 
the temperature
$T>T_{\rm cb}^s$, rather than for $T<T_{\rm cb}^s$. As the consequence of the strong
boson-fermion-antifermion interaction, the number of fermion degrees of
freedom is dramatically increased that, on the other hand,
results in the increase of the boson
abundance.
Boson degrees of freedom feel 
a lack of the
phase space for energies and momenta $\sim T$
and a part of them is forced to
occupy the coherent condensate state, thereby.

Let us now show that the quasiparticle
approximation for bosons, which we have assumed to be valid exploiting (\ref{branch}),
is, indeed, fulfilled in a wide temperature range.
For that
let us evaluate $\im \Sigma_b^{R}$. Within the validity of
 the non-relativistic approximation for fermions, one has
$2\sqrt{J_s}\ll m_f$. 
Then from (\ref{sigm-R0-first1}) and
(\ref{mainsol-lov}) we find
\be\label{sigm-R0-im1}
&&\im \Sigma_b^{R}(q)\simeq -\frac{g^2_s}{8m_f^4 J_s^2}
\int \frac{d^4 p}{(2\pi)^4}\sqrt{2\gamma_s^2 -\left((p_0 +q_0)^2 -
\epsilon_{\vec{p}+\vec{q}}^2 \right)^2}\nonumber\\
&\times& \left(p^2 +m_f^2 +pq \right)
\sqrt{2\gamma_s^2 -\left(p^2_0 -\epsilon_{p}^2 \right)^2}\nonumber\\
&\times&\left[ \frac{1}{\mbox{exp}[(p_0 -\mu_f )/T]+ 1}-
\frac{1}{\mbox{exp}[(p_0 +q_0 -\mu_f )/T]+ 1}\right] .
\ee
We see that in the critical point of the HBC ($q_0 =0$)
the squared bracketed term vanishes and, thereby, 
the boson width also vanishes. 
Thus in the problem of the determination of the critical point of the HBC
one, indeed, may use the quasiparticle approximation for bosons.

Now let us consider finite but rather small values of $q_0$ and $|\vec{q}|$.
For $q_0 \ll T$, $|\vec{q}|\ll \sqrt{m_f T}$ we may drop the
$q$-dependence everywhere except
particle occupation factors.
Separating fermion particle and antiparticle contributions, with the help of
the replacement $p_0 =\epsilon_p -\frac{2m_f \sqrt{J_s}}{\epsilon_p}+Ty$,
we obtain
\be\label{sigm-R0-exp2} 
&&\im \Sigma_b^{R}(q)\simeq 
-\frac{2g^2_s T^2\left[
\mbox{exp}(\mu_f /T)+\mbox{exp}(-\mu_f /T)\right]
e^{-q_0 /T} \left( e^{q_0 /T}-1\right) 
}{\pi  J_s^{3/2}}\nonumber\\
&\times&\int \frac{d^3 p}{(2\pi)^3} e^{-\epsilon_p /T}
\mbox{exp}\left(\frac{2m_f \sqrt{J_s}}{ \epsilon_p T}
\right)I_4 \left(\frac{4m_f \sqrt{J_s}}{\epsilon_p T}\right),
\ee
\be\label{I4}
I_4 (x)=\int^{x}_{0}e^{-y}dy \left( y-y^2/x\right)=e^{-x}
\left(1+\frac{2}{x}\right)+1-\frac{2}{x} ,
\ee
\be\label{I4ap}
I_4 &\simeq& \frac{x^2}{2}\,,\quad \mbox{for} \quad x\ll 1,\nonumber\\
I_4 &\simeq& 1-\frac{2}{x}\,,\quad \mbox{for} \quad x\gg 1.
\ee
Finally, we find
\be\label{sigm-R0-imfin} 
\im \Sigma_b^{R}&\simeq&-\frac{1}{2}\alpha_s 
T^{1/2} J_s^{1/4}m_f
I_4 \left(\frac{4\sqrt{J_s}}{T}\right)\nonumber\\
&\times&
I_2^{-1} \left(\frac{4\sqrt{J_s}}{T}\right)e^{-q_0 /T} \left( e^{q_0 /T}-1
\right) . 
\ee
Comparing (\ref{sigm-R0-imfin}) and (\ref{sigm-R0-fin}) we see that in 
the high temperature limit under consideration, for
$x\simeq \frac{4\sqrt{J_s}}{T}\gg 1$, and for $q_0^2 \ll T\sqrt{J_s}$
one has $|\im \Sigma_b^{R}|\ll
|\re \Sigma_b^{R}|$. 
The  inequality $|\im \Sigma_b^{R}|\ll
|\re \Sigma_b^{R}|$ also holds 
for $q_0 \ll \sqrt{J_s}$ at $\frac{4\sqrt{J_s}}{T}\ll 1$.
Thus, {\em the quasiparticle approximation is, indeed, valid
for bosons} in a wide temperature range of our interest.
For typical values $q_0\sim T$
we have $|\im \Sigma_b^{R}|\ll
|\re \Sigma_b^{R}|$ in both limits $x\gg 1$
and $x\ll 1$. Thereby, we justified that above
we correctly used the quasiparticle
approximation to calculate $J_s$.
On the other hand, the 
quasiparticle approximation fails 
for $|\im \Sigma_b^{R}(q_0=m_b^{*}(T))|
\gsim
m_b^{*2}(T)$, that takes place in a narrow vicinity of the HBC critical point, but not
in the critical point itself, where $\im \Sigma_b^{R}=0$.

\subsubsection{  
Hot Bose condensation in the regime of  blurred fermion continuum.}
\label{Bose-rel}
For $J_s > J_s^{\rm bl.f}$ the fermion
energy region $p_0 \ll m_f$ is permitted. Due to that there appears an
additional term in the boson self-energy. To find it
let us consider the limit $J_s - J_s^{\rm bl.f}\ll J_s^{\rm bl.f}$, for
$J_s > J_s^{\rm bl.f}$.
Then still one has $G_1 \simeq G_2$ with a reasonable accuracy.
With the help of eq. (\ref{tblrel}) from (\ref{sigm-R0-fap})
within the same set of approximations, 
which we have used to obtain eq. (\ref{rhobl}),
we find 
\be\label{sigm-R0-finlim2rel} 
\delta \re \Sigma_b^{R}(q)&\simeq& 
-\frac{36(\mbox{ln}2)  g_s^2 \delta\rho_{f,\bar{f}}(p_0 \lsim T) }{
\pi^2 T_{\rm bl.f}^s}\left( 1-\frac{9}{4}\frac{q_0^2}{m_f^2}+
\frac{3}{4}\frac{\vec{q}^{\,2}}{m_f^2}\right), 
\ee
where $\delta\rho_{f,\bar{f}}(p_0 \lsim T)=\delta\rho_{f,(+ )}+
\delta\rho_{f,(-)}$, see eq. (\ref{rhobl}).
Eq. (\ref{sigm-R0-finlim2rel}) yields the correction term to the
effective boson mass (\ref{mbr-l})
\be\label{mbr-lcor}
m_b^{*2}(T)\simeq 
m_b^2 (1-\alpha_s)^{-1}\left(1-
\frac{2g_s^2 \rho_{f,\bar{f}}}{m_b^2\sqrt{J_s}}-
\frac{36(\mbox{ln}2)  g_s^2 \delta\rho_{f,\bar{f}}(p_0 \lsim T) }{
\pi^2 T_{\rm bl.f}^s m_b^2}\right) ,
\ee
where we dropped a numerically small contribution  to the wave function
renormalization from the region $p_0 \lsim T$.

If the 
effective boson mass $m_b^{*}(T)$ achieves zero in the regime $T>T_{\rm bl.f}^s$,
$T-T_{\rm bl.f}^s \ll T_{\rm bl.f}^s$,
then the critical point of HBC  is determined by the condition
\be\label{Jcbrel}
m_b^2 -\frac{2g_s^2 \rho_{f,\bar{f}}(p_0 \sim m_f )}{\sqrt{J_s}}-
\frac{36\mbox{ln}( 2)  g_s^2 \delta\rho_{f,\bar{f}}(p_0 \lsim T) }{
\pi^2 T_{\rm bl.f}^s} =0,
\ee
where we used eqs (\ref{sigm-R0-finlim2}),
(\ref{branch}) and (\ref{sigm-R0-finlim2rel}). 

Applying results for $N\bar{N}\sigma$ we should replace  
$\rho_{f,\bar{f}}\rightarrow 2\rho_{f,\bar{f}}=\rho_{N\bar{N}}$ and
$\delta\rho_{f,\bar{f}}\rightarrow 2\delta\rho_{f,\bar{f}}=
\delta\rho_{N\bar{N}}$ in boson 
self-energy terms to
take into account two fermion species, i.e., neutrons and protons
in the given case.
Numerical estimate shows that for 
$m_{\sigma}\simeq 600$~MeV and $g_s \simeq 10$, the critical temperature of
the HBC is
$T_{\rm cb}^s\simeq 114$~MeV,
$J_{cs}^{(b)} (T_{\rm cb}^s )\simeq 1.5 J^{\rm bl.f}_s (r_s =1)$,
$\alpha_s (T_{\rm cb}^s )\simeq 0.3$.
As we mentioned, if the limit $T\gg m_{\sigma}^* (T)$ were fulfilled for
$T\simeq T_{\rm bl.f}^{s}$, 
using (\ref{Jexp-lim2})
we would come back to the estimate
$T_{\rm bl.f}^{s,0}\simeq m_f \beta_s^{3/4} /g_s \simeq 93$~MeV. However
for such a temperature $m_{\sigma}^*$ is still
close to the bare mass and one needs to use  the opposite
limit expression (\ref{Jexp-lim1}). 
Thereby, the value $T_{\rm bl.f}^s$ should be somewhat larger. 
Within a coupled channel
estimate we find that the renormalized value
$T_{\rm bl.f}^{s} \gsim m^*_{\sigma}(T_{\rm bl.f}^{s} )$.
Finally we find that $T_{\rm bl.f}^{s}\simeq 105$~MeV, that 
is only slightly less than $T_{\rm cb}^s$.
Using instead of $g_s$ a smaller value of the
effective coupling constant $g_s^*\simeq 0.7 g_s$, as we estimate it below,
we obtain higher values $T_{\rm bl.f}^{s}\simeq 150~$MeV and $T_{\rm
cb}^s \simeq 160$~MeV. 
For smaller value $m_{\sigma}$ (e.g., for $500$~MeV
instead of $600$~MeV) we would get smaller values of
$T_{\rm bl.f}^{s}$ and $T_{\rm cb}^s$. With these variations we see
that in all relevant cases
$T_{\rm bl.f}^{s}$ and $T_{\rm cb}^s$ 
remain to be somewhere in the  vicinity
of the pion mass ($\sim 140$~MeV).

Please notice that in order to find a possible
relation between the HBC and the chiral symmetry 
restoration
one would need to consider both possibilities in the framework of the 
very same model, e.g., the linear
$\sigma$-model, introducing meson self-interaction terms
and the spontaneous symmetry breaking for the $\sigma$
vacuum at $T=0$. As an intriguing circumstance, let us mention that replacing
$g_{\sigma}\simeq m_N /f_{\pi}$ (for $\sigma =f_{\pi}$, as it follows from the
$\sigma$ model)
into (\ref{Tcss}) and (\ref{Tcssn})
we find 
$T_{\rm bl.f}^s =f_{\pi}$ and  $T_{\rm bl.f}^{s,\rm n.rel}=\sqrt{3}f_{\pi}$
for $\beta_s \ll 1$. 
Namely in this range of temperatures one expects the chiral
restoration phase transition, cf. \cite{RW}. 
We postpone a more detailed discussion of these questions
to the future work. 

For $T>T_{\rm cb}^s$, in the mean field
approximation, the classical scalar field $\phi_{sc}$
is determined by the equation
\be\label{phicl}
&&m_{\rm MF}^{2}\phi_{sc} +\lambda_s^{\rm ef} \phi_{sc}^3 =0,\\ 
&&m_{\rm MF}^{2}= m_b^{2}+\re \Sigma_b^{R}(q\rightarrow 0)+
\delta \re \Sigma_b^{R}(q\rightarrow 0),\nonumber
\ee
which has the solution $\phi_{sc}^2 
\simeq - (m_{\rm MF}^{2}/\lambda_s^{\rm ef} )\Theta (
-m_{\rm MF}^{2})$
for $T>T_{\rm cb}^s$. Here the value 
$\lambda_s^{\rm ef} >0$ is the effective boson self-interaction
coupling constant,
\be\label{lambphi}
L_{int} =-\lambda_s^{\rm ef} \phi_{sc}^4 /4 . 
\ee
It arises since for $T>T_{\rm cb}^s$
one needs to add the condensate
dependent terms to the one term $\Phi$-diagram. With inclusion of 
this interaction excitations become stable in presence of the
condensate.  In reality one also has an extra term in the Lagrangian,
$L_{int} =-\lambda_{\rm vac} \phi_{sc}^4 /4$, related to the vacuum boson-boson
self-interaction. We for simplicity suppressed the latter.

The condensate field is the static field corresponding to the absence of real scalar
particles with zero momentum.
The contribution of the condensate to the free energy density becomes
\be\label{fr-en}
\delta F (V,T)=- \frac{m_{\rm MF}^{4}(T)}{4\lambda_s^{\rm ef} }\Theta
(T-T_{\rm cb}^s )\propto -(T-T_{\rm cb}^s )^2 \Theta
(T-T_{\rm cb}^s ),
\ee
demonstrating typical second order phase transition behavior, 
however for $T>T_{\rm cb}^s $ rather than for  $T<T_{\rm cb}^s$, as it would
take place
for
the ordinary phase transition.

\subsubsection{Boson abundance in matter and at infinity}

The boson population in the medium
is greatly enhanced with the increase of the temperature
due to the decrease of the effective boson mass.
Accordingly, {\em{the distribution function (\ref{3oc}) 
has a sharp peak at small momenta}} $|\vec{q}|^2 \lsim
m_b^{*\,2}(T)/\beta_s$: 
\be\label{nbos}
\frac{dN_{s}^{\rm med}}{d^3 q /(2\pi)^3}=\mbox{Tr}
\,\widehat{n}_{s,(\pm)}(\vec{q})=
\frac{(1-\alpha_s )^{-1}}{e^{\omega_{s} (\vec{q},T)/T}-1},
\quad \mbox{for} \quad T<T_{\rm cb}^s ,
\ee
with $\omega_{s} (\vec{q},T)$ given by (\ref{branch}).
Distributions of particles at infinity might be significantly different
from distributions inside the matter. This depends on the scenario of the 
breakup stage. If breakup were sudden, then particle distributions
at infinity would be given by  \cite{VS89,MSTV}
\be\label{nbos-en}
\frac{dN_{s}^{\infty}}{d^3 q /(2\pi)^3}\simeq \frac{\sqrt{m_s^2 +
\vec{q}^{\,2}}}{\omega_{s} (\vec{q},T)(1-\alpha_s )}
\frac{dN_{s}^{\rm med}}{d^3 q /(2\pi)^3}
,
\quad \mbox{for} \quad T<T_{\rm cb}^s .
\ee
For sudden change of the system the particle momentum is conserved, 
whereas the
particle energy might change. However the total energy is certainly conserved.
Thus eq. (\ref{nbos-en}) should be still
supplemented by the requirement  of the
conservation of the total energy. The energy mismatch that
arises at the breakup stage
is compensated by  the
change of the energy of the particle collective flow.

For $T>T_{\rm cb}^s$, in the vicinity of the critical point
the stable spectrum of excitations
is determined by the equation
\be\label{branch-b}
(1-\alpha_s )\om^2_{s} (\vec{q},T)
\simeq m_{\rm MF}^{2}(T) +3\lambda_s^{\rm ef} (T)\phi_{sc}^2 +(1-\frac{1}{2}
\alpha_s (T))\vec{q}^{\,2} , 
\ee
for $\alpha_s <1$, $|\vec{q}|>0$. Here we used eq. (\ref{sigm-R0-fin})
and added the contribution of the classical condensate field. The latter
is found with the help of eq. (\ref{lambphi}), if one does there the variable replacement
$\phi_{sc}\rightarrow \phi_{sc}+\phi^{\prime}$ to recover
fluctuation-condensate coupling terms.
To avoid a more cumbersome expression
we dropped here a numerically 
small correction term to the wave function
renormalization from the region $p_0\lsim T$.

For the temperature in the vicinity of $T_{\rm cb}^s$
the boson behaves as almost massless particle, 
$\omega^2 \simeq \vec{q}^{\,2}$ (for $\alpha_s \ll 1$).
This results in the enhancement of  the  production of soft bosons
and in the corresponding enhancement of the total
particle yield compared to that would be for the originally massive bosons.

Observation of a $\delta$-function-like peak in the meson distribution
at zero momentum, if occurred, 
could be interpreted, as the fulfillment of the condition
$T_{\rm b.up}\geq T_{\rm bc}$, where $T_{\rm b.up}$ is the
temperature reached at the  breakup stage.
A significant enhancement of the meson distribution at small momenta 
$|\vec{p}|\lsim T$ can be interpreted as a signal of
the closeness of $T_{\rm b.up}$
to the value $T_{\rm bc}$ (for $T_{\rm b.up}< T_{\rm bc}$), see also \cite{V94}.

The HBC may appear only, if $T_{\rm cb}^s$ is less than the 
critical temperature for
the deconfinement,
$T_{\rm dec}$, since 
for $T>T_{\rm dec}$ there would occur a complete 
breakdown of the hadron vacuum. As we 
estimated $T_{\rm cb}^s \sim m_{\pi}$, and as we
argued, at relevant rather small density the deconfinement is probably
delayed up to a  high temperature. Please notice that 
to simplify the consideration
we disregarded in our analysis
the
quark-gluon contribution to hadron quantities, e.g., an extra decrease of 
meson masses due to the change of the quark condensate with the 
increase of the temperature. 
 
Concluding, {\em{ 
we treated the fermion-boson problem self-consistently. Fermions
due to rescatterings
on bosons acquire broad widths and, as the reaction on that, bosons decrease
their masses.}}

\begin{figure*}
\includegraphics[clip=true,width=9cm]{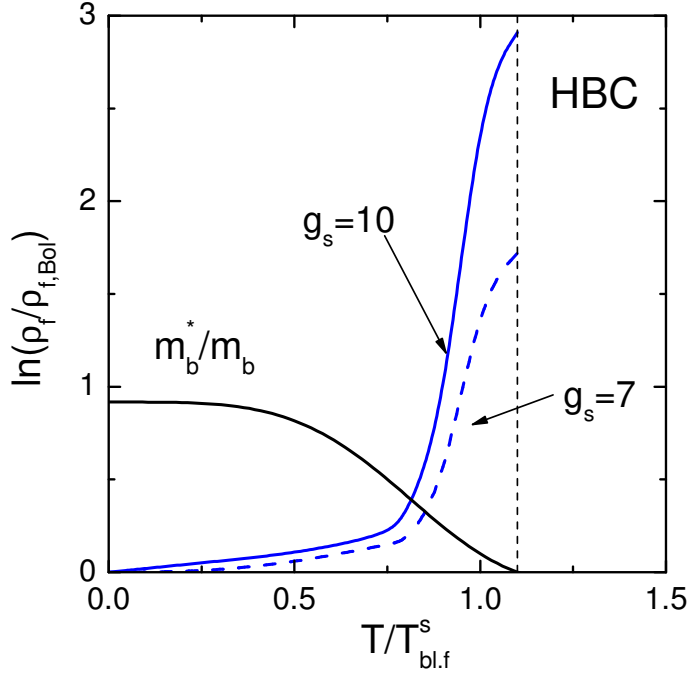}
\caption{Logarithm of the ratio of the fermion density to the corresponding
Boltzmann quantity for $g_s =10$ and $g_s =7$,
and the effective scalar boson mass
as functions of the temperature in units of $T_{\rm bl.f}^s$. The vertical
dash-line indicates the area of the HBC.
}
\label{fig:dens-S.eps}
\end{figure*}

The typical 
behavior of the logarithm of the ratio of the fermion density to the corresponding
Boltzmann quantity as function of the temperature is shown in
Fig. \ref{fig:dens-S.eps} for $T<T_{\rm cb}^s$.
Moreover Fig. \ref{fig:dens-S.eps} demonstrates 
the temperature dependence of the effective scalar
boson mass.
We see a huge enhancement of $\mbox{ln}(\rho_f /\rho_{f,\rm Bol})$ and a drastic
decrease of $m_{b}^* /m_{b}$ in the
vicinity of $T=T_{\rm bl.f}^s$.  At the critical point $m_{bt}^* /m_{b}$
reaches zero demonstrating possibility of the second order phase transition to
the HBC state.

\subsection{Contribution of baryon resonances}
Above we considered a simple example of the system consisting 
of heavy fermions of one species interacting with
one kind of less massive scalar bosons. However
at finite temperature together with nucleon/antinucleon states
the high-lying baryon/antibaryon  
resonances, like $\Delta$ isobar, hyperons, etc, are also populated 
with some probability. Resonances interact with each other by
boson exchanges, as well as by a residual interaction (including repulsive
baryon-baryon correlations).
To describe the multi-component system of the baryon/antibaryon  resonances
interacting with mesons we need to know coupling constants
between different particle species. In general, Dyson equations for Green
functions of different particles are coupled and the problem proves to be
very complicated. 
Different meson exchanges may significantly contribute. 
E.g., for the $\Delta$
isobar the coupling in the pion channel is the dominant one. However the
scalar channel might be also important. 
According to \cite{WBB89} $\sigma$ mesons
interact with $\Delta$-isobars with the same universal coupling constant 
$g_s\simeq 10$.
In case of 
$N^{*} (1440)$ one gets
 \cite{DVF03} $g_{\sigma NN^{*}}/g_{\sigma NN}
\simeq 0.47$, $g_{\pi NN^{*}}/g_{\pi NN}\simeq 
0.53 g_{\sigma NN^{*}}/g_{\sigma NN} \simeq 0.25$.

\subsubsection{Low temperature limit}
As above, let us for simplicity first assume that baryon resonances couple 
by an exchange of only a scalar neutral boson ($s$).
This simplification
is  sufficient to find particle distributions in the low temperature
limit, $T\ll \mbox{min}\{ m_b , T_{\rm bl.f}\}$. 
Then we may still use eqs (\ref{3oc}),
(\ref{den}), (\ref{Af-qp}) for the given baryon resonance, however with 
$\widehat{
\Lambda}_f^0$ operators being
different in dependence on the spin  of the baryon
species.
Calculating the resonance width we need to take into account in 
(\ref{sigm-R0-sim}) various
possible intermediate states, since the given resonance may decay
to the virtual boson and to an 
another baryon resonance then absorbing back
the
virtual (off-mass shell) boson. Notice that we discuss only temperature effects.
Just to simplify the consideration 
{\em{we artificially suppressed
the widths terms surviving for $T= 0$.}}

Let us present the density of the baryon resonance/antiresonance states
of the fixed species $B^*$.
In the low temperature limit 
using  (\ref{gamf1-gen}), (\ref{r-rat}),
(\ref{r-Bol})
we obtain
\be\label{r-r}
&&\rho_{B^* ,(\pm)}\simeq N_{B^*}\left(\frac{m_{B^*}T}{2\pi}\right)^{3/2}
\mbox{exp}\left[ -\frac{m_{B^*} \mp \mu_{B^*}}{T}\right] \\
&&+N_{B^*}\sum_{i}\left(\frac{m_{B^*_i}T}{2\pi}\right)^{3/2}
\mbox{exp}\left[ -\frac{m_{B^*_i} \mp \mu_{B^*_i}}{T}\right]
\frac{g_{s B^* B^*_i}^2 }{4\pi^2 }I_{0s} \left(\frac{m_{B^*_i}}
{m_b}\right)\nonumber
. \ee
$N_{B^*}$ is the degeneracy factor (e.g., $N_n =N_p =2$).
The summation is performed over all possible
states including the given baryon state ($B^*$),
$m_{B^*}$ is the mass of the
$B^*$ baryon resonance.
For non-strange baryons with the same baryon number, as for
the nucleon,  we have 
$\mu_{B^*}=\mu_N$, if there is a permitted reaction
channel: $B^* \leftrightarrow N + b^{virt}$. Here $b^{virt}$ is the 
virtual boson (in our model example it is a scalar boson, whereas in reality 
it also could be the pion, $\sigma$, $\omega$, etc). 
Second term in (\ref{r-r}) is due to the diagram
\be\label{phibar}
\includegraphics[width=3cm,clip=true]{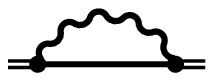}
\ee
where external double-lines correspond to the given resonance $B^*$ and
internal solid line relates to the resonance $B^*_i$, including the given
resonance $B^*$ and the nucleon state. In the latter case the self-energy
term is the same as in eq. (\ref{selfzf}).

Thus, 
the baryon resonance density is
substantially increased compared to the Boltzmann value. The term in eq. (\ref{r-r}) 
relating to the decay of the given baryon resonance to the nucleon and 
the virtual boson ($i=n$ or $p$, solid line in (\ref{phibar})),
has no  suppression factor, as $e^{-(m_{B^*}-m_N )/T}$. The latter factor would
arise, if one worked  in the framework
of the quasiparticle picture. 
If $N_{B^*}g^2_{s B^* N}$ were larger than 
$N_N g^2_{s N N}$, the density of a high-lying resonance 
could be even higher than the nucleon density,
showing a {\em laser effect.} Opposite, if
$g_{s B^*_i B^*_j}$ were negligible for $i\neq j$, the laser
enhancement would disappear and 
a high-lying resonance state would be less populated than the nucleon one. 
Nevertheless, in any case {\em{the resonance state
proves to be more populated compared to the value determined by the 
corresponding Boltzmann expression.}}

\subsubsection{High temperature limit}

To proceed in the high temperature limit let us additionally assume that 
$g_{sB^*_i B_j^*}\neq 0$ only for $i = j$. Then in the non-relativistic
approximation
for the resonance,
the density of a $B^*$ baryon resonance (and its anti-partner) is  
found with the help of eq.
(\ref{ratd1}). We obtain
\be\label{ratd-res}
\rho_{B^* ,(\pm)}\simeq \frac{N_{B^*}
  m_{B^*}^{3/2}T^{3}}{2\sqrt{2}\pi^{5/2}J_{s,B^*}^{3/4} }
\mbox{exp}\left[ -\frac{m_{B^* ,(\pm)}^* (T)}{T}
\right] I_2 \left( \frac{4\sqrt{J_{s,B^*}}}{T}
\right) .
\ee
Here, in agreement with eq. (\ref{sigmJ-s})
the intensity of the multiple scattering is
\be\label{sigmJ-sstar}
&&J_{s,B^*} =2g_{\sigma B^* B^*}^2
\int \frac{d^3 q}{(2\pi)^3}\int_{0}^{\infty} \frac{d q_0}{2\pi}
A_{s ,(+)}(q) n_{s ,(+)}(q_0 ),
\ee
where we used that for scalar neutral bosons $A_{s ,(+)}(q)=A_{s ,(-)}(q)$
and $n_{s ,(+)}(q_0 )=n_{s ,(-)}(q_0 )$.
The effective 
mass of the baryon
resonance follows from expression (\ref{effermm}):
\be\label{effermmstar}
m_{B^* ,(\pm)}^* (T) =m_{B^*} \left(1\mp \frac{\mu_N}{m_{B^*}} -\frac{2
\sqrt{J_{s,B^*}}}{m_{B^*}}\right),\,\,\, \mbox{for} \,\,\, 
2\sqrt{J_{s,B^*}}\ll m_{B^*}.
\ee
Analogously, one rewrites expressions (\ref{rhobl-1}), (\ref{rhoblart})
for the density of the baryon resonance in the relativistic energy region.
Within the quasiparticle approximation for the
boson  using (\ref{sigm-R0-finlim2}),
(\ref{sigm-R0-finlim2rel}) we obtain the effective boson mass: 
\be\label{mbrstar}
&&m_{b}^{*2}(T)\simeq (1-\alpha_s)^{-1} \\
&&\times \left[m_{b}^2 -
\sum_i \left(
\frac{2g^2_{s B^*_i B^*_i }
\rho_{B^*_i }(p_0 \sim m_{B^*_i } ) }
{\sqrt{J_{s,B^*_i}}}+\frac{36(\mbox{ln}2)
g^2_{s B^*_i B^*_i }\delta \rho_{B^*_i }(p_0 \lsim T)}{\pi^2 T^s_{\rm bl.f}}
\right)\right],
\nonumber \\
&&\alpha_s =\sum_i 
\frac{2g^2_{s B^*_i B^*_i }\rho_{B^*_i }(p_0 \sim m_{B^*_i } )}
{J_{s,B^*_i}m_{B^*_i }},
\ee
The summation is over all baryons and antibaryons, $\rho_{B^*_i }$ is the
corresponding baryon or antibaryon density. 
As in eq. (\ref{mbr-lcor}),
we dropped a numerically 
small correction term to the wave function
renormalization from the region $p_0\lsim T$.

In the limiting cases of high and low temperatures we recover 
eqs (\ref{Jexp-lim2}), (\ref{Jexp-lim1}), now with the coupling constant
$g_{\sigma B^* B^*}$ standing instead of $g_s$
and $\beta_{B^*}$, instead of $\beta_s$,
\be\label{Jexp-lim2star}
J_{s,B^*}= \frac{g_{\sigma B^* B^*}^2  T^2}
{12\beta_{B^*}^{3/2}}\,,
\quad \mbox{for} \quad T\gg m_{b}^{*}(T) ,
\ee
and
\be\label{Jexp-lim1star}
J_{s,B^*}= \frac{g_{\sigma B^* B^*}^2 
T^{3/2}\sqrt{m_{b}^{*}} }
{2^{3/2} \pi^{3/2}\beta_{B^*}^{3/2}} 
\mbox{exp}{\left( -m_{b}^{*}/T\right)}\,, \quad \mbox{for} \quad T\ll 
m_{b}^{*}(T).
\ee
We see that for $T\gg m_{b}^{*}(T)$ the intensity of the multiple scattering,
$J_{s,B^*}$, of a high-lying baryon resonance would
exceed that for the nucleon, $J_{s,N}$, if $g_{s B^* B^*}^2$ were larger
than  
$g_{s NN}^2$. As follows from eq. (\ref{effermmstar}),
for $g_{s B^* B^*}^2 /m_{B^*}^2 >
g_{s NN}^2/m_{N}^2 $, 
the baryon continuum for the given
high-lying baryon resonance would be blurred at a smaller temperature than 
for the nucleon. Please notice that such a relation between coupling constants
is not fulfilled for the realistic hyperon-$\sigma$-nucleon interaction, 
cf. \cite{KV03}.
Nevertheless, using estimates \cite{KV03}
we conclude that
resonances contribute essentially to the total baryon-antibaryon density for
$T\sim T_{\rm bl.f}$. Moreover, we artificially suppressed 
all couplings except $g_{sB^*_i B_j^*}$ for $i = j$
that is certainly not the case in the reality. Also in reality 
$g_{sB^*_i B_j^*}\neq g_{sB^*_j B_i^*}$ for $i \neq j$
since $N_{B^*_i}\neq N_{B^*_j}$, that may stimulate
in some cases {\em{ a laser effect}}. A high-lying state might be more
populated than a low-lying state
(Note that, e.g., in case of 
the $\Delta$ isobar, i.e.  $(\frac{3}{2} ,\frac{3}{2})$ 
resonance, one has $N_{\Delta}/N_N =4$, that works in favor of the laser effect).

Concluding, indeed, we deal with the 
{\em{ hadron resonance porridge at a sufficiently large temperature.}}

\subsection{Evaluation of correlation effects}

Above we discussed properties of the system
described within the simplest $\Phi$-derivable approximation with only
one
diagram (\ref{phi}). Exact fermion and boson self-energies are 
determined by 
diagrams (\ref{selfzfex}) and (\ref{selfzbex})
with one free and one exact vertices. 
{\em{In the low temperature limit vertex
corrections are negligible.}} Thereby, we further consider the high temperature
limit $T\gsim m_b^* (T)$.
The equation for the vertex can be greatly simplified within the ladder
re-summation:
\be\label{fulvert}
\includegraphics[width=7cm,clip=true]{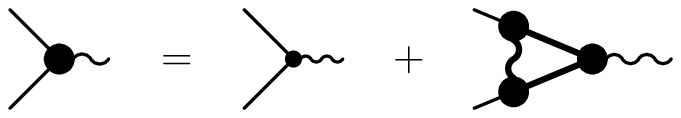}
\ee
that reads as
\be\label{vert-lad}
&&\widehat{V}_{\rm lad} (q_1 ,p, p+q_1 )=\widehat{V}_{0} (q_1 )-
\int 
\widehat{G}_f (p-q)\widehat{G}_f (p+q_1 -q)\nonumber\\
&&\times i\widehat{V}_{\rm lad} 
(q, p+q_1 -q, p+q_1)G_b (q)
\widehat{V}_{\rm lad}
(-q ,p, p-q )\nonumber\\
&&\times\widehat{V}_{\rm lad}
(q_1  ,p-q ,p+q_1 -q)
\frac{d^4 q}{(2\pi)^4}  ,
\ee
the corresponding matrix indices are implied.
We need $-,-,-$ and $+,+,+$ vertex functions 
of the same signs, since in the formalism that
uses full Green functions, cf. \cite{KV96}, any extra
full Green function $G^{-+}$
or $G^{+-}$ corresponds to the 
scattering process involving extra particle in the
initial and the final state. These processes  are suppressed, 
if the number of
fermion-antifermion pairs is not too large. 
In the STL approximation, 
which we now exploit,
we find 
\be\label{vert-lad1}
V_{\rm lad}^{- - -} \simeq \frac{g_s}{1+J_s (V_{\rm lad}^{- - -})
(p^2 +m_f^2 +2\FMslash{p}m_f )
\left[ \re G_1 (p)\right]^2 },
\ee
where we also used the spin structure of fermion Green functions 
(\ref{spinstr}), within the ansatz $G_3 =0$,
assuming $G_1 \simeq G_2$. The quantity
$J_s (V_{\rm lad}^{- - -})$ is $J_s$ with the
bare
vertices $g_s$  replaced to the full vertices, see corresponding fat dots
in (\ref{fulvert}).
Thus, one may restrict the consideration to the first diagram
(\ref{phi}) only if $(p^2 +m_f^2 ) J_s (V_{\rm lad}^{- - -})
\left[ \re G_1 (p)\right]^2 \ll 1$.

At $T\leq T_{\rm bl.f}^{s,0}$ using (\ref{tblrel}) we evaluate
$(p^2 +m_f^2 ) J_s  \left[ \re G_1 (p)\right]^2 \leq 1/2$. For 
artificially large temperatures $T\gg T_{\rm bl.f}^s$ we would have 
$(p^2 +m_f^2 ) J_s  \left[ \re G_1 (p)\right]^2 \simeq m_f^2 /(2J_s)\ll 1$.

In the full series of vertex diagrams, beyond the ladder approximation,
there are graphs with crossed boson lines.
In the STL approximation each diagram that
includes   the same number of
full scalar boson Green functions yields the very same
contribution independently on where the boson lines are placed inside 
the diagram.
Counting the number of boson lines with full vertices in 
first diagrams of $\Sigma_f$ we find
\be\label{vert-j}
&&V^{- - -} \simeq \frac{g_s}{1+J_s  (V^{- - -})(p^2 +m_f^2 +2\FMslash{p}m_f )
\left[ \re G_1 (p)\right]^2 }
\\
&&+\frac{g_s J_s^2  (V^{- - -})(p^2 +m_f^2 +2\FMslash{p}m_f )^2
\left[ \re G_1 (p)\right]^4}
{1+6J_s  (V^{- - -})(p^2 +m_f^2 +2\FMslash{p}m_f )
\left[ \re G_1 (p)\right]^2 }
.\nonumber
\ee
Thus, the ladder approximation yields an appropriate estimate 
of the full vertex up to rather
high temperatures.
 
For rough estimates we, as before, may consider
only one diagram of $\Phi$  but 
with an effective
coupling constant $g^*_s$ instead of the bare vertex $g_s$. For
low temperatures we have $g^{*}_s \simeq g_s$.
The vertex suppression factor increases with the temperature
reaching the value 
$g^{*}_s \simeq 0.7 g_s$
for $T=T_{\rm bl.f}^s$ and again $g^{*}_s \rightarrow  g_s$ for $T\gg 
T_{\rm bl.f}^s$.  
Note that the vertex suppression factor essentially depends on the
structure of the fermion -- boson interaction. For the $N\pi$
interaction the corresponding vertex would be 
less suppressed,
cf. \cite{D93} and a discussion in subsection \ref{cor-pi}.

Note that we discussed just a model example.
We suppressed a possible
boson-boson self-interaction. Inclusion of the latter complicates the
consideration yielding a repulsion \cite{VM83,Dyug83,MSTV,RW94}. Also
simplifying we considered only the baryon interaction with the scalar meson
disregarding the baryon interaction with other meson species. A residual
baryon-baryon interaction has been also dropped out as well as an interaction
between different meson species.  Further we will proceed step by step 
studying relevant models, thus permitting
different types of interactions.

\section{System  of heavy fermions and less massive vector bosons. }
\label{vector}

Now we will consider another example, the fermion -- vector boson 
system with the coupling
given by 
\be
L_{int} =-g_v \bar{\psi}\gamma_{\mu}\phi^{\mu}\psi.
\ee
The bare vertex is  $\widehat{V}_0 =g_v\gamma_{\mu}$. Then we will apply 
results to the $N\bar{N}\omega$ and $N\bar{N}\rho$ systems.
Again we first solve a model problem disregarding other relevant couplings of
vector mesons with other mesons, tensor coupling with nucleons, etc.

\subsection{Spin structure of vector boson propagator}

In the medium the Green function of the vector boson 
changes as follows
\be\label{frvec-m}
(\widehat{G}_b^{R})_{\mu \nu} &=&\frac{L_{\mu\nu}}{q^2 -m_b^2 -
{\Sigma}_b^l}
+\frac{T_{\mu\nu}}{q^2 -m_b^2 -{\Sigma}_b^t}-\frac{h_\mu
    h_\nu}{m_b^2},\\ 
L_{\mu\nu}&=&l_\mu l_\nu /l^2  \,,\quad T_{\mu\nu}=g_{\mu\nu}-h_\mu h_\nu
-L_{\mu \nu} ,\nonumber \\
l^{\mu}&=&(q\cdot u)q^\mu -u^\mu q^2 ,
\quad h^\mu = q^\mu /\sqrt{q^2},\quad q^2 >0,
 \nonumber 
\ee
$u$ is, as before, 
the 4-velocity of the frame. In the rest frame one has $u =(1,\vec{0})$.
The retarded self-energy of the vector boson
is subdivided to the longitudinal ($\Sigma_b^l$)
and the transversal ($\Sigma_b^t$) parts
\be\label{tr-l}
\Sigma_b^{\mu\nu}&=&\Sigma_b^l L^{\mu\nu}+\Sigma_b^t T^{\mu\nu}.
\ee
One usually assumes $h_\mu 
\Sigma_b^{\mu\nu}=0$. For $\Sigma_b^l , \Sigma_b^t \rightarrow 0$,
eq. (\ref{frvec-m}) coincides with  (\ref{frvec}).

From (\ref{frvec-m}) we find
\be\label{tenstr-gen}
&&(\Sigma_b^{R})^{l}(q)=-\frac{q^2}{\vec{q}^{\,2}}
(\Sigma_b^{R})^{0}_0 ,\quad 
(\Sigma_b^{R})^{t}(q)=\frac{1}{2}T^{i}_{j}(\Sigma_b^{R})_{i}^j ,
\quad i,j =1,2,3,
\ee
\be\label{tenstr}
 &&\sum_{i}(\Sigma_b^{R})^{i}_i (q_0 =0, |\vec{q}|\rightarrow 0)=
2(\Sigma_b^{R})^{t}(q_0 =0, |\vec{q}|\rightarrow 0), \quad i=1,2,3,
\nonumber\\
&&(\Sigma_b^{R})^{0}_0 (q_0 =0, |\vec{q}|\rightarrow 0)
=(\Sigma_b^{R})^{l}(q_0 =0, |\vec{q}|\rightarrow 0).
\ee

\subsection{Low temperature limit}

The low temperature limit, $T\ll \mbox{min}\{ m_b , T_{bl.f}\}$, 
is considered quite similar to that for 
scalar bosons. We replace in (\ref{gamf1-gen}) 
\be
&&\widehat{S}^0 =\widehat{V}_0 (q)
\widehat{\Lambda}_f^0 (p+q)
\widehat{V}_0 (-q)\widehat{\Lambda}_b^0 (q)\nonumber\\
&&= g_{v}^2  \left(2\FMslash{p}+2\FMslash{q}-4m_f +\frac{2pq 
\FMslash{q}}{m_b^2}-\frac{\FMslash{p}q^2}{m_b^2}+\frac{\FMslash{q}q^2}{m_b^2}+
\frac{m_f q^2}{m_b^2}
\right) .
\ee 
Assuming that $p_0 \simeq m_f$, $q^2 =q_0^2 -\vec{q}^{\,2}\simeq 
m_b^2$ and dropping linear  terms in $\vec{\gamma}$ and the term $\propto
\vec{p}\vec{q}$, which do not contribute,
we get
\be\label{vecap}
&&\widehat{S}^0 \simeq 3g_{v}^2 m_f (\gamma_0 -1) +
g_{v}^2\frac{2m_f \vec{q}^{\,2} \gamma_0}{m_b^2}+3g_{v}^2 q_0 \gamma_0 .
\ee
Then from
(\ref{gamf1-gen}) we find 
\be\label{gamf1v}
&&\widehat{\Gamma}_{f}^{q.p} (p_0 -\epsilon_{p})\simeq
\frac{g^2_v }{4\pi } 
\mbox{exp}\left[-(\epsilon_{p}-p_0 )/T\right]
\frac{\mbox{sinh }(y\eta)}{y\eta}e^{-y^2 /2}
\sqrt{(p_0 -\epsilon_{p})^2 -m_b^{\,2} }\nonumber\\
&&\times
\left\{ 3(\gamma_0 -1)
+\frac{2\gamma_0 }{m_b^2}
[(\epsilon_p -p_0 )^2 -m_b^2 ]+3\gamma_0 (\epsilon_p -p_0 )m_f^{-1}\right\},
\ee
for $\epsilon_p >p_0$,
$y$ and $\eta$ are the same as in (\ref{gamf1}).

Using (\ref{gamf1v}) and  (\ref{foclow}) 
the fermion distribution is presented as follows
\be\label{3ocU-v}
&&\widehat{n}_{f,(\pm)} (\vec p )\simeq \frac{(\epsilon_{p}+ m_f\gamma_0)}
{2\epsilon_{p}}n_{{\rm Bol},(\pm)}(\vec p )\nonumber \\
&&+\frac{g_v^2 }{8
\pi^2}\left[
I_{0v}^0  (\frac{\epsilon_p }{m_b})+\frac{3m_b}{2m_f}I_{0v}^1
(\frac{\epsilon_p }{m_b})
\right]
n_{{\rm Bol},(\pm)}(\vec p ), 
\ee
\be\label{i0v}
I_{0v}^0 (x)= \int_{1}^{x}\frac{\mbox{sinh }(y\eta)}{y\eta}e^{-y^2 /2}
(z^2 -1)^{3/2} \frac{dz}{z^2},
\ee
\be\label{i0v1}
I_{0v}^1 (x)= \int_{1}^{x}\frac{\mbox{sinh }(y\eta)}{y\eta}e^{-y^2 /2}
(z^2 -1)^{1/2} \frac{dz}{z}.
\ee
Here variables $y$ and $\eta$ are determined as in (\ref{I0}).
Cutting off integrals at $z$ corresponding to
$y\sim y\eta \sim 1$ we estimate
\be\label{vap}
I_{0v}^0 (x)\sim 
-\frac{(\bar{x}^2 -1)^{3/2}}{\bar{x}}+\frac{3}{2}\bar{x}(\bar{x}^2 -1)^
{1/2}-\frac{3\mbox{ln}
(\bar{x}+(\bar{x}^2 -1)^{1/2})}{2\bar{x}^{1/2}},
\ee
\be\label{vap1}
I_{0v}^1 (x)\sim 
(\bar{x}^2 -1)^{1/2}+\mbox{arcsin}\frac{1}{\bar{x}}-\frac{\pi}{2},
\ee
with $\bar{x}\simeq 
(1+m_f T /m_b^{\,2} )^{1/2}$ the same, as in (\ref{sap}), and
$I_{0v}^0 (x\rightarrow \infty)\rightarrow \bar{x}^2 /2$,
$I_{0v}^0 (x\rightarrow 1)\rightarrow \frac{2^{5/2}}{5}(\bar{x}-1)^{5/2} $,
$I_{0v}^1 (x\rightarrow \infty)\rightarrow \bar{x}$,
$I_{0v}^1 (x\rightarrow 1)\rightarrow \frac{2^{3/2}}{3}(\bar{x}-1)^{3/2}$.

Finally, the fermion particle and antiparticle densities are
\be\label{r-rat-f}
\frac{\rho_{f,(\pm)}}{\rho_{{\rm Bol},(\pm)}}\simeq
1+\frac{g_v^2}{4\pi^2}\left[ 
I_{0v}^0\left(\frac{m_f }{m_b}\right) 
+\frac{3m_b}{2m_f}I_{0v}^1
(\frac{m_f }{m_b})\right]. 
\ee
{\em{Correlation effects are negligible in the low temperature limit.}}
For
$T\ll m_b^3 /m_f^2$ the term $\propto I_{0v}^1$ is the dominating
term. Compared to the
scalar boson case, see eq. (\ref{r-rat}), here
boson distributions are $3g_v^2 m_b /(2g_s^2 m_f)$ times suppressed 
(for $g_v^2 \sim 
g_s^2$, $m_b \ll m_f$), and
$\frac{\rho_{f,(\pm)}}{\rho_{{\rm Bol},(\pm)}}
\simeq 1+\frac{g_v^2 m_f^{1/2} T^{3/2}}
{8\pi^2 m_b^2}$.
For $T\gg m_b^2 /m_f$
the term $\propto I_{0v}^0$ is the dominating term and 
$\frac{\rho_{f,(\pm)}}{\rho_{{\rm Bol},(\pm)}}\simeq 1+\frac{g_v^2 m_f T}
{8\pi^2 m_b^2}$. 
Thus the enhancement is here higher than in the scalar boson case
(again for $g_v \sim g_s$ and $m_f \gg m_b$).

One can easily show that,
as for scalar bosons,  in the vector boson case
{\em the quasiparticle approximation fails 
for the description of the warm hadron liquid of a small fermion chemical 
potential for the relevant value $g_v \sim 10$, $m_f \gg m_b$.}

For the $N\bar{N}\om$  system of zero total baryon number we have
$m_{\om}\simeq 782$~MeV. The  $g_\om$ coupling is less known.
Its evaluation used in the relativistic mean field models \cite{Glen}
yields 
$g_\om \simeq 8\div 10$. The mass of $\om$ is rather high.
Thereby, the limit $m_b \gg T\gg m_b^2 /m_f$
is not realized. For $T\sim m_{\pi}/2$ (such a temperature is in the range of {\em{a slightly
    heated hadron liquid}} for $\omega$) and for $g_\om \simeq 10$
we estimate 
$\rho_{p}^{sym}=\rho_{n}^{sym}=\rho_{\bar{p}}^{sym}=\rho_{\bar{n}}^{sym}
\simeq 1.1\rho_{\rm Bol}^{sym}$.

\subsubsection{Vector-isospin vector boson -- fermion system}

For the vector-isospin-vector boson - fermion coupling 
($N\rho$ sub-system)  the interaction term of the Lagrangian density is given by
\be\label{lintv}
L_{int} =-g_{i.v.} 
\bar{\psi}\gamma_{\mu}\vec{\tau}\vec{\phi}^{\mu}\psi .
\ee
We suppress possibility of a tensor coupling. The latter has been discussed
in \cite{SH}, where the nucleon-antinucleon  loop diagram has been studied for
the case of cold nuclear matter.
Quite similar to the vector boson case we obtain
\be\label{3ocU-r}
&&\widehat{n}_{f,(\pm)} (\vec p )\simeq \frac{(\epsilon_{p}+ m_f\gamma_0)}
{2\epsilon_{p}}n_{{\rm Bol},(\pm)}(\vec p ) \\
&&+\frac{N_{i.v}g_{i.v}^2 }{8
\pi^2}\left[
I_{0v}^0  (\frac{\epsilon_p }{m_b})+\frac{3m_b}{2m_f}I_{0v}^1
(\frac{\epsilon_{p}}{m_b})\right]
n_{{\rm Bol},(\pm)}(\vec p ).\nonumber
\ee
In comparison with
eq. (\ref{3ocU-v}), here appeared extra isospin vector 
degeneracy factor $N_{i.v}=3$.

The fermion particle and antiparticle densities are
\be\label{r-rat-fvi}
\frac{\rho_{f,(\pm)}}{\rho_{{\rm Bol},(\pm)}}\simeq
1+\frac{N_{i.v}g_{i.v}^2}{4\pi^2}\left[ 
I_{0v}^0\left(\frac{m_f }{m_b}\right) 
+\frac{3m_b}{2m_f}I_{0v}^1
(\frac{m_f }{m_b})\right] . 
\ee
The $\rho$ meson mass is $769$~MeV and
the coupling constant is $g_\rho \simeq 3$, cf. \cite{SH}. 
With these values we obtain one fourth  of the $\omega$ meson
density, for $g_{\om} =10$. 
Summing up contributions of
$\sigma$, $\om$ and $\rho$ at $g_s =g_v =10$, $g_{i.v}=3$, $T\sim m_{\pi}/2$,
$m_\sigma \simeq 600~$MeV,
we arrive at the estimation
$\rho_p^{sym} \simeq \rho_n^{sym} \simeq\rho_{\bar{p}}^{sym} 
\simeq\rho_{\bar{n}}^{sym} 
\simeq 1.2\rho_{{\rm Bol}}^{sym}$.

\subsection{High temperature limit}
\subsubsection{Analytic solution for fermion Green function in 
STL approximation}
For high temperatures, $T\gsim m_{bt}^* (T)$, where
$m_{bt}^* (T)$ is the effective mass
for the transversal vector boson,
we may use the same STL
approximation, as we exploited for the scalar boson. In the latter case
we first considered the problem within one diagram of $\Phi$, i.e. without
inclusion of correlations, and then estimated the contribution of 
correlation diagrams. For vector bosons 
  within the STL approximation
we may solve the problem in general case 
using the Ward -- Takahashi  identity \cite{Kapusta}:
\be\label{WT}
\frac{\partial\widehat{G}_f^{-1}}{\partial p_\mu}=\widehat{V}^{\mu} (q=0,p,p),
\ee
where $\widehat{V}^{\mu}$ is the full vertex function. 

The full
Dyson equation in the STL approximation
takes the form
\be\label{Dys-v}
\widehat{G}_f^R =\widehat{G}_f^{0,R}-\widehat{G}_f^{0,R}J_v^{\mu\nu}\gamma_\mu
\widehat{G}_f^R\frac{d\widehat{G}_f^{R -1}}{d p^\nu}
\widehat{G}_f^R ,
\ee
or equivalently 
\be\label{Dys-veq}
\left(\widehat{G}_f^{0,R}\right)^{-1}\widehat{G}_f^R 
=1+
J_v^{\mu\nu}\gamma_\mu
\frac{d\widehat{G}_f^{R}}{d p^\nu} .
\ee
Here $J_v^{\mu\nu}$ demonstrates the intensity of the multiple scattering,
which is now the tensor function,
\be\label{Jvec}
&&J_v^{\mu\nu}=g_v^2 \int \frac{d^3 q}{(2\pi)^3}
\int_{0}^{\infty} \frac{d q_0}{2\pi}
\left[ A_{b,(+)}^{t}n_{b,(+)} (q_0 )T^{\mu\nu}
+ A_{b,(+)}^{l}n_{b,(+)} (q_0 )L^{\mu\nu}\right]\nonumber\\
&+&g_v^2 \int \frac{d^3 q}{(2\pi)^3}
\int_{0}^{\infty} \frac{d q_0}{2\pi}
\left[ A_{b,(-)}^{t}n_{b,(-)} (q_0 )T^{\mu\nu}
+ A_{b,(-)}^{l}n_{b,(-)} (q_0 )L^{\mu\nu}\right], 
\ee
and there appear transversal and longitudinal spectral functions
\be\label{A-G-v}
A_{b,(\pm)}^j  =\frac{-2 \im \Sigma_{b,(\pm)}^{j,R}  }
{ \left[ q^2 -m_b^2 -\re 
\Sigma_{b,(\pm)}^{j,R}
 \right]^2 +\left[\im \Sigma_{b,(\pm)}^{j,R}\right]^2 },\,\,\,j=\{t,l\},
\,\,\,q_0 >0.
\ee
The
general structure of the Green function is determined by
eq. (\ref{spinstr}), where we again assume $G_3 =0$. Using it  we work out 
the tensor structure of (\ref{Dys-veq}):
\be\label{aux0}
&&\left( A_{b,(\pm)}^t n_{b,(\pm)}T_{\mu}^{\nu} +A_{b,(\pm)}^l n_{b,(\pm)}
L_{\mu}^{\nu}\right)\gamma^{\mu} 
\left[ \frac{dG_1}{dp^2}2p_\nu \FMslash{p} +\frac{dG_2}{dp^2}m_f 2p_\nu
+G_1 \gamma_\nu \right]\nonumber\\
&&=A_{b,(\pm)}^t n_{b,(\pm)}
(\delta_{\mu}^\nu -h_{\mu}h^\nu )\gamma^\mu 
\left[ \frac{dG_1}{dp^2}2p_\nu \FMslash{p} +\frac{dG_2}{dp^2}m_f 2p_\nu
+G_1 \gamma_\nu \right] 
\nonumber\\
&&-(A_{b,(\pm)}^t - A_{b,(\pm)}^l )n_{b,(\pm)}L_{\mu}^{\nu}\gamma^\mu 
\left[ \frac{dG_1}{dp^2}2p_\nu \FMslash{p} +\frac{dG_2}{dp^2}m_f 2p_\nu
+G_1 \gamma_\nu \right] .
\ee
Here
\be\label{aux}
&&(\delta_{\mu}^\nu -h^{\nu} h_{\mu})\gamma^\mu 
\left[ \frac{dG_1}{dp^2}2p_\nu \FMslash{p} +\frac{dG_2}{dp^2}m_f 2p_\nu
+G_1 \gamma_\nu \right]\\
&&=  3G_1 +2 (p^2 -
\frac{\FMslash{q}\FMslash{p}\,(pq)}{q^2})
\frac{dG_1}{dp^2} +2m_f \left(\FMslash{p}-\frac{\FMslash{q}(pq)}{q^2}\right)
\frac{dG_2}{dp^2} ,\nonumber 
\ee
\be\label{aux1}
&&L_{\mu}^{\nu}\gamma^\mu 
\left[ \frac{dG_1}{dp^2}2p_\nu \FMslash{p} +\frac{dG_2}{dp^2}m_f 2p_\nu
+G_1 \gamma_\nu \right]=G_1  \\
&&+2\frac{(qu \cdot qp -up \cdot q^2 )}
{q^2 [q^2 -(qu)^2 ]}
\left[ \frac{dG_1}{dp^2}(qu \cdot \FMslash{q}\FMslash{p} -\FMslash{u}\FMslash{p} \cdot q^2 )+
\frac{dG_2}{dp^2}m_f (qu \cdot \FMslash{q} -\FMslash{u} q^2 )\right] . 
\nonumber
\ee
To avoid more cumbersome expressions we used a symbolic notation 
$\frac{dG_{1,2}}{dp^\nu}=
2p_{\nu}\frac{dG_{1,2}}{dp^2}$, whereas in general case $G_{1,2}$ depend
separately on $p_0$ and $\vec{p}$. Approximate equality holds only
for non-relativistic fermions and for $T\gg m_{bt}^{*2}(T)/m_f$.

In the rest frame for the hadron vacuum case ($\mu_f =0$), 
dropping the term $\propto \vec{\gamma}$ and the linear term
$\propto \vec{q}\vec{p}$, which do not contribute to the particle densities,
we rewrite eq. (\ref{aux1}) as follows
\be\label{aux11}
&&L_{\mu}^{\nu}\gamma^\mu 
\left[ \frac{dG_1}{dp^2}2p_\nu \FMslash{p} +\frac{dG_2}{dp^2}m_f 2p_\nu
+G_1 \gamma_\nu \right]\nonumber\\
&&\simeq G_1  -2\frac{p_0 \vec{q}^{\,2}}
{q^2 }
\left[ \frac{dG_1}{dp^2_0}p_0 +
\frac{dG_2}{dp^2_0}m_f \gamma_0 \right] . 
\ee
To simplify expressions we also suppressed contributions $\propto 
\vec{p}^{\,2}\frac{dG_{1,2}}{dp^2_0}$ which are small for $T\gg m_{bt}^{* 2} (T)/m_f$.
In these assumptions from (\ref{Dys-veq})
we derive two coupled equations
\be\label{Gsyst}
p^2 G_1 -m_f^2 G_2 &\simeq&1+(3J_0^v +J_{01}^v )
G_1 -2(J^v_1 +J^v_{11})p_0^2\frac{dG_1}{dp^2_0}
,\nonumber\\
G_2 -G_1 &=&-2(J^v_1 +J^v_{11}) \frac{dG_2}{dp^2_0}. 
\ee
We introduced 4 types of intensities of the multiple scattering
\be\label{Jv}
J_0^v &=& g_v^2 
\int \frac{d^3 q}{(2\pi)^3}
\int_{0}^{\infty} \frac{d q_0}{2\pi}
\left[ A_{b,(+)}^{t}n_{b,(+)} (q_0 )
+ A_{b,(-)}^{t}n_{b,(-)} (q_0 )\right],\\
J_1^v &=& g_v^2 
\int \frac{\vec{q}^{\,2}}{q^2}\frac{d^3 q}{(2\pi)^3}
\int_{0}^{\infty} \frac{d q_0}{2\pi}
\left[ (A_{b,(+)}^{t}n_{b,(+)} (q_0 )+ 
A_{b,(-)}^{t}n_{b,(-)} (q_0 )\right],\nonumber\\
J_{01}^v &=& g_v^2 
\int \frac{d^3 q}{(2\pi)^3}
\int_{0}^{\infty} \frac{d q_0}{2\pi}\nonumber\\
&&\times
\left[ (A_{b,(+)}^{l}- A_{b,(+)}^{t})n_{b,(+)} (q_0 )
+ (A_{b,(-)}^{l}-A_{b,(-)}^{t})n_{b,(-)} (q_0 )\right],\nonumber\\
J_{11}^v &=& g_v^2 
\int \frac{\vec{q}^{\,2}}{q^2}\frac{d^3 q}{(2\pi)^3}\int_{0}^{\infty} 
\frac{d q_0}{2\pi}\nonumber\\
&&\times
\left[ (A_{b,(+)}^{l}- A_{b,(+)}^{t})n_{b,(+)} (q_0 )
+ (A_{b,(-)}^{l}-A_{b,(-)}^{t})n_{b,(-)} (q_0 )\right].\nonumber
\ee
Compare these expressions with eq. (\ref{sigmJ-s}).
The consideration is simplified, if one assumes 
$A_{b,(\pm )}^{t}\simeq A_{b,(\pm )}^{l}$, being correct in case of a small change of the
effective boson mass. Then $J_{01}^v$ and  $J_{11}^v$ can be dropped out.

Let us assume the validity of the quasiparticle approximation for
vector bosons and
 use the spectrum
\be\label{brv}
\om^2 ( \vec{q}, T)=m_{bt}^{*2} (T)+\beta_v  (T)\vec{q}^{\,2}+O(\vec{q}^{\,4})
\ee 
for transversal modes
with $m_{bt}^{*2} (T)>0$, $\beta_v >0$, cf. (\ref{branch}).
Then in the limiting cases $T\ll m_{bt}^* (T)$ and 
$T\gg m_{bt}^* (T)$
we find  
\be\label{J1h}
J_1^v =\frac{\pi^2  g_{v}^2 T^4}{30 \beta_v^{5/2} m_{bt}^{*\,2}}, \quad \mbox{for} 
\quad T\gg m_{bt}^* (T) ,
\ee
\be\label{J1l}
J_1^v =\frac{3 g_v^2 T^{5/2}}{2^{3/2} \pi^{3/2}\beta_v^{5/2}
m_{bt}^{*\,1/2}}\mbox{exp}
(-m_{bt}^* /T),
\quad \mbox{for} \quad T\ll m_{bt}^* (T).
\ee
\begin{figure*}
\includegraphics[clip=true,width=8cm]{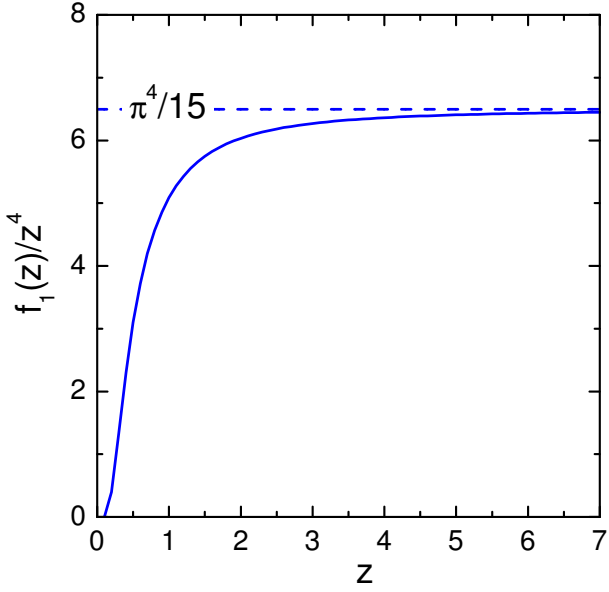}
\caption{$f_1 (z)/z^4$, cf. eq. (\ref{f1}). Dash line demonstrates asymptotic behavior for $z\gg 1$.}
\label{fig:f1.eps}
\end{figure*}
Numerical evaluation of the intensity $J_1^v$ is demonstrated in Fig. \ref{fig:f1.eps},
\be\label{f1}
J_1^v =\frac{\pi^2  g_{v}^2 (m_{bt}^*)^2 z^4}{30 \beta_v^{5/2}}r_v (z), \quad
r_v (z)=\frac{15f_1 (z)}{\pi^4 z^2},
\quad z=\frac{T}{m_{bt}^*}.
\ee
The horizontal line shows the asymptotic behavior (\ref{J1h}), 
$r_v \rightarrow 1$ for $z\rightarrow \infty$.
Comparing  (\ref{J1h}), (\ref{J1l}) with expressions (\ref{Jexp-lim2}),
(\ref{Jexp-lim1})  for the
scalar boson case we see that in the high temperature limit 
($T\gg m_{bt}^* (T)$)
the intensity of the multiple scattering $J_1^{v}$
is much higher in the
vector case (for $g_v \sim g_s$ and for $m_{bt}^* \sim m_s^*$). At these
conditions
the quantity
$J_0^{v}$
is  of the same order of magnitude
as $J_0^{s}$. Please notice that $J_1^{v}\rightarrow \infty$
for $m_{bt}^* (T)\rightarrow 0$. Thus we deal here with {\em{a strong critical opalescence.}}

Replacing $G_1$ from the second equation (\ref{Gsyst}) to the first one we 
obtain one
differential equation
\be\label{Jv-tot}
&&(p^2_0  -\epsilon_p^2 -3J_0^v -J_{01}^v )   G_2 -1+2(J_1^v + J_{11}^v )
(2p^2_0 
-3J_0^v -J_{01}^v )\frac{dG_2}{dp^2_0}\nonumber \\
&&
+4(J_1^{v} +J_{11}^v )^{2}p_0^2\frac{d^2 G_2}{(dp^2_0)^2}
=0.
\ee
Equation (\ref{Jv-tot}) can be solved perturbatively only for very small 
values of $J^v_0$, $J^v_1$, $J^v_{01}$, $J^v_{11}$.
For typical values $p_0^2 -\epsilon_p^2 \ll m_f T$ permitted within
the quasiparticle approximation,  the perturbative regime would hold
only for $J^v_1 , J^v_{01}, J^v_{11}\ll T^2$. 
On the other hand, as we have mentioned,
at rather low temperature the STL approximation is not applicable
anymore.  Even in the low temperature limit for the regime of
{\em{a warm hadron liquid}} the quasiparticle approximation (for $g_v \sim 10$)
and
the perturbative approach fail, as we have shown it above.

\subsubsection{Non-relativistic fermion distributions and 
blurring of the fermion continuum}

In non-relativistic approximation for fermions,
as we shall see it below (cf. eqs (\ref{Jv-int}), (\ref{f-d-v})),
typical fermion 4-momenta of our interest are given by
\be\label{typ}
p^2 \simeq m_f^2 +O\left(\sqrt{J_1^v}\,m_f ; J_1^v m_f T^{-1}\right) .
\ee
For such momenta 
one may drop the higher order derivative term in (\ref{Jv-tot})
Then eq. (\ref{Jv-tot}) is simplified as
\be\label{Jv-sim}
&&\frac{(\gamma^{\rm n.rel}_v )^2}{2} \frac{dG_2}{dz}+zG_2 -1 =0,
\quad z\simeq p^2_0  
-\epsilon_p^2 ,\\
&&(\gamma^{\rm n.rel}_v )^2 \simeq 8(J_1^v +J_{11}^v )m_f^2 .\nonumber
\ee
The 
form of eq. (\ref{Jv-sim}) coincides 
with that have been used for the description of the blurring 
of the electron gap in semiconductors due to the
electron-phonon interaction \cite{E}. 
Heavy fermions (e.g., nucleons)
play the same role, as electrons, whereas
light vector bosons (e.g., $\omega$
with an effective mass $m^*_{\om t}(T)\ll m_N$) play the role of phonons.

Solution of (\ref{Jv-sim}) satisfying appropriate condition
$G_2 \rightarrow 1/z$ for $z\rightarrow \infty$ 
($z\gg \gamma_v^{\rm n.rel}$ to be more concrete)
can be presented in the 
integral
form 
\be\label{Jv-int}
G_2 (z)=\frac{1}{\sqrt{\pi}\gamma_v^{\rm n.rel} }\int^{\infty}_{-\infty}
(z-\eta )^{-1}e^{-(\eta^2/(\gamma^{\rm n.rel}_v)^2)}d\eta .
\ee
Now we may 
recover the explicit 
expression for the full vertex $\widehat{V}^{\mu} (q=0, p,p)$.
From (\ref{Gsyst}) and (\ref{Jv-sim})
we find 
\be\label{G1G2}
\widehat{G}_f \simeq G_2^f (\FMslash{p}+m_f)+\FMslash{p}\frac{1-z G_2 }{2m_f^2},
\ee
i.e. $\widehat{G}_f \simeq G_2^f (\FMslash{p}+m_f)$ for typical energies 
and momenta of our
interest. Using (\ref{WT}) and (\ref{Jv-sim}) we obtain
\be\label{vertap}
\widehat{V}^{\mu} (q=0,p,p)\simeq 
  -\frac{2 p^{\mu}(\FMslash p -m_f )[2(1-zG_2 )z+
G_2 (\gamma^{\rm n.rel}_v)^2]}{(\gamma^{\rm n.rel}_v)^2
z^2 G_2^2 }+\frac{\gamma^{\mu}}{z G_2} .
\ee

In order to calculate fermion distributions and the
density we  
need to know $\im G$. From (\ref{Jv-int}) and (\ref{G1G2}) we find
\be\label{Img2}
\im G_2 =-\frac{\sqrt{\pi}}{\gamma_v^{\rm n.rel}}e^{-z^2 /
(\gamma^{\rm n.rel}_v)^2},\quad 
\im G_1 =\im G_2 \left( 1-\frac{z}{2m_f^2}\right).
\ee
We see that 
$\im G_1 \simeq \im G_2$ for $|z| \ll m_f^2$ of our interest. 
The imaginary part of the
Green function given by (\ref{Img2}) satisfies the sum rule (\ref{fsum-r})
for $|z| \ll m_f^2$.

For 
typical energies given by eq. 
(\ref{typ}) with the help of (\ref{3oc}) we obtain
\be\label{f-d-v}
&&\widehat{n}_{f,(\pm)} \simeq \frac{\gamma_0 +1}
{2\sqrt{\pi}\,\gamma_v^{\rm n.rel}}
\mbox{exp}\left[-\frac{
(\epsilon_p \mp \mu_f)}{T}\right]\int_{-\infty}^{\infty}dz
\mbox{exp}\left[-\left(\frac{
z^2}{(\gamma^{\rm n.rel}_v)^2}+\frac{z}{2\epsilon_p T}
\right)\right]\nonumber\\ 
&&\simeq\frac{\gamma_0 +1}{2}\mbox{exp}\left[-\frac{
(\epsilon_p\mp \mu_f)}{T}+\frac{(\gamma^{\rm n.rel}_v)^2}
{16\epsilon_p^2 T^2}\right]. 
\ee
Compare this result with (\ref{hdist1}).
For $\sqrt{J_1^v +J_{11}^v} \ll  T$ (low temperatures)
eq. (\ref{f-d-v}) yields only a slightly corrected
Boltzmann distribution. However  (\ref{f-d-v}) becomes incorrect in the low temperature
limit
due to the violation of the STL approximation in this
limit. In opposite limit case, $\sqrt{J_1^v +J_{11}^v} \gg  T$,
the 3-momentum fermion distribution is exponentially enhanced. 

For the fermion-antifermion density we obtain
\be\label{f-d-vr}
\rho_{f,(\pm)}=\rho_{{\rm Bol},(\pm)}\mbox{exp}\left(
\frac{J_1^v +J_{11}^v}{2T^2}\right). 
\ee
The value 
\be\label{vefm}
m_f^* =m_f - \mu_f -\frac{J_1^v +J_{11}^v}{2T} ,
\ee
as it is extracted from (\ref{f-d-v}),
plays the role of the effective fermion mass.

The temperature $T_{\rm bl.f}^{v,{\rm n.rel}}$
is estimated from the condition $m_f^* (T_{\rm bl.f}^{v,{\rm n.rel}})=0$. 
Then from (\ref{vefm}), (\ref{f1})
we evaluate
\be\label{Tcv}
T_{\rm bl.f}^{v,{\rm n.rel}}\simeq 
\left[\frac{60 m_{bt}^{*\,2} (m_f -\mu_f)}{ \pi^2  g_v^2 r_v 
(T_{\rm bl.f}^{v,{\rm n.rel}})}\right]^{1/3}.
\ee
Simplifying, we suppressed the contribution $J_{11}^v$ in this estimate.
Comparing  (\ref{Tcv})  with (\ref{Tcssn}) for scalar bosons for $r_v \simeq
r_s \simeq 1$
we see that 
in the heavy fermion limit ($m_f \gg m_b^{*}$) and for $g_s \sim g_v$ we have
$T_{\rm bl.f}^{v,{\rm n.rel}} <
T_{\rm bl.f}^{s,{\rm n.rel}}$, i.e. in this limit
vector bosons contribute more in the
blurring of the fermion continuum than scalar bosons. Please notice that
deriving eq. (\ref{Tcv}) we assumed $A_{b,(\pm )}^{t}\simeq A_{b,(\pm )}^{l}$.
The latter condition is not fulfilled for $m_{bt}^{*}\ll m_b$. Thus
(\ref{Tcv})
can be considered only as a rough estimate.

For the $\omega$ meson from (\ref{Tcv}) 
we would obtain $T_{\rm bl.f}^{v,0,{\rm n.rel}} \simeq 2.3 m_{\pi}$,
for $g_\om =10$, $r_v \simeq 1$ and for the mass
$m_{bt}^{*}=m_{\om}=782$~MeV. 
But then $T_{\rm bl.f}^{v,0,{\rm n.rel}}\ll m_{bt}^{*}
(T_{\rm bl.f}^{v,0,{\rm n.rel}})$ and the condition
for the use of the high temperature limit is violated ($r_v \ll 1$). By index zero we again
indicate that we will refuse of this estimate. Thus one should take
into account the decrease of $m_{\omega t}^{*}$ for $T\sim T_{\rm bl.f}^{v,{\rm n.rel}}$.
Taking  $m_{bt}^{*}\simeq m_{\pi}$ we would get $T_{\rm bl.f}^{v,{\rm n.rel}} \simeq 120~$MeV.
However, as we have demonstrated in case of the $\sigma$ meson, fully correct
estimate should be done within  
self-consistent account of both the fermion and the boson coupled channels.
Finally we may roughly evaluate $T_{\rm bl.f}^{v,{\rm n.rel}}\lsim m_{\pi}$.

We supplied the value 
$T_{\rm bl.f}^{v,{\rm n.rel}}$ with the index $''{\rm n.rel}''$,
since the estimate was done in the non-relativistic approximation for fermions.
In pure relativistic case one needs to solve 
much more involved
eq. (\ref{Jv-tot}). 
Without additional calculations 
we may only conclude that 
for  relativistic fermions the value $T_{\rm bl.f}^v$
is still less than $T_{\rm bl.f}^{v,{\rm n.rel}}$.
However in our model example we did not take into account interaction
with other mesons, e.g. scalar boson
correlation terms. Such terms suppress vertices 
that may result in an  increase of the actual value of $T_{\rm bl.f}^v$.

\subsubsection{Vector-isospin  vector boson -- fermion
system} 
The vector-isospin vector boson -- fermion
system (the interaction term
$L_{int} =-g_{i.v}
\bar{\psi}\gamma_{\mu}\vec{\tau}\vec{\phi}^{\mu}\psi$)
is considered in the same manner, as the vector boson
-- fermion system just discussed.
All expressions 
obtained for vector bosons continue to
hold also for vector-isospin vector bosons, if one 
replaces $g_v\rightarrow g_v^* \simeq \sqrt{3}g_{i.v.}$. 

For $\omega$ and $\rho$ mesons one has $g_{\omega}\simeq 8\div 10$ and 
$g_{\rho}\simeq 2.5\div 3$, from where we estimate that the 
effective $\rho$ meson coupling is $\sim 0.5$ of that  for the
$\omega$.

\subsubsection{Comparison with
example of one diagram of $\Phi$ (no correlations) }

To compare our general result with the result one would
obtain  without inclusion of
correlations let us 
consider the problem 
with only one diagram of $\Phi$ (first diagram (\ref{phi})).
In this case one should replace the vertex (\ref{WT}) by the bare vertex
$\gamma^\mu$. 
In the STL approximation the
Dyson equation (\ref{Dys-v}) then takes the form
\be\label{Dys-vA}
\widehat{G}_f^R =\widehat{G}_f^{0,R}-\widehat{G}_f^{0,R}J_v^{\mu\nu}\gamma_\mu
\widehat{G}_f^R \gamma_{\nu}
\widehat{G}_f^R ,
\ee
with $J_v^{\mu\nu}$ given by eq. (\ref{Jvec}). 
Again, to simplify the consideration we will 
assume
that $A_{b,(\pm)}^t \simeq A_{b,(\pm)}^l$, being 
correct in the case of a small change of the effective vector boson mass.
This assumption allows us to omit values $J_{01}^v$ and $J_{11}^v$.

Now we may work out the tensor structure of (\ref{Dys-vA}). For that we
first calculate the auxiliary quantity
\be\label{aux111}
&&(\delta_{\mu}^\nu -\frac{q_\mu q^\nu }{q^2} )\gamma^\mu 
(G_1 \FMslash{p} +G_2 m_f )\gamma_{\nu}(G_1 \FMslash{p} +G_2 m_f )
\\
&&=-G_1^2 p^2 +2m_f \FMslash{p}G_1 G_2 +3m_f^2 G_2^2 - 
2\frac{(pq)\FMslash{q}\FMslash{p}}{q^2}G_1^2 -
2m_f \frac{\FMslash{q}(pq)}{q^2}G_1 G_2 ,\nonumber
\ee
see eq. (\ref{aux}).

Let us again use the non-relativistic approximation for fermions.
Then we may suppress
terms containing $\vec{p}\,\vec{q}$ compared to the corresponding
term $\propto p_0 q_0 \sim m_f  q_0$. 
Also we drop the term  $\propto \vec{\gamma}$, 
which does not contribute to the particle density. Then
we obtain two coupled equations
\be\label{Gsyst1}
p^2 G_1 -m_f^2 G_2 &=&1+J_0^v (G_1^2 p^2 -3G_2^2 m_f^2 )+2J^v_2 m_f^2
G_1^2 ,\nonumber\\
G_2 -G_1 &=&2J^v_1 G_1 G_2 ,
\ee
with
\be\label{Jv2}
J_2^v = g_v^2 
\int \frac{q_0^2}{q^2}\frac{d^3 q}{(2\pi)^3}
\int_{0}^{\infty} \frac{d q_0}{2\pi}
\left[ A_{b,(+)}^{t}n_{b,(+)} (q_0 )
+ A_{b,(-)}^{t}n_{b,(-)} (q_0 )\right] .
\ee
Within the quasiparticle approximation for bosons
$J_2^v$ is expressed through $J_0^v$ and $J_1^v$.

We replace $G_1$ from the second eq. (\ref{Gsyst1}) to the first one and
retain there only quadratic terms in $G_2$, where we put $p^2 \simeq m_f^2$,
and we use $J^v_1 \ll m_f^2$. Approximations are correct 
for typical momenta of our interest given by (\ref{typ}). Then we find 
\be\label{mainsol-lovv}
G_2^R \simeq \frac{p^2 -m_f^2
\pm i\sqrt{2(\gamma^{{\rm n.rel}}_v)^2 -(p^2 -m_f^2 +i0)^2}
}{(\gamma^{{\rm n.rel}}_v)^2} .
\ee
For $|p^2 -m_f^2|<\sqrt{2}\gamma^{{\rm n.rel}}_v$ the upper sign solution
should be dropped.
This result is nicely 
matched with that in the scalar case, cf. (\ref{mainsol-lov}). The
difference is 
only in the value of the intensity of the multiple 
scattering ($J_1^v$ in the vector boson case
instead of $J_{s}$ in the scalar boson case). 
Imaginary part of the Green function (\ref{mainsol-lovv})
satisfies the
exact sum rule. However we note that
the Green function (\ref{mainsol-lovv})
is  quite different from that given by eqs (\ref{Jv-int}), (\ref{Img2}), which 
exploit the full vertex. 

We conclude that {\em in the vector boson
case 
correlations qualitatively change the picture 
modifying expressions for the effective fermion mass
and  the value of the temperature $T_{\rm bl.f}^v$}.

\subsection{Hot vector boson condensation}

Now let us consider a possibility of the HBC  for vector bosons.
To simplify the consideration we 
will not discuss a modification of the vector boson 
spectrum rather we will only estimate the critical temperature for 
the HBC in the transversal channel.
For that we need only to evaluate the quantity $\re (\Sigma_b )^{i}_i (q_0 =0,
|\vec{q}|\rightarrow 0)$, cf. (\ref{tenstr}).

Using (\ref{sigm-R0-first}), 
where we replace one bare vertex to the full vertex (\ref{WT}), we obtain
\be\label{sigm-R-first} 
&&\re (\widehat{\Sigma}_b^{R})^{\nu\mu}(q_0 =0, |\vec{q}|\rightarrow 0) 
\simeq -4 g_v^2 \mbox{Tr}
\int \frac{d^3 p}{(2\pi)^3}\int_{0}^{\infty}\frac{d p_0}{2\pi}
\gamma^{\nu} \re \widehat{G}_f^{R} (p)
\nonumber \\
&\times&
 \re\frac{\partial \widehat{G}_f^{-1}}{\partial p_\mu}
\im \widehat{G}_{f}^R (p)n_{f}(p_0 ) +\{f\rightarrow \bar{f}\}.
\ee
With the help of eq. (\ref{vertap}), where only second term contributes,
assuming $G_1 \simeq G_2$ (non-relativistic nucleons) we
find 
\be\label{sigm-R-se} 
&&\re (\widehat{\Sigma}_b^{R})^{i}_{i}(q_0 =0, |\vec{q}|\rightarrow 0) 
\simeq 48 g_v^2
\int \frac{d^3 p}{(2\pi)^3}\int_{0}^{\infty}\frac{d p_0}{2\pi}
\nonumber \\
&&\times\im G_2 n_f
(p_0)+\{f\rightarrow \bar{f}\} . 
\ee
Using (\ref{3oc}), (\ref{den}) and (\ref{tenstr}) we obtain
\be\label{sigm-R-tr} 
\re (\Sigma_b^{R})^{t}(q_0 =0, |\vec{q}|\rightarrow 0) 
\simeq -6 g_v^2\rho_{f,\bar{f}} /m_f . 
\ee
Here 
$\rho_{f,\bar{f}}$ is, as before, the density of fermion-antifermion pairs
of one fermion species.  In reality other fermion species (e.g.,
as neutrons and protons) may also contribute to $\widehat{\Sigma}_b$.

The critical temperature and the fermion-antifermion density
for the HBC of vector bosons in the transversal channel under consideration
would be found
from the condition 
\be
m_{bt}^{*\,2} =m_b^2 +\re (\Sigma_b^{R})_{tr}(q_0 =0, |\vec{q}|\rightarrow 0)=0 ,
\ee
if it had the solution.
However, since the intensity of the multiple scattering $J_1^v$
tends to infinity for $m_{bt}^{*\,2}\rightarrow 0$, see
(\ref{J1h}), the density of fermion-antifermion pairs given by eq. 
(\ref{f-d-vr}) is
anomalously
increased. This motivates the possibility of the
first order phase transition to the HBC state with a jump of  
$m_{bt}^{*\,2} (T_{\rm cb}^v )$ from a positive to a negative value 
in the critical point.

Let us assume that this 
jump occurs at $m_{bt}^{*\,2} =\zeta m_{bt}^{2}$, $0<\zeta <1$.
Then with the help of eq. (\ref{sigm-R-tr})  we find
\be\label{cB}
\rho_{f,\bar{f}}^{\rm cb} (T_{\rm cb}^v )\simeq \frac{m_f m_b^2 (1-\zeta )}{6g_v^2}.
\ee
Note that eq. (\ref{f-d-vr}) was derived in the non-relativistic approximation
for fermions. It should be replaced to the relativistic expression
at temperatures, when $m_f^* (T)$ becomes to be $\lsim T$. Thus 
eq. (\ref{cB}) yields a correct estimate only, if $m_f^*  (T)\gsim T$.
To further study the first order phase transition one needs to
calculate the contribution to the thermodynamic potential in both phases.
This requires a more 
detailed analysis. Only for a rough estimate of the value
$T_{\rm cb}^v$ one may use (\ref{f-d-vr}), (\ref{J1h}), (\ref{cB}).

If the system consists of two fermion species (e.g., neutrons and protons)
it can be taken into account 
by the replacement $\rho_{f,\bar{f}}^{\rm cb}\rightarrow
2\rho_{f,\bar{f}}^{\rm cb}=
\rho_{N,\bar{N}}^{\rm cb}$
in the boson self-energy.
For the $N\bar{N}\omega $ (two fermion species)
and for $g_v \simeq 10$, $m_{\om}\simeq 782~$MeV,
$m_N \simeq 933~$MeV, assuming $m_{bt}^{*} (T_{\rm cb}^v )\simeq m_{\pi}$
we estimate
$\rho_{N,\bar{N}}^{\rm cb}
\simeq 0.7 \rho_0$.  According to (\ref{f-d-vr}) 
and (\ref{f1}), 
the density $\rho_{N,\bar{N}}^{\rm cb}
\simeq 0.7 \rho_0$
corresponds to the
temperature $T\simeq 110~$MeV.
The  value $0.7 \rho_0$ at such a low temperature  is  well below the
deconfinement density. Thus quark effects seem to be still unimportant.  
However note that values $T_{\rm bl.f}^{\rm v,n.rel}
\simeq 120~$MeV and $T_{\rm cb}^v \simeq 110~$MeV
could be underestimations, as the result of 
crude approximations, which we have done.  As we have mentioned the
non-relativistic limit
eq.
(\ref{f-d-vr}) 
might become quantitatively incorrect
in the vicinity of $T_{\rm bl.f}^{\rm v,n.rel}$ yielding an overestimation
of the density and, thus, an underestimation of $T_{\rm cb}^v$.
Another reason might be that
in reality $g_v$ is less than $10$. For $g_v \simeq 8$ we would get higher values 
$T_{\rm bl.f}^{\rm v,n.rel}
\simeq 140~$MeV, $T_{\rm cb}^v \simeq 130~$MeV. 
Nevertheless, within a reasonable variation of parameters
we always obtain that the blurring of the nucleon
continuum and the HBC of $\omega$ and $\rho$ mesons
could occur already for $T<m_{\pi}$. 
Concluding this discussion we
again stress that we considered a model problem and disregarded many relevant
couplings
that might yield an effective repulsion.

\begin{figure*}
\includegraphics[clip=true,width=10cm]{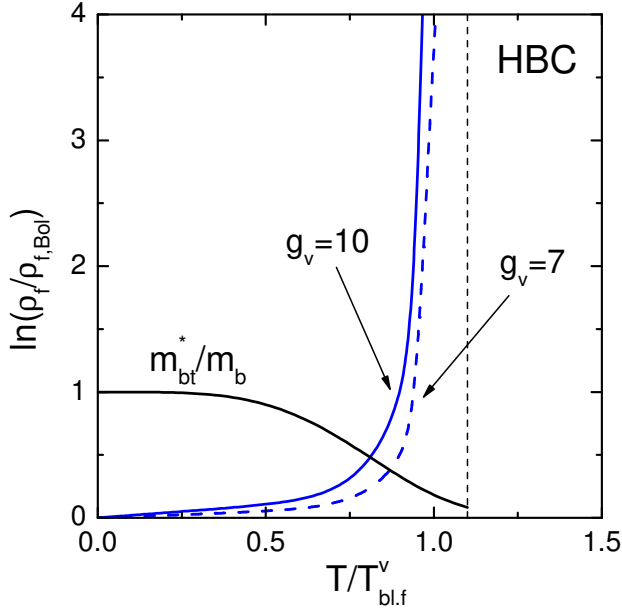}
\caption{Logarithm of the ratio of the fermion density to the corresponding
Boltzmann quantity for $g_v =10$ and $g_v =7$,
and the effective transverse vector boson mass
as functions of the temperature in units of $T_{\rm bl.f}^v$. The vertical
dash-line indicates the area of the HBC.
}
\label{fig:dens-V.eps}
\end{figure*}

The typical 
behavior of the logarithm of the ratio of the fermion density to the corresponding
Boltzmann quantity as function of the temperature is shown in
Fig. \ref{fig:dens-V.eps} for $T<T_{\rm cb}^v$.
Moreover Fig. \ref{fig:dens-V.eps} demonstrates 
the temperature dependence of the effective transverse vector
boson mass.
We see a huge enhancement of $\mbox{ln}(\rho_f /\rho_{f,\rm Bol})$ and a drastic
decrease of $m_{bt}^* /m_{b}$ in the
vicinity of $T=T_{\rm bl.f}^v , T_{\rm cb}^v$. 
At the critical point   the first order phase transition to
the HBC state may occur. The value $m_{bt}^* /m_{b}$
does not reach zero at this point.

\section{Heavy fermions and less massive 
pseudo-scalar bosons (pseudo-vector coupling) 
}\label{pseudo-vector}
Let us now consider pseudo-scalar  boson -- spin $\frac{1}{2}$ fermion system
interacting via the pseudo-vector coupling 
\be\label{lintp}
L_{int} =-ig_{p.v}m_b^{-1}
\bar{\psi}\widehat{q}\gamma_5 \vec{\tau}\vec{\phi}\psi .
\ee 
In realistic case bosons are $\pi^+ ,\pi^- ,\pi^0$ and 
fermions are $n$ and $p$. The $\pi NN$ coupling is $g_{p.v} \simeq 1$.

\subsection{Low temperature limit}
Again we start with the simplest $\Phi$ 
given by the first
diagram (\ref{phi}).
We calculate the value
\be\label{Snonr}
&&\widehat{S}^0 =\widehat{V}_0 (q)\widehat{\Lambda}_f^0 (p+q)
\widehat{V}_0  (-q)
\widehat{\Lambda}_b^0 (q)\nonumber\\
&&=N_{p.v}g_{p.v}^2 m_b^{-2} [2 p q\FMslash{q}-
(m_f +\FMslash{p})q^2 +q^2\FMslash{q}], 
\ee 
where $N_{p.v} =3$ is the isospin degeneracy factor ($\pi^+ ,\pi^{-},\pi^{0}$).
Within the non-relativistic approximation for fermions 
replacing $p_0 \simeq m_f$, $q^2 \simeq
m_b^2$, using $q_0 \ll m_f$, $|\vec{q}|\ll m_f$, 
and retaining only zero components of
vectors (dropping linear  terms in $\vec{\gamma}$, which do not further
contribute)
we simplify (\ref{Snonr}) as follows
\be\label{pvecap}
&&\widehat{S}^0 \simeq N_{p.v}
g_{p.v}^2 m_b^{-2}m_f \left[ \vec{q}^{\,2} (1 +\gamma_0 )
-q_0^2 (1 -\gamma_0 )\right].
\ee
With the help of eqs (\ref{pvecap}), (\ref{foclow-0}), (\ref{foclow}) 
we calculate the fermion 3-momentum 
distribution
\be\label{3ocU-pv}
\widehat{n}_{f,(\pm)} (\vec p )\simeq \left[\frac{(\epsilon_{p}+ m_f\gamma_0)}
{2\epsilon_{p}}
+\frac{(1 +\gamma_0 )N_{p.v} g_{p.v}^2 }{8
\pi^2}I_{0v}^0  (\frac{\epsilon_p }{m_b})\right]
n_{{\rm Bol},(\pm)}, 
\ee
where
$I_{0v}^0 (x)$ is determined by eq. (\ref{i0v}).

Finally, the fermion particle and antiparticle densities are
\be\label{r-rat-fpv}
\frac{\rho_{f,(+)}}{\rho_{{\rm Bol},(+)}}=\frac{\rho_{f,(-)}}
{\rho_{{\rm Bol},(-)}}\simeq
1+\frac{N_{p.v} g_{p.v}^2}{4\pi^2}I_{0v}^0 \left(\frac{m_f }{m_b}\right). 
\ee
This result essentially differs from eq. (\ref{r-rat})
for  scalar neutral bosons,
since $I_{0v}^0$ and $I_{0s}$
are different. The difference with the result for
the vector-isospin vector boson system for the case of
{\em{a warm hadron liquid}}, where $I_{0v}^1$ term is suppressed
(see eq. (\ref{r-rat-f})), is hidden in the different coupling constants
and spin-isospin dependent coefficients.

For  $T\sim m_{\pi}/2$,
the regime of {\em{the warm hadron liquid}}
is realized, $m_b \gg T\gg m_b^2 /m_f$. We find
$\rho_{p}^{sym}=\rho_{n}^{sym}=\rho_{\bar{p}}^{sym}=\rho_{\bar{n}}^{sym}
\simeq 1.16\rho_{\rm Bol}^{sym}$,
that yields rather small correction
due to a moderate value of the coupling constant, 
$g_{\pi N} \simeq 1\ll g_{\sigma N}\simeq 10$. Nevertheless this correction is
of the order of that for $\sigma$ and $\omega$ mesons.
The reason is that the pion is lighter than $\sigma$ and $\omega$. Thereby,
$I_{0v}^0$
is enhanced, see eq. (\ref{vap}). Moreover for $T\sim m_{\pi}/2$ for 
$\sigma$ and $\omega$ still the regime of {\em{the slightly heated hadron liquid}}
occurs rather than the regime of {\em{the warm hadron liquid}}, as for the pion.

{\em{Contribution of the correlation diagrams is negligible in the 
low temperature
limit, as in the scalar and vector boson cases.}}

\subsection{High temperature limit}
\subsubsection{Analytic solution for fermion Green functions 
in STL approximation}
We continue operate  with the single diagram of $\Phi$
(the first
diagram (\ref{phi})).
Now we calculate the value
\be
&&\widehat{S}=\widehat{V}_0 (q)\widehat{\Lambda}_f(p)\widehat{V}_0  (-q)
\widehat{\Lambda}_b^0 (q)\widehat{\Lambda}_f(p)\nonumber
\\
&&= 3g_{p.v}^2 m_b^{-2}\left(-p^2q^2 +2(pq)  \FMslash{q}\FMslash{p}\right)
(G_1^R)^2  \nonumber
\\
&&- 3g_{p.v}^2 m_b^{-2}\left(
m_f^2 q^2(G_2^R)^2 +2 m_f (q^2 \FMslash{p} - qp\cdot \FMslash{q})G_1^R G_2^R
\right), 
\ee
entering eq. (\ref{sigmJ}). We used eq. (\ref{spinstr}), where we
have put $G_3 =0$.
Dropping terms
linear in 
$\vec{\gamma}$ and $\vec{q}\vec{p}$
and using (\ref{spinstr}), 
we may rewrite the Dyson equation (\ref{Dscat}) for the fermion sub-system,
as the system of coupled equations
\be\label{Dys-coup1pvsys}
&&p^2 G_1^R -m_f^2 G_2^R = 1+ [(J_1^{p.v}-J_2^{p.v})p^2 
+2p_0^2 J_2^{p.v} ]
(G_1^R)^2
\nonumber\\
&&+(J_1^{p.v}-J_2^{p.v})m_f^2 (G_2^R)^2 ,\nonumber\\
&&G_2^R =G_1^R +2J_1^{p.v}G_1^R G_2^R ,
\ee
where now 
\be\label{sigmJ-pv}
&&J^{p.v}_1 =\frac{3 g_{p.v.}^2}{m_b^2}
\int \frac{\vec{q}^{\,\,2}d^3 q}{(2\pi)^3}\int_{0}^{\infty} \frac{d q_0}{2\pi}
\left[ \widehat{A}_{b,(+)} n_{b,(+)}(q_0 )+
\widehat{A}_{b,(-)} n_{b,(-)}(q_0 )\right],
\ee
\be\label{sigmJ-pv2}
&&J^{p.v}_2 =\frac{3 g_{p.v.}^2}{m_b^2}
\int \frac{q^{\,\,2}_0 d^3 q}{(2\pi)^3}\int_{0}^{\infty} \frac{d q_0}{2\pi}
\left[ \widehat{A}_{b,(+)} n_{b,(+)}(q_0 )+
\widehat{A}_{b,(-)} n_{b,(-)}(q_0 )\right].
\ee
Setting 
$G_1 \simeq G_2$ in quadratic terms ($(G^R)^2$) we
arrive at the Dyson equation
\be\label{Dys-coup1pv}
&&G_1^R = G_f^{0,R}+ G_f^{0,R} [(p_0^2 +3m_f^2 )J^{p.v}_1 
-
(J^{p.v}_1 -J^{p.v}_2 )\vec{p}^{\,2} ](G_1^R)^2  \nonumber\\
&&+J^{p.v}_2 (G_1^R)^2.
\ee
This equation already allows for the analytic solution.

\subsubsection{Intensity of multiple scattering}
Within the quasiparticle approximation for bosons and
antibosons the spectrum of boson excitations becomes
\be\label{branchpv}
\om_{(\pm)}^2 (\vec{q},T) &\simeq&m_b^2 +\beta_{p.v}(T)\vec{q}^{\,2}+
\widetilde{\beta}_{p.v}(T)\vec{q}^{\,4}+...,\nonumber\\
\ee
cf. subsection \ref{pvHBC}. Due to the p-wave nature of the interaction
(\ref{lintp})
in eq. (\ref{branchpv}) there appear the bare boson mass and a significantly
corrected $\vec{q}^{\,2}$ term. Even one may have $\beta_{p.v}<0$ in some
temperature interval. For cases of sufficiently small $\beta_{p.v}$ and for
$\beta_{p.v}<0$ one should keep in (\ref{branchpv}) the higher order terms, at
least the $\vec{q}^{\,4}$ term.

Using this spectrum 
we may evaluate the intensity of the multiple scattering $J^{p.v}_1$.
For $\beta_{p.v}>0$, assuming it is not as small, dropping the term 
$\widetilde{\beta}_{p.v}\vec{q}^{\,4}$ in (\ref{branchpv}),
with the help of eq. (\ref{sigmJ-pv})
we find 
\be\label{Jexppv}
J^{p.v}_1&=&\frac{6g_{p.v}^2}{m_b^2} 
\int \frac{\vec{q}^{\,\,2}d^3 q}{(2\pi)^3}\int_{0}^{\infty}
\frac{d q_0 \delta \left(q_0^2 -m_b^2 -\vec{q}^{\,\,2} -\re 
\Sigma_b^{R,(0)}(q^2)
\right)}{e^{q_0 /T}-1}\\
&=&\frac{3g_{p.v}^2}{2\pi^2 m_b^2}\int_{0}^{\infty}\frac{ \vec{q}^{\,\,4} 
d |\vec{q}|} 
{ \left(
m_b^2 +\beta_{p.v}\vec{q}^{\,\,2}
\right)^{1/2}}
\frac{1}{\mbox{exp}\left[\left(
m_b^2 + \beta_{p.v}
\vec{q}^{\,\,2}\right)^{1/2}
/T\right]- 1} ,\nonumber
\ee
and in limiting cases
\be\label{Jexppv-limh}
J^{p.v}_1&=& 
\frac{\pi^2 g_{p.v}^2  T^4}{10\beta_{p.v}^{5/2}m_b^2}
\,,\quad \mbox{for} \quad T\gg m_b ,
\ee
\be\label{Jexppv-lim}
J^{p.v}_1&=& \frac{9g_{p.v}^2 T^{5/2} }
{2^{3/2} \pi^{3/2}\beta_{p.v}^{5/2} m_b^{1/2}} 
\mbox{exp}\left( -m_b /T\right) \,,\quad \mbox{for} \quad T\ll m_b .
\ee
Compare these expressions with the corresponding eqs (\ref{J1h}),
(\ref{J1l}) for the vector boson case.

Using (\ref{sigmJ-pv2})
we obtain (for $\beta_{p.v}>0$)
\be\label{Jexppvj2}
J^{p.v}_2
&=&\frac{3g_{p.v}^2}{2\pi^2 m_b}\int_{0}^{\infty}\frac{ \vec{q}^{\,\,2} 
d |\vec{q}| 
\left(
m_b^2 +\beta_{p.v}\vec{q}^{\,\,2}
\right)^{1/2}}{\mbox{exp}\left[\left(
m_b^2 +\beta_{p.v}
\vec{q}^{\,\,2}\right)^{1/2}
/T\right]- 1} ,
\ee
and in limiting cases
\be\label{Jexppv-limhj2}
J^{p.v}_2&=& 
\frac{\pi^2 g_{p.v}^2  T^4}{10\beta_{p.v}^{3/2}m_b^2}
\,,\quad \mbox{for} \quad T\gg  m_b ,
\ee
\be\label{Jexppv-limj2}
J^{p.v}_2&=& \frac{3g_{p.v}^2 T^{3/2}  m_b^{1/2}}
{2^{3/2} \pi^{3/2}\beta_{p.v}^{3/2} } 
\mbox{exp}\left( -m_b /T\right) \,,\quad \mbox{for} \quad T\ll m_b .
\ee
For $\beta_{p.v}\rightarrow 0$ intensities of the multiple
scattering
anomalously increase.
However in reality they do not tend to infinity, if one incorporated the
term $\widetilde{\beta}_{p.v}\vec{q}^{\,\,4}$
in eq. (\ref{branchpv}). 

The maximum value of the density that can be achieved in this regime can be
estimated as (\ref{ratd1-max}), now using eq. (\ref{Jexppv-limhj2}):
\be\label{ratd1-maxps}
\rho_{f,(\pm)}^{\rm max}
\simeq \frac{m_f^{3/2}T^{3}}{2^{3/2}\pi^{2} J_s^{3/4}}
\rightarrow \frac{5^{3/4}\beta_{p.v}^{9/8} m_f^{3/2}m_b^{3/2}}{2^{3/4}
\pi^{7/2} g_{p.v}^{3/2}} .
\ee

As we shall see in subsection \ref{pvHBC}, the quantity $\beta_{p.v}$
decreases with the increase of the temperature. For $T=T_{{\rm cb},1}^{p.v}$
the value $\beta_{p.v}$ reaches zero and for $T>T_{{\rm cb},1}^{p.v}$ it becomes
negative. In refs \cite{Dyug83,V93} the density/temperature, when
$\beta_{p.v}$
crosses zero, was called the critical density/temperature for the appearance of
{\em{the liquid phase of the pion condensation}}. 
There is no yet a long-ranged
order for $T_{\rm cb}^{p.v}>T>T_{{\rm cb},1}^{p.v}$. However there arise
many virtual boson excitations carrying finite momentum $|\vec{q}_c|$.
Actual
values of particle momenta  are near $|\vec{q}_c|$, and directions of the
momenta are randomly distributed.
For $\beta_{p.v} (T)<0$, the quantity
\be
\widetilde{\om}^2 (\vec{q},T)
=m_b^2 +\re \Sigma_b^R (q_0 =0, \vec{q},T),
\ee
that can be called {\em{the effective boson
    gap}}, gets the minimum, 
cf. \cite{MSTV,V93}.
At low $q_0$, the boson spectrum
has a non-quasiparticle nature.
As we will demonstrate it below, see eqs
(\ref{sigm-R0-finpv}), (\ref{sigm-R0-imfinpv}), the dispersion relation 
is then given by
\be\label{brim}
i\beta_1 (|\vec{q}|)\om \simeq \om_0^2 +\beta_0 (|\vec{q}|-|\vec{q}_c|)^2 ,
\quad |\vec{q}_c|\neq 0,\,\om_0^2 ,\,\beta_1 ,\,\beta_0 >0 ,
\ee
where
\be\label{omze}
\om_0^2 &=&\widetilde{\om}^2 (\vec{q}_c )
=m_b^2 +\re \Sigma_b^R (q_0 =0, \vec{q}=\vec{q}_c )\\
&\simeq& m_b^2 -
\beta_{p.v}^2 /(4\widetilde{\beta}_{p.v}).\nonumber
\ee
The value
\be
|\vec{q}_c | \simeq \sqrt{-\beta_{p.v}/(2\widetilde{\beta}_{p.v})}
\ee 
corresponds to the minimum of $\widetilde{\om}^2 (|\vec{q}|)$,
$\beta_0 \simeq -2\beta_{p.v}$.
The form of the spectrum (\ref{brim})
coincides with that previously 
used in the description of the dense baryon matter in the pion
condensation problem,
cf. \cite{MSTV}.
The boson spectral function has the form
\be\label{vec-nonq}
\widehat{A}_{b,(\pm)}\simeq \frac{2\beta_1 q_0}{
[\om_0^2+\beta_0 (|\vec{q}|-|\vec{q}_c|)^2]^2 +\beta_1^2 q_0^2}.
\ee
Replacing (\ref{vec-nonq})  in eqs (\ref{sigmJ-pv}), (\ref{sigmJ-pv2})
and using for simplicity the limit $\beta_1 T\gg \om_0^2$,
$\beta_1 \equiv \beta_1 (\vec{q}_c )$, we calculate
\be\label{relJk0}
J^{p.v}_1 \simeq \frac{3g_{p.v}^2 \vec{q}_c^{\,4} T}{2\pi m_b^2 
\sqrt{\beta_0}\om_0 },\quad J^{p.v}_2 \ll J^{p.v}_1 .
\ee
One has $J^{p.v}_2 
\ll J^{p.v}_1$, since $J^{p.v}_2$ has no singularity for $\om_0
\rightarrow 0$. We find an anomalous increase of the intensity of multiple
scattering for $\om_0
\rightarrow 0$, i.e. the critical opalescence. In the given case, as for
vector bosons, it
is the precursor of the first order phase
transition to the HBC. However in difference with the
vector boson case here the phase transition may occur
to the crystalline or liquid crystalline state, see further discussion in
subsection \ref{pvHBC}. 

\subsubsection{Non-relativistic fermions}

In non-relativistic approximation for fermions we may put
$|\vec{p}|\ll p_0$  and $p_0
\simeq m_f$.
Then  (\ref{Dys-coup1pv}) simplifies as
\be\label{Dys-coup1pvnr}
G_1^R = G_f^{0,R}+ G_f^{0,R} 4m_f^2 J^{p.v}_1 (G_1^R)^2 .
\ee
Its solution takes the form
\be\label{mainsol-lovpv}
&&\re G_1^R \simeq \re G_2^R \simeq \frac{p^2 -m_f^2}{\gamma_{p.v}^2},
\quad \gamma_{p.v}^2 =8J_1^{p.v}
m_f^2 ,
\nonumber\\
&&\im G_1^R \simeq \im G_2^R\simeq -
\frac{\sqrt{2\gamma_{p.v}^2 -(p^2 -m_f^2 +i0)^2}}{\gamma_{p.v}^2},
\ee
for $2\gamma_{p.v}^2 >(p^2 -m_f^2 )^2$.
The only difference with eq. (\ref{mainsol-lov}) (the latter equation
is valid for the scalar boson)
is that in  (\ref{mainsol-lovpv}) 
enters $\gamma_{p.v}$ instead of $\gamma_{s}$.
Therefore, after the replacement
$J_s \rightarrow J^{p.v}_1 $ {\em{expressions 
(\ref{hdist}) -- (\ref{Tcssn}) 
hold also in the pseudo-scalar case.}}

The temperature of the blurring of the fermion continuum evaluated within the
non-relativistic approximation for fermions ($T_{\rm bl.f}^{p.v ,\rm n.rel}$)
follows from the relation
$J^{p.v}_1 =m_f^2 /4$ (for $\mu_f =0$), from where using 
(\ref{Jexppv-limh}) (at the condition $\beta_{p.v}>0$) we  estimate,
cf. \cite{D93}, 
\be\label{tblps}
T_{\rm bl.f}^{p.v ,\rm n.rel}=\frac{5^{1/4}\beta_{p.v}^{5/8}(m_f m_b )^{1/2}
}{2^{1/4}\pi^{1/2}g_{p.v.}^{1/2}}. 
\ee

\subsubsection{Relativistic fermions}
Let us consider the  contribution of the energy region $p_0\ll m_f$. Then  
from (\ref{den}), (\ref{nrelzero}), (\ref{Dys-coup1pv}) we find the 
fermion-antifermion density (for one species
of fermion):
\be\label{rhoblpv}
&&\delta\rho_{f,(\pm)}(p_0 \lsim T)
=\frac{T^2}{6\pi}\int_0^{|\vec{p}_{\rm max}|}\vec{p}^{\,2} 
d|\vec{p}| \nonumber\\
&&\times\frac{\sqrt{12m_f^2 J_1^{p.v} -4J_1^{p.v} 
\vec{p}^{\,2} -4J_2^{p.v}m_f^2
-(\vec{p}^{\,2} 
+m_f^2 )^2}}
{2m_f^2  
[3J_1^{p.v}-J_2^{p.v}-J_1^{p.v}
\vec{p}^{\,2}/m_f^2 ]},
\\
&&\vec{p}_{\rm max}^{\,2} =-m_f^2 -2J_1^{p.v}
+\sqrt{4(J_1^{p.v})^2
+16J_1^{p.v} m_f^2 -
4J_2^{p.v} m_f^2},\nonumber
\ee
cf. (\ref{rhobl}).
We see that for 
$J_1^{p.v}-\frac{1}{3}J_2^{p.v}
> J_{p.v}^{\rm bl.f}=\frac{m_f^2 }{12}$ (positive square root in 
(\ref{rhoblpv}) for $\vec{p}\simeq 0$)
{\em the fermion sub-system represents a rather
dense packing of
fermion-antifermion pairs.}
The condition 
\be\label{blpv}
J_{p.v} \equiv J_1^{p.v}-\frac{1}{3}J_2^{p.v}
= J_{p.v}^{\rm bl.f}=m_f^2 /12
\ee
determines the typical temperature $T_{\rm bl.f}^{p.v}$ of the
blurring of the fermion continuum. 
Assuming
$\beta_{p.v}>0$, and using (\ref{blpv}) and
high temperature limit estimates (\ref{Jexppv-limh}) 
and (\ref{Jexppv-limhj2}),
we obtain 
\be
(J_1^{p.v})^{\rm bl.f}\simeq \frac{m_f^2}{4 (3-\beta_{p.v})},
\quad 
\ee
and
\be\label{tblps-rel}
T_{\rm bl.f}^{p.v}=\frac{T_{\rm bl.f}^{p.v ,\rm non.rel}}{
(3-\beta_{p.v})^{1/4}}
=\frac{5^{1/4}(m_f m_b )^{1/2}
\beta_{p.v}^{5/8}}{2^{1/4}\pi^{1/2}g_{p.v.}^{1/2}(3-\beta_{p.v})^{1/4}}. 
\ee
For pions we find $T_{\rm bl.f}^{p.v}\simeq 215$~MeV for
$\beta_{p.v}=1$. As a more realistic estimation, taking 
$\beta_{p.v}\simeq 0.5$ we obtain
$T_{\rm bl.f}^{p.v}\simeq 132$~MeV.

For
$T\geq T_{\rm bl.f}^{p.v}$, small energies $p_0 \lsim T$ 
essentially contribute
to the filling of
the particle and antiparticle Fermi seas.
For $T$ in the vicinity of  $T^{p.v}_{\rm bl.f}$, 
using that $|\vec{p}_{\rm max}|=0$ for 
$J_{p.v} =J_{p.v}^{\rm bl.f}$ and expanding $|\vec{p}_{\rm max}|$  in 
$0<J_{p.v} -J_{p.v}^{\rm bl.f}\ll J_{p.v}^{\rm bl.f}$, from (\ref{rhoblpv}) 
we obtain 
\be
|\vec{p}_{\rm max}|\simeq 
\frac{(36 -26\beta_{p.v}+4\beta_{p.v}^2 )^{1/2}}{(7 -\beta_{p.v})^{1/2}}
\sqrt{
J^{p.v}_1 -(J^{p.v}_1)^{\rm bl.f}}, 
\ee
and 
\be\label{rhobl-1pv}
&&\delta\rho_{f,,(\pm)}(p_0 \lsim T)\simeq\frac{(3-\beta_{p.v})^{7/2} T^2 
\left[ J_1^{p.v} -(J_1^{p.v})^{\rm bl.f}\right]^2}
{3(7 -2\beta_{p.v})^{3/2}m_f^3}\,.
\ee

For artificially large values $J_1^{p.v}, J_2^{p.v}\gg m_f^2$ 
from (\ref{rhoblpv})
we evaluate
\be\label{rhoblart-pv}
\rho_{f,(\pm)}(p_0 \lsim T)\simeq \frac{T^2 m_f^2 (3-\beta_{p.v})}
{24 (J_1^{p.v})^{1/2}},\quad \vec{p}_{\rm max}^{\,2} \simeq
(3-\beta_{p.v})m_f^2 .
\ee
Using (\ref{Jexppv-limh}) we find
\be\label{rhoblart-pvfin}
\rho_{f,(\pm)}(p_0 \lsim T)\simeq \frac{10^{1/2}\beta_{p.v}^{5/4}
m_f^2 m_b (3-\beta_{p.v})}
{24 \pi g_{p.v}} .
\ee
If $\beta_{p.v}$ were independent of $T$, we would get a saturation
of $\rho_{f,(\pm)}$ with increase of $T$. However $\beta_{p.v}$
decreases with the temperature increase, see next subsection.
Maximum available density of fermions can be estimated equating
(\ref{rhobl-1pv}) plus (\ref{ratd1}) (after the replacement $J_s \rightarrow
J_1^{p.v}$) 
and, on the other hand, (\ref{rhoblart-pv}). With the further increase
of the temperature,  $\beta_{p.v}(T)$ diminishes and $\rho_{f,(\pm)}(p_0 \lsim T)$
decreases. Eqs (\ref{rhobl-1pv}), (\ref{rhoblart-pv}) continue to hold for 
$\beta_{p.v} <0$, if one sets in these equations $\beta_{p.v} =0$ (since
$J_2^{p.v}$ is suppressed for $\beta_{p.v} <0$). 
For $J_1^{p.v}$ then one should use the 
estimate (\ref{relJk0}).

\subsection{Medium modification  of boson excitations. HBC}\label{pvHBC}
Now we are at the position to evaluate the boson self-energy.
Taking the trace in (\ref{sigm-R0-first})
we obtain
\be
\mbox{Tr}\widehat{V}_0 (q)(\widehat{p}+m_f )\widehat{V}_0 (-q)
(\widehat{p}+m_f )=-\frac{4g_{p.v}^2}{m_b^2}[q^2 (p^2 +m_f^2 )-2p_0^2 q_0^2 ],
\ee
where we dropped small terms $\propto \vec{p}\vec{q}$.
Further for the sake of simplicity we will use
the non-relativistic approximation for fermions and we put
$G_1 \simeq G_2 $, $G_3
=0$.  Also,
restricting ourselves by consideration of small $q^\mu$,
let us for simplicity omit a $q$ dependence of the fermion Green function yielding
the wave function renormalization terms. Then from (\ref{sigm-R0-first})
we find
\be\label{sigm-R0-fappv} 
\re \Sigma_b^{R}(q) &\simeq&-32 N_{bar} g^2_{p.v} m_b^{-2}\vec{q}^{\,\,2} m_f^2
\int \frac{d^3 p}{(2\pi)^3}\int_{0}^{\infty} \frac{dp_0}{2\pi}
\re G_2^{R} (p) 
\nonumber\\
&\times&\im G_2^{R} (p)[n_{f,(+)}(p_0) +n_{f,(-)}(p_0)] .
\ee
$N_{bar}=2$ 
takes into account two type of fermions (neutrons and protons.
Quite similar to  that how we derived eq. (\ref{sigm-R0-fin})
we obtain
\be\label{sigm-R0-finpv} 
&&\re \Sigma_b^{R}(q)\simeq 
-\frac{2N_{bar}
g^2_{p.v}\vec{q}^{\,\,2} \rho_{f,\bar{f}}}{m_b^2 \sqrt{J^{p.v}_1}
}\nonumber \\ 
&&\times \left[ 1
-\frac{T}{2\sqrt{J^{p.v}_1}}
I_3 \left(\frac{4\sqrt{J^{p.v}_1}}{T}\right)
I_2^{-1} \left(\frac{4\sqrt{J^{p.v}_1}}{T}\right)\right].
\ee
Here $\rho_{f,\bar{f}}$ is, as above,
the density of fermion-antifermion pairs for
one fermion species.
With this expression we are able to recover the quantity
\be\label{alpv}
\beta_{p.v} &=&1-\frac{2N_{bar}
g^2_{p.v} \rho_{f,\bar{f}}}{\sqrt{J^{p.v}_1} m_b^2},
\ee
entering the boson spectrum (\ref{branchpv}). This equation becomes invalid
for rather small $\beta_{p.v}$, since we did not take into account
$\vec{q}^{\,4}$ terms, see  (\ref{branchpv}).

Instead of (\ref{sigm-R0-imfin}), that we had for the scalar boson,
for the given case of the pseudo-scalar boson we obtain
\be\label{sigm-R0-imfinpv} 
\im \Sigma_b^{R}&\simeq&-
\frac{N_{bar}
g^2_{p.v}\vec{q}^{\,\,2}\rho_{f\bar{f}}T^{1/2} }{(J^{p.v}_1)^{3/4}m_b^2}
I_4 \left(\frac{4\sqrt{J^{p.v}_1}}{T}\right)\nonumber\\
&\times&
I_2^{-1} \left(\frac{4\sqrt{J^{p.v}_1}}{T}\right)e^{-q_0 /T} 
\left( e^{q_0 /T}-1
\right) . 
\ee
Eq. (\ref{sigm-R0-fin} ) (after dropping there the wave function 
renormalization terms)
and eq. (\ref{sigm-R0-finpv}), and also (\ref{sigm-R0-imfin}) and  
(\ref{sigm-R0-imfinpv}) transform to each other with the help of the
replacements 
$g_s^2\rightarrow N_{bar}g_{p.v}^2 \vec{q}^{\,2}/m_b^2$ and $J_s \rightarrow
J_1^{p.v}$. From (\ref{sigm-R0-imfinpv}) for $q_0 \ll T$ we find 
$\im \Sigma_b^{R}\simeq -\beta_1 (\vec{q})q_0$, where
the value 
\be
\beta_1 (\vec{q})
=\frac{2N_{bar}g^2_{p.v} \rho_{f,\bar{f}}\vec{q}^{\,\,2} }{\pi^{1/2}m_b^2 
(J^{p.v}_1)^{3/4}T^{1/2}},\quad \mbox{for} \quad 4\sqrt{J^{p.v}_1}\gg T,
\ee
controls the low-energy part of the spectrum (\ref{brim}).
In the boson self-energy for $\beta_{p.v}>0$ we retained
the term $\propto\vec{q}^{\,\,2}$ and
dropped terms of the higher order 
$\propto \vec{q}^{\,\,4}$.
In
order to find an appropriate spectrum
for small $\beta_{p.v}>0$ and for
$\beta_{p.v}<0$  
one needs to incorporate $\propto \vec{q}^{\,\,4}$ terms.

The  second order phase transition to the p-wave HBC state could occur 
for $\beta_{p.v}<0$,
if the effective boson gap $\om_0$  reached zero at some finite
momentum $\vec{q}_c$. However the eq. (\ref{omze}) 
has no solution for $\om_0 =0$
at least in the region of the validity of the non-relativistic approximation
for nucleons. 
This is due to the fact that
the 
intensity of the
multiple scattering entering equations for the density $\rho_{f,\bar{f}}$
and $\om_0$ is 
$J_1^{p.v}\propto 1/\om_0$, see eqs (\ref{omze}),
(\ref{relJk0}) and (\ref{sigm-R0-finpv}).  
The second order phase transition is, thus, excluded. 
{\em{The phase transition to the p-wave HBC state is then the first order
transition.}} 
The latter possibility, however, needs a 
more
detailed analysis 
(including the comparison of thermodynamic potentials of two phases). 
It is worthwhile to mention
that  the pion condensation in dense nuclear medium also
arises by the first order phase transition at finite temperature, namely
as the
consequence of 
thermal fluctuations \cite{MSTV,VM83}. 
The latter are described by the same tadpole diagram,
as the intensity of the multiple scattering discussed here.

\begin{figure*}
\includegraphics[clip=true,width=9cm]{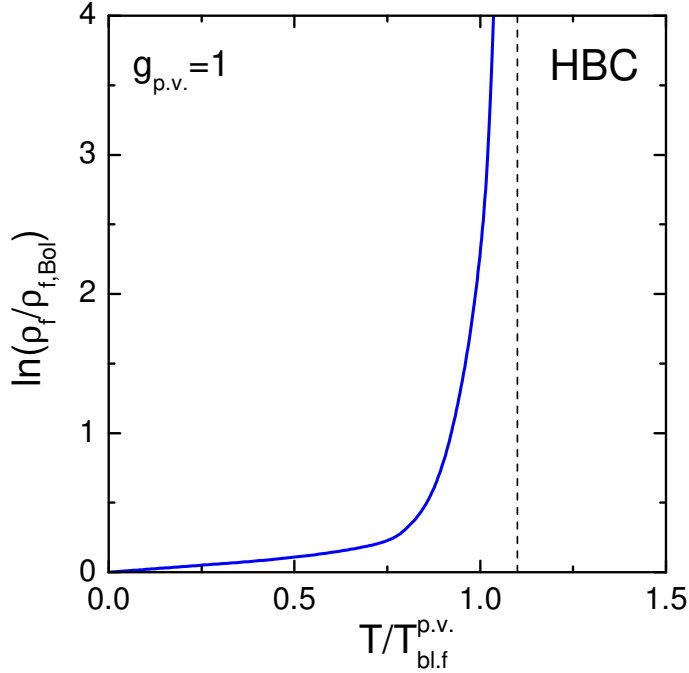}
\caption{Logarithm of the ratio of the fermion density to the corresponding
Boltzmann quantity for $g_{p.v} =1$
as function of the temperature in units of $T_{\rm bl.f}^{p.v}$. The vertical
dash-line indicates the area of the HBC.
}
\label{fig:dens-PV.eps}
\end{figure*}

The typical 
behavior of the logarithm of the ratio of the fermion density to the corresponding
Boltzmann quantity as function of the temperature is shown in
Fig. \ref{fig:dens-PV.eps} for $T<T_{\rm cb}^{p.v}$.
Here  the effective 
boson mass does not significantly deviate from the bare mass
but the p-wave boson-fermion interaction changes substantially $\vec{q}^{\,2}$
terms in the spectrum.
As in scalar and
vector boson cases,
we see a huge enhancement of $\mbox{ln}(\rho_f /\rho_{f,\rm Bol})$ in the
vicinity of $T=T_{\rm bl.f}^v$.  At the critical point   the first order phase transition to
the HBC state may occur.

\subsection{Correlation effects}\label{cor-pi}
The first diagram of $\Phi$, that we have studied, sums up 
all perturbative diagrams for the fermion Green function 
without crossing of the boson lines, cf. eq. (\ref{ladpr}). 
For the fermion-boson coupling under discussion,
within the STL approximation,
each diagram with the
one crossed boson line (see eq. (\ref{cr}))
brings the suppression factor $\nu =\nu_{\tau}\nu_{\sigma}$ compared to the 
corresponding
diagram with the same number of boson lines but without their crossing,
cf. \cite{D93,Dyug83}.
Indeed, in the non-relativistic approximation for nucleons,
$\nu_{\tau}=1/3$ is due to the 
non-commutation of the isospin $\tau$ matrices and
in non-relativistic approximation for nucleons
$\nu_{\sigma} =1/3$ is due to the 
non-commutation of the spin $\sigma$ matrices. Each product $\tau_i \tau_i 
\tau_j \tau_j$ yields factor 9, whereas 
each product $\tau_i \tau_j \tau_i \tau_j$ yields  factor 3. The same
statement is valid for $\sigma$ matrices. 
Concluding, we may retain only the first
diagram (\ref{phi}) in $\Phi$ with an appropriate accuracy.

\section{Towards the description of the state of
hadron porridge}\label{Porridge}
We discussed the behavior of model systems 
with a small fermion chemical potential,
assuming for the sake of simplicity
only one type
of the interaction in each case, like the case of spin $\frac{1}{2}$
fermions (e.g., $N$) coupled with the scalar boson ($\sigma$), or with the
vector boson ($\om$ or $\rho$), or pseudo-scalar
boson ($\pi$). This is, of course,  a gross oversimplification.
In reality nucleons couple with all mesons: $\sigma$,
$\pi$, $\omega$, $\rho$, $K$, etc. Also high lying baryon
resonances,
like $\Delta$-isobar, $N^* (1440)$, hyperons, etc., interact with the nucleon
and with each other.  Moreover, mesons interact with each other.
We may call such a system state {\em 
the hadron porridge}, bearing in mind that {\em{already at sufficiently low
temperatures the quasiparticle approximation fails to describe baryons
and that for $T\gsim m_{\pi}$ the continuum is blurred for all 
relevant hadrons,}} as we have shown it above on different concrete
examples.

\subsection{Hadron porridge in the low temperature limit}
In 
the low temperature limit contributions of different mesons are summed up
yielding resulting baryon resonance distribution and the total density of
baryon resonances. The  density of the
spin $\frac{1}{2}$
baryon resonance of given species is easily recovered with the help of 
eq. (\ref{r-r}):
\be\label{r-rtot}
&&\rho_{B^* ,(\pm)}=N_{B^*}\left(\frac{m_{B^*}T}{2\pi}\right)^{3/2}
\mbox{exp}\left[ -\frac{m_{B^*} \mp \mu_{B^*}}{T}\right] \\
&&+N_{B^*}\sum_{i}\left(\frac{m_{B^*_i}T}{2\pi}\right)^{3/2}
\mbox{exp}\left[ -\frac{m_{B^*_i} \mp \mu_{B^*_i}}{T}\right]
\frac{1 }{4\pi^2 }\nonumber\\
&&\times\left[ g_{\sigma B^* B^*_i}^2 I_{0s} \left(\frac{m_{B^*_i}}
{m_{\sigma}}\right)+3g_{\pi B^* B^*_i}^2 I^0_{0v} \left(\frac{m_{B^*_i}}
{m_{\pi}}\right)\right.\nonumber\\
&&\left. +g_{\omega B^* B^*_i}^2 \left(
I^0_{0v} \left(\frac{m_{B^*_i}}
{m_{\omega}}\right)+\frac{3}{2}\frac{m_{\omega}}
{m_{B^*_i}}I^1_{0v} \left(\frac{m_{B^*_i}}
{m_{\omega}}\right)\right)\right.\nonumber\\
&&\left. +3g_{\rho B^* B^*_i}^2 \left(
I^0_{0v} \left(\frac{m_{B^*_i}}
{m_{\rho}}\right)+\frac{3}{2}
\frac{m_{\rho}}{m_{B^*_i}}I^1_{0v} \left(\frac{m_{B^*_i}}
{m_{\rho}}\right)\right)+...
\right]\nonumber
. 
\ee
The summation is performed over all possible $s_{B^*} =1/2$
states 
including the given baryon state ($B^*$). Dots mean contributions of
other meson exchanges and transitions between baryon states with different
spins in initial and intermediate states 
in the diagram (\ref{selfzf}), e.g. ($N-K-\Lambda ; \Lambda -K-N$),
($N-\pi-\Delta (\frac{3}{2},\frac{3}{2}); \Delta (\frac{3}{2},
\frac{3}{2} )-\pi -N$),
 etc. Correlation effects are unimportant in the low temperature limit.

Numerically, already the contribution given by
terms, which are explicitly presented in eq. (\ref{r-rtot}), yields
$R_N =\rho_{N }/\rho_{\rm Bol}^{N }\simeq 1.5$
for $T\sim m_{\pi}/2$ and for values of coupling constants, which we have used
above; $\rho_{\rm Bol}^{N }$
is the nucleon density calculated with the help of the
Boltzmann distribution.
With taking into account of other contributions one may expect a further
enhancement of the ratio.  

We would like to pay attention to {\em{the enhancement of the 
population of baryon resonances, e.g., of the
strange baryons}},
since contributions like ($\Delta -\pi -N; N-\pi -\Delta$),
($\Lambda-K-N; N-K -\Lambda$), etc, 
have no any additional exponential suppression
compared to ($N-\pi-N; N-\pi -N$) terms, see eq. (\ref{phibar})
and the corresponding discussion. Thus, {\em{the ratio of the densities
of the
given resonance and the nucleon,
$R_{B^*/N} =\rho_{B^* }/\rho_{N }$, can be $\sim 1$
for the low lying baryon resonances, like
$\Delta$, $\Lambda$,  etc, for which $B^*$-meson-$N$ coupling
constants are not suppressed.}}

\subsection{Hadron porridge in the high temperature limit}

Description of the hadron porridge in the high temperature limit, 
$T\gsim m_{\pi}$,
is much more involved. Even in the STL approximation
one gets coupled system of Dyson equations for different baryon resonance Green
functions. Intensities of the multiple scattering entering the
Dyson equation for the given resonance describe
the coupling of this resonance to other resonances via
the radiation and the absorption 
of different type of mesons. Correlation effects
contribute to each vertex. The main correlation effect comes
from scalar meson and vector meson terms.
Moreover, boson effective masses and the $|\vec{q}|$ dependent terms 
of the boson spectrum are
changed. In addition to all has been said the problem should be solved 
self-consistently. 

Assume for a moment that  correlation
effects are suppressed. Then one deals with the sum of
one-diagram $\Phi$-terms, see first diagram (\ref{phi}),
where the boson line corresponds to
any relevant meson and the fermion line relates to one of baryon
resonances. 
Summation is over mesons and baryon resonances,
including nucleons.
Assume also that baryon resonance states decouple, i.e., coupling constants
$g_{\mu B^*_i ,B^*_j}=0$, 
for $i\neq j$, $\mu$ is one of mesons. Then, 
above results obtained for the system with
one type of baryons and one type of mesons
are easily generalized to include all above mentioned 
types of the interaction. The
contribution of the given baryon resonance $B^*$ in
non-relativistic approximation for baryon resonances is obtained
with the help of the replacement of the
corresponding $J$ to the sum of relevant
$J$'s, e.g.  
\be\label{sum-int}
J_s \rightarrow J_{\sigma ,B^* }+J^{\omega ,
B^* }_1 +
J^{\rho ,B^* }_1 +J^{\pi ,B^*}_1 +...
\ee

Inclusion of cross-channel couplings 
($g_{\mu B^*_i ,B^*_j}\neq 0$) and correlations
make the problem much more involved. {\em{Correlations may yield a significant
suppression of the resulting intensity of the 
multiple scattering}} compared to that given by (\ref{sum-int}).
Moreover, going beyond the
non-relativistic ansatz for baryons, that is necessary to do 
for $T\gsim T_{\rm bl.f}$,
additionally complicates the consideration.
We postpone a more detailed discussion of properties of the
hadron porridge  (including 
baryon resonances with a cross-channel coupling) to the subsequent paper.

\section{Conclusion}
The description of the hadron system with a small baryon chemical potential,
$|\mu_{\rm bar} |\ll T$, at the finite temperature
is essentially different from the description of the dense baryonic system
$|\mu_{\rm bar} - m_N |\gg T$, cf. \cite{MSTV}. The 
former regime might be relevant for  heavy ion collisions at
RHIC and LHC, whereas the latter one, for SIS energies. 

We treated the fermion-boson problem self-consistently considering first one heavy
fermion
and one lighter boson species. Then we estimated a role of correlation effects.

\begin{itemize}
\item
We found that
already at low temperatures,
$T\ll T_{\rm bl.f}\sim m_{\pi}\simeq 140$~MeV, the nucleon 3-momentum distributions are  enhanced
compared to ordinary Boltzmann distributions. Quasiparticle approximation
fails
already in regime of {\em{a warm hadron liquid}} ($m_b^2 /m_f \lsim T\ll
\mbox{min}\{m_b , T_{\rm bl.f}\}$), if this limit is indeed
realized for the given particle species and if the coupling constant is $g\gg 1$.
\item
For a higher temperature (typical value is $T\sim T_{\rm bl.f}\sim m_{\pi}$)
the hadron continuum is totally blurred.  
Effective masses of all hadrons are decreased and baryon widths are
dramatically increased due to baryon rescatterings on bosons. 
Baryons play a role of a glue for mesons.
The hadron liquid comes to the state of {\em{a hadron
porridge.}}
\item
Scalar mesons, like $\sigma$,
may undergo second order phase transition to the s-wave
hot Bose
condensate state for $T>T^s_{\rm cb}\sim T_{\rm bl.f}\simeq m_{\pi}$. 
\item
Vector mesons, like $\om$, $\rho$, 
may undergo first order phase transitions to s-wave hot Bose condensate
states for $T>T^v_{\rm cb}\sim T_{\rm bl.f}\simeq m_{\pi}$. 
In the vicinity of the transition point the intensity
of the multiple scattering is significantly increased leading to a critical
opalescence. 
\item
Pions  
may form a liquid hot Bose condensate  for $T>T_{{\rm cb},1}^{p.v}\simeq m_{\pi}$ and 
with a subsequent increase of the temperature
may undergo first order phase transition to the p-wave
hot Bose
condensate state. As in the vector case the intensity
of the multiple scattering is significantly increased in the vicinity of the 
critical point.
\item
For $T>T_{\rm bl.f}$
fermions fill a rather dense packing state. 
\item
The number of
baryon resonances, like $\Lambda$, should be 
not much less than the number of nucleons. In particular cases, like for 
$\Delta$ isobar, one even may expect a higher abundance than for nucleons. 
The number of kaons, 
which interact with nucleons and hyperons,
should be also large. One may expect the same effects for them, as for pions
and $\sigma$'s, due to a strong s- and p-wave $KN\Lambda$ interactions.
\end{itemize}

When scalar meson mass reaches zero, 
the alternative state 
to the hot Bose
condensate one  could be 
the chiral phase state. 
To conclude, which state is preferable, one needs to study
the chiral model, e.g., the linear sigma model, and one should
include quark effects.

For the low temperature regime, as a guide, we used the Urbach rule.
As a relevant approximation to simplify the self-consistent analysis of the
hot fermion-boson system 
($T\gsim m_{\pi}$), we exploited {\em{the soft thermal loop 
approximation.}}
The latter  approximation
allowed us to solve the problem. The self-consistency of the
approach is manifested in the fact 
that {\em{fermion Green functions 
completely loose their former pole shape.  It is the
consequence
of multiple collisions of the fermion with bosons. On the other hand, 
bosons drop their masses  interacting with modified fermions,
that  further stimulates the enhancement of the fermion width.}}
Please notice that  with only one $\Phi$ diagram we estimated $T_{\rm
  bl.f}\lsim m_{\pi}$. Attractive interactions of the nucleon with other meson
species, being taken into account, 
could still decrease the value $T_{\rm
  bl.f}$.
However we have shown that correlation effects in the scalar boson channel 
increase the value $T_{\rm
  bl.f}$. In the regime that we called ``the hadron porridge'' interactions
between many resonance and meson species come into play. Scalar bosons
contribute to vertices with participation of other mesons effectively 
suppressing vertices.
We considered only
some of interactions. Therefore it is hard to realistically estimate the
actual value of $T_{\rm
  bl.f}$. Our rough estimates show that $T_{\rm
  bl.f}$ is somewhere in the vicinity of the
pion mass ($140$~MeV). However further more detailed analysis is 
necessary.

We described the system in terms of only hadron degrees of freedom
and we found a state of {\em{a hadron porridge}} for $T\gsim m_{\pi}$.
In realistic situation  
quark-gluon degrees of freedom are also excited for
$T\gsim m_{\pi}$. These effects may yield an additional
diminishing
of effective hadron masses  due to the decrease  of the quark condensate
density with the increase of the temperature.
{\em{The role of quark-gluon fluctuations 
rises with increase of the temperature.}}
More likely, the system comes to 
{\em{a strongly
correlated hadron-quark-gluon state or, in another words, 
a boiled hadron-quark-gluon porridge, (HQGP).}} Deconfinement, in a standard 
meaning, as the pure quark-gluon state, 
is probably delayed up to a higher temperature.

Our aim was to discuss qualitative effects. Thereby, we focused our study 
on simplified models.
In order to quantify the
consideration and to do an appropriate  fit of experimental results, cf. 
\cite{EXP,RL,BR},
one needs to include $\Delta$ and other
nucleon resonances and hyperons. In the meson sector one has to incorporate
kaons. Meson-meson interaction, short-range correlations
and other effects may also contribute essentially, as we have argued. 
Thermodynamical characteristics,
like the pressure, energy, entropy still should be computed. 
Comparison with lattice results, cf. \cite{Fodor,KRT},
is then required.
Moreover, one should 
supplement the 
consideration of in-medium properties of the system by a relevant model of the
freeze out stage
to match in-medium particle distributions to those distributions
one observes at 
infinity. For sure
the realistic consideration  of strongly interacting heated nuclear system
is a very complicated problem that needs
a further study.
We hope to return to these questions in forthcoming publications.

\section*{Acknowledgement}

The author thanks N.O. Agasian, 
Yu.B. Ivanov, J. Knoll,  E.E. Kolomeitsev, M.F.M. Lutz
and V.D. Toneev
for the discussions. He acknowledges
the hospitality and support of GSI Darmstadt. The work has been
supported in part by DFG (project 436 Rus 113/558/0-2), and by
RFBR  grant NNIO-03-02-04008.

\section*{APPENDICES}
\appendix
\section{Formalism}\label{Formalism}

\subsection{Green functions and self-energies}

In matrix $\{-,+\}$
notations there exist
exact relations between
two-point functions  $F$
(Green functions and self-energies), 
cf. \cite{IKV}, 
\be\label{rel}
\widehat{F}^R &=&\widehat{F}^{--}-\widehat{F}^{-+}=\widehat{F}^{+-}-
\widehat{F}^{++}, \,\,\,\nonumber\\ 
\widehat{F}^A &=&\widehat{F}^{--}-\widehat{F}^{+-}
=\widehat{F}^{-+}-\widehat{F}^{++},
\ee
$\widehat{F}^R$ and  $\widehat{F}^A$ 
are retarded and advanced two-point functions (operators in spin space). 
The  co-contour functions are $\widehat{F}_{jk}=\sigma_{jl}
\sigma_{km}\widehat{F}^{lm}$, 
$\sigma_{j}^k=\delta_{j}^k$,  $\sigma_{jk}=
\sigma^{jk}$, are standard $\sigma_3$ Pauli matrices,  space indices
$j,k,l,m=1,2,3$,
$\delta_{j}^k =1$ for $j=k$ and $\delta_{j}^k =0$ for $j\neq k$.
\footnote{We use notations of ref. \cite{IKV}, where 
Green functions $\widehat{G}^{ij}$ (here $i,j =\{+,-\}$) are the same, as 
in refs. \cite{LP81,KV96}, and $\widehat{\Sigma}^{-+}$, 
$\widehat{\Sigma}^{+-}$ differ by the sign from quantities
used in refs \cite{LP81,KV96}.
Self-energies 
$\widehat{\Sigma}_{ij}$ coincide with  $\widehat{\Sigma}^{ij}$ of 
\cite{LP81,KV96}.} 

The free fermion retarded Green function is
\be\label{zer-f}
\widehat{G}_f^{0,R}(p) =\frac{\widehat{\Lambda}_f^0 (p)}{p^2 -m^2_f +i0}
\equiv G_f^{0,R}(q)\widehat{\Lambda}_f^0  (p), 
\ee
and the free boson retarded Green function is
\be\label{zer-b}
\widehat{G}_b^{0,R}(q) =\frac{\widehat{\Lambda}_b^0 (q)}{q^2 -m^2_b +i0}
\equiv G_b^{0,R}(q)\widehat{\Lambda}_b^0  (q), 
\ee
$\widehat{\Lambda}_f^0$ and  
$\widehat{\Lambda}_b^0$ 
are corresponding spin structure operators of fermions and bosons,
$p,q$ are 4-momenta. 
For the spin  $1/2$ fermions one has $\widehat{\Lambda}_f^0= 
\FMslash{q}+m_f$ and  for the scalar neutral boson  one has
$\widehat{\Lambda}_b^0 =1$, $\FMslash{q}\equiv \gamma^{\mu}q_{\mu}$, $\gamma^{\mu}$
are Dirac matrices, $\mu =0,1,2,3$. 

The spin structure
of the free massive vector-boson propagator
is given by
\be\label{frvec}
\widehat{\Lambda}_b^0 =
(g^{\mu\nu}-q^\mu q^\nu /m_b^2).
\ee

Using (\ref{Dyson}) it is convenient to introduce one-particle 
spectral and width functions (operators), 
\be\label{A-G}
\widehat{A}_i =-2\im \widehat{G}_i^R (q)=-2\im\frac{1}
{ \widehat{M}_i  +i\widehat{\Gamma}_i /2},\,\,\,
\widehat{\Gamma}_i =-2\im 
\widehat{\Sigma}_i^R \,.
\ee
The quantity
\be\label{M}
\widehat{M}_i =(\widehat{G}_i^{0,R})^{-1}-\re 
\widehat{\Sigma}_i^R
\ee
demonstrates the deviation from the mass shell, $\widehat{M}=0$ on the
mass shell in the matter, the relation that introduces quasiparticles.

At equilibrium all non-equilibrium Green functions, 
$\widehat{G}^{- -}$, $\widehat{G}^{+ +}$, $\widehat{G}^{- +}$, and 
$\widehat{G}^{+ -}$, are expressed through the retarded Green function,
cf. \cite{IKV}, e.g.,
\be\label{mpeqG}
i\widehat{G}_i^{-+}=\mp n_i (q_0 ) \widehat{A}_{i}(q),\,\,\,
i\widehat{G}_i^{+-}=(1\mp n_i (q_0 )) \widehat{A}_{i}(q).
\ee
The upper sign corresponds to fermions, the lower, to bosons,
\be
n_i (q_0 )=\frac{1}{\mbox{exp}[(q_0 -\mu_i )/T]\pm 1}
\ee
are fermion/boson occupations, $\mu_i$ are chemical potentials.
Relations for $\widehat{\Sigma}_i^{-+}$ and
$\widehat{\Sigma}_i^{+-}$ are similar to eqs
(\ref{mpeqG}):
\be\label{mpeqS}
i\widehat{\Sigma}_i^{-+}&=&-i\widehat{\Sigma}_{-+,i}=
\mp n_i (q_0 ) \widehat{\Gamma}_{i}(q),\,\,\,\nonumber \\
i\widehat{\Sigma}_i^{+-}&=&-i\widehat{\Sigma}_{+-,i}
=(1\mp n_i (q_0 )) \widehat{\Gamma}_{i}(q) .
\ee
 
\subsection{Particle-antiparticle separation}

To deal with Green functions of physical particles and
antiparticles we introduce the decomposition of  Green functions
into particle ($+$) and antiparticle ($-$) parts
\be\label{G-R}
\widehat{G}_i^{R}(q)&=&\Theta (q_0)\widehat{G}_{i,(+)}^{R}(q)
+\Theta (-q_0)\widehat{G}_{i,(-)}^{A}(-q),\nonumber\\
\label{G-pm}
\widehat{G}_i^{-+}(q)&=&\Theta (q_0)\widehat{G}_{i,(+)}^{-+}(q)
+\Theta (-q_0)\widehat{G}_{i,(-)}^{+-}(-q)\,,
\ee
\be
\widehat{A}_i (q)=
\Theta (q_0)\widehat{A}_{i,(+)}(q)
-\Theta (-q_0)\widehat{A}_{i,(-)}(-q),
\ee
where $\Theta (x)=1$ for $x>0$ and $\Theta (x)=0$ for $x\leq 0$.
These equations allow to reduce the integration in $q_0$  
in self-energy diagrams to positive energies.  

Spectral functions 
satisfy sum rules:
\be\label{fsum-r}
\frac{1}{N_f} \mbox{Tr} \int^{\infty}_0 
\gamma_0  \left[\widehat{A}_{f,(+)} (q_0 ,\vec{q})+
\widehat{A}_{f,(-)} (q_0 ,-\vec{q})\right] 
\frac{d
  q_0}{2\pi}=2,\,\,\,
\ee
\be\label{bsum-r}
\frac{1}{N_b}\mbox{Tr} \int^{\infty}_0 
q_0 \left[\widehat{A}_{b,(+)} (q_0 ,\vec{q})+\widehat{A}_{b,(-)} 
(q_0 ,-\vec{q})\right] 
\frac{d
  q_0}{2\pi}=1 .
\ee
The trace is taken over spin degrees of freedom.
$N_f =(2s_f +1)$ is the fermion degeneracy factor 
and  $N_b =(2s_b +1)$ is the boson degeneracy factor.
For vector bosons sum rules are fulfilled separately for transversal 
($N_b =3$, $\widehat{A}_{b,(\pm)}\rightarrow \widehat{A}_{b,(\pm)}^t$) and
longitudinal ($N_b =1$, $\widehat{A}_{b,(\pm)}\rightarrow 
\widehat{A}_{b,(\pm)}^l$) degrees of freedom, see eqs (\ref{frvec-m}), 
(\ref{tr-l}).

At the thermal equilibrium 
we may introduce occupations of physical particles and antiparticles as
\be\label{rel-oc}
n_{i,(+)} (q_0 )=n_i (q_0 )\Theta (q_0 ), \,\,\,
n_{i,(-)} (q_0 )=-[n_i (-q_0 )\mp 1]\Theta (q_0 ),
\ee
\be\label{rel-oc1}
n_{i,(+)} (q_0 )&=&
\frac{\Theta (q_0 )}{\mbox{exp}[(q_0 -\mu_i )/T]\pm 1},\,\,\,\nonumber\\
n_{i,(-)} (q_0 )&=&\frac{\Theta (q_0 )}{\mbox{exp}[(q_0 +\mu_i )/T]\pm 1} \,.
\ee
In-medium
3-momentum distributions of particles and antiparticles are obtained with
the help of (\ref{rel-oc1}) from the definition of the zero 
component of the Noether 4-current, cf. 
\cite{IKV},
\be\label{3oc}
\widehat{n}_{i,(+)} (\vec q )&=&\int^{\infty}_{0}
\frac{\widehat{v}^0_i 
\widehat{A}_{i,(+)}(q)}{\mbox{exp}[(q_0 -\mu_i )/T]\pm 1}
\frac{d q_0}{2\pi},\nonumber \\
\widehat{n}_{i,(-)} (\vec q )&=&\int^{\infty}_{0}
\frac{
\widehat{v}^0_i\widehat{A}_{i,(-)}(q)}{\mbox{exp}[(q_0 +\mu_i )/T]\pm 1}
\frac{d q_0}{2\pi},
\ee
$\widehat{v}^{0}_i =\gamma_0$ 
for fermions and $\widehat{v}^{0}_i =2q_0$ for
bosons. The chemical potential of the antiparticle differs by the sign from
the chemical potential of the particle. 
For the non-strange
bosons the boson number is not conserved and the chemical potential
$\mu_b =0$. 

For a large size system we may neglect particle radiation from the surface and
the baryon 
chemical potential is determined using the baryon
number conservation
\be\label{barden}
\rho_{\rm bar}=\rho_{f,(+)}-\rho_{f,(-)}.
\ee
The total Noether 4-current presented
as the sum of partial contributions is conserved
quantity \cite{IKV}. It is expressed through the standard $A_i$ spectral
functions. As it has been shown in \cite{IKV03}, in thermal equilibrium the Noether current
coincides with the conserved effective  current (that can be derived from the
Kadanoff--Baym kinetic equation),
provided $\Phi$ derivable
approximations are used. Refs \cite{IKV} have demonstrated thermodynamical
consistency of the approach. Thus the zero component of the Noether current coincides 
with the corresponding
quantity obtained with the help of the standard thermodynamical relation, e.g.
for the baryon density 
\be
\rho_{\rm bar} =\frac{\partial \Omega}{\partial \mu_{\rm bar}}.
\ee
Using the latter expression ref.  \cite{Weinhold} studied the nucleon -- delta
isobar -- pion system in the virial limit of Dashen, Ma and Bernstein \cite{DMB}.

Particle and antiparticle Noether densities are
\be\label{den}
\rho_{i,(\pm)}=\mbox{Tr}\int \widehat{n}_{i,(\pm)} (\vec q )
\frac{d^3 q}{(2\pi)^3}.
\ee

In the paper body we focus on the discussion of 
only
slightly 
particle-antiparticle asymmetric systems
and, as the limiting case, {\em{the particle-antiparticle symmetric system 
(heated
hadron vacuum)}}, bearing in mind possible 
applications to 
heavy ion collisions at RHIC and LHC energies, respectively.
We assume $\rho_{\rm bar} \ll \rho_{f,(\pm)}$
and, thereby, $|\mu_i |/T \ll 1$. Then one may simplify above equations
as 
\be\label{oc-sim}
n_{i,(+)} (q_0 )=n_{i}^{sym} (q_0 )\left[ 1+ \frac{\mu_i}{T}(1\mp 
n_{i}^{sym} (q_0 ))\right],
\ee
\be\label{oc-simqa}
n_{i,(-)} (q_0 )=n_{i}^{sym} (q_0 )\left[ 1- \frac{\mu_i}{T}(1\mp 
n_{i}^{sym} (q_0 ))\right],
\ee
where  the fermion/boson  quantities are  supplied
by index ``sym'' for the 
particle-antiparticle symmetric system. For $\rho_{\rm bar} \gsim 
\rho_{f,(-)}$ particle-hole effects become important.

\section{Non-relativistic heavy fermions}\label{non-rel}
In the paper body we have shown that in the low temperature limit 
heavy fermions can be treated as
non-relativistic particles, since both the energy deviation from the 
bare fermion mass and the fermion momentum are much smaller than $m_f$.
In the high temperature limit the
non-relativistic approximation for fermions also holds in some temperature
interval, namely, if conditions 
(\ref{typf}), (\ref{typ}) are
fulfilled, the former for the case of the
scalar or pseudo-scalar boson exchange, and the latter for the vector boson
exchange.  
The relativistic study performed in the paper body justifies the validity
of the non-relativistic approach for given cases.
In order to come to non-relativistic expressions from
relativistic ones we need to expand $G_1 \simeq G_2$ near the fermion
mass-shell, $p^2 =m_f^2$,
and to introduce the fermion non-relativistic self-energy (\ref{nonr})
averaged over spin
degrees of freedom. 

Suppose, conditions for
the non-relativistic approximation for fermions are fulfilled.
Then the consideration is simplified.
The free fermion particle/antiparticle Green function  
obeys the equation
\be\label{zer-fnrel}
[G_f^{0,R}(p_0 ,\vec{p})]^{\rm n.rel} =\frac{1}{p_0  -\vec{p}^{\,2} 
/(2m_f ) +i0},
\ee
instead of eq. (\ref{zer-f}) in relativistic case.
The spectral function (instead of (\ref{A-G}), (\ref{M})) is given by
\be\label{Asfnrel}
A_{f}^{\rm n.rel}=\frac{\Gamma_{f}^{\rm n.rel}}{[M_{f}^{\rm n.rel}]^2 +
[\Gamma_{f}^{\rm n.rel}]^2 /4},
\ee
with $M_{f}^{\rm n.rel}=p_0  -\vec{p}^{\,2} 
/(2m_f )-[\re\Sigma^R_f ]^{\rm n.rel}$. 
The sum rule is changed accordingly.
Fermion particle and antiparticle distributions and densities are 
\be\label{3ocnrel}
n_{f}^{\rm n.rel} (\vec q )&=&\int^{\infty}_{-\infty}
\frac{
A_{f}^{\rm n.rel}(q)}{\mbox{exp}[(q_0 \mp \mu_f )/T]+ 1}
\frac{d q_0}{2\pi},
\ee
\be\label{dennrel}
\rho_{f}^{\rm n.rel}=2\int n_{f}^{\rm n.rel} (\vec q )
\frac{d^3 q}{(2\pi)^3},
\ee
cf. eqs (\ref{3oc}) and (\ref{den}). 
Further to shorten notations we 
suppress index $''{\rm non-rel}''$. 

Thermodynamic quantities 
are easily found. 
The Noether energy--momentum tensor \cite{IKV}
%
\begin{eqnarray}
\label{E-M-new-tensork}
\Theta^{\mu\nu}/V
&=& 
\mbox{Tr} 
\left(\frac{1}{2}\right)_{\rm{n.b.}}\sum_{i}
\int (\mp)i G^{-+}_i v^{\mu}_i p^{\nu}_i 
\frac{d^4 p}{(2\pi )^4}\nonumber \\ 
&+& g^{\mu\nu}\left(
{\cal E}^{\rm{int}} -{\cal E}^{\rm{pot}}
\right)+{\cal E}_{\rm{(der.)}}^{\mu\nu}
\end{eqnarray}
%
is {\em exactly} conserved within the $\Phi$ derivable scheme. Here $V$ is the
system volume, $i=\{f,b\}$ indicates 
that sum is over all non-relativistic fermions and relativistic
bosons.
Extra factor $\left(\frac{1}{2}\right)_{\rm{n.b.}}$ occurs
only for neutral bosons. The velocity $v_i^{\mu}=\partial G^{0,R}_i /
\partial p_{\mu}$.
The  
potential
energy ${\cal E}^{\rm{pot}}$, which a probe particle with Wigner density 
$G^{-+}$ would have due to the interaction with all other particles in
the system, is 
%
\begin{eqnarray}
\label{eps-potk-K-B}
{\cal E}^{\rm{pot}}
= 
\mbox{Tr}
\left(\frac{1}{2}\right)_{\rm{n.b.}}\sum_i
\int
\frac{d^4 p_i}{(2\pi )^4}
 \left[(\mp)iG^{-+}_i
\re \Sigma^R_i    
+ (\mp)i \Sigma^{-+}_i\re G^R_i 
\right],   
\end{eqnarray}
%
and the interaction energy ${\cal E}^{\rm{int}}$ is simply related to 
${\cal E}^{\rm{pot}}$, 
%
\begin{eqnarray}\label{int-spec}
{\cal E}^{\rm{int}}=\frac{2}{\alpha}{\cal E}^{\rm{pot}},
\end{eqnarray}
%
in case all the interaction vertices of the theory have the same number of 
lines attached to them. For the two-fermion--one-boson 
interaction, see (\ref{phi}),  one has $\alpha=3$, that results in
${\cal E}^{\rm{int}}=\frac{2}{3}{\cal E}^{\rm{pot}}$. 
The additional term 
${\cal E}_{\rm{(der.)}}^{\mu\nu}$ appears in
eq. (\ref{E-M-new-tensork}) only in the case of derivative coupling,
cf. \cite{IKV03}.
The energy and the pressure at the quasi-equilibrium are given by 
\begin{eqnarray}
E=\Theta^{00},\,\,\,\,
P=-\Omega /V =\frac{1}{3}\left( \Theta^{11}+\Theta^{22}+\Theta^{33}\right),
\end{eqnarray}
\begin{eqnarray}
&&\Omega =V\mbox{Tr} \sum_i \int \frac{d^4 p_i}{(2\pi )^4}n_i (p_{0i})\left\{
-2\mbox{Im}\,\,\mbox{ln}\left[ -G_i^R (p_{0i} +i0, \vec{p}_i
)\right]\right.\nonumber\\
&&\left. -\mbox{Re}G_i^R \Gamma_i -A_i \mbox{Re}\Sigma_i^R\right\}+\Phi_T
,\quad \Phi_T =-iT\Phi ,
\end{eqnarray}
$\Phi$ is the functional expressed in terms of non-equilibrium contour Green
functions,
cf. \cite{IKV}.
The expression for the entropy is as follows
%
\begin{eqnarray}
\label{Scorn} 
s^0 &=&S/V 
= -
\mbox{Tr} \sum_i \int \frac{d^4 p_i}{(2\pi )^4}
\frac{\partial n_i (p_{0i})}{\partial T}\nonumber \\
&\times&
\left[- 2\mbox{Im}\;\ln\left[-G_i^R (p_{0i}+i 0,\vec{p}_i )\right]
- \mbox{Re}G_i^{R}\Gamma_i -A_i
\mbox{Re}\Sigma_i^{R}\right]
-\frac{\partial \Phi_T}{\partial T}.
\end{eqnarray} 
%

This way one may  recover all necessary thermodynamic characteristics 
of the system.

\end{fmffile}
\end{document}